\documentclass[12pt]{article}
\pdfoutput=1
\usepackage[centertags]{amsmath}
\allowdisplaybreaks[1]
\usepackage{array,multirow}
\numberwithin{equation}{section}
\usepackage{amssymb}
\usepackage{float}
\usepackage{graphicx}
\usepackage{color}
\usepackage{relsize}
\usepackage{verbatim}
\usepackage{mathtools}
 
\usepackage{epsfig}

\def\Or[#1]{{\text{O}}\left({#1}\right)}
\def\dotl[#1,#2]{\left\langle #1, #2 \right\rangle}
\def\dotlb[#1,#2]{[ #1, #2 ]}
\def\dotp[#1,#2]{(#1) \cdot (#2)}
\def\aff[#1,#2]{\hat{#1}(#2)}
\def\n4sym{{\cal N}=4 SYM}
\def\>{\rangle}
\def\<{\langle}
\def\weight[#1,#2,#3]{\{(#1),#2,#3\}}
\def\ads[#1]{$\text{AdS}_{#1}$}

\newcommand{\ba}{\begin{eqnarray}}
\newcommand{\ea}{\end{eqnarray}}
\newcommand{\be}{\begin{eqnarray}}
\newcommand{\ee}{\end{eqnarray}}
\newcommand{\bq}{\begin{equation}}
\newcommand{\eq}{\end{equation}}
\newcommand{\benn}{\begin{equation*}}
\newcommand{\eenn}{\end{equation*}}
\newcommand{\bi}{\begin{itemize}}  
\newcommand{\ei}{\end{itemize}}

\newcommand{\CA}{{\cal A}}
\newcommand{\CS}{{\cal S}}

\newcommand{\CL}{{\cal L}}

\newcommand{\CN}{{\cal N}}
\newcommand{\CO}{{\cal O}}
\newcommand{\CP}{{\cal P}}

\newcommand{\CV}{{\cal V}}
\newcommand{\nn}{\nonumber}

\newcommand\oo\infty
\newcommand\s\sigma
\newcommand\de\delta
\newcommand\De\Delta

\newcommand\f\phi
\newcommand\g\gamma
\newcommand\x\times

\usepackage{jheppub}

\makeatletter
\def\@fpheader{\vspace{-.1cm}}
\makeatother

\title{The Bulk-to-Boundary Propagator \\ in Black Hole Microstate Backgrounds}

\author[a]{Hongbin Chen,} 
\author[b]{A. Liam Fitzpatrick,} 
\author[a]{Jared Kaplan,} 
\author[a]{Daliang Li}
\affiliation[a]{Department of Physics and Astronomy,  Johns Hopkins University, \\
Charles Street, Baltimore, MD 21218, USA} 
\affiliation[b]{Department of Physics, Boston University, \\
Commonwealth Avenue, Boston, MA 02215, USA}

\abstract{ 
First-quantized propagation in quantum gravitational AdS$_3$ backgrounds can be exactly reconstructed using CFT$_2$ data and Virasoro symmetry.  We develop methods to compute the bulk-to-boundary propagator in a black hole microstate, $\< \phi_L \CO_L \CO_H \CO_H\>$, at finite central charge.  As a first application, we show that the semiclassical theory on the Euclidean BTZ solution sharply disagrees with the exact description, as expected based on the resolution of forbidden thermal singularities, though this effect may appear exponentially small  for physical observers.  
} 

\arxivnumber{}
 
\begin{document} 
      
\maketitle
\flushbottom
 
\section{Introduction} 

Perturbative gravitational physics in AdS$_3$ is largely determined by the Virasoro algebra of CFT$_2$ \cite{HartmanLargeC, DongGravityRenyi, Fitzpatrick:2014vua, Roberts:2014ifa, Fitzpatrick:2015zha, Fitzpatrick:2015foa, Anous:2016kss, Asplund:2015eha, Asplund:2014coa, TakayanagiExcitedStates, Beccaria, KrausBlocks, Hijano:2015qja, Alkalaev:2015wia, Alkalaev:2015lca, Chen:2016dfb, Lashkari:2017hwq, Maxfield:2017rkn, Kusuki:2018nms, Hikida:2018dxe,  Kraus:2017kyl, Cotler:2018zff}.  But one can go further, and  explicitly compute many nonperturbative quantum gravitational effects  \cite{Fitzpatrick:2016ive, Chen:2017yze, Kusuki:2018wcv, Fitzpatrick:2016mjq, Kusuki:2018wpa, Kraus:2018zrn} as well.  These  include a prescription for bulk reconstruction that  incorporates the  exchange of all multi-graviton states \cite{Anand:2017dav}, and has led to a quantitative prediction for the breakdown of bulk locality at the non-perturbative level in $G_N$ \cite{Chen:2017dnl}.  In this work we will study the heavy-light bulk-boundary correlator 
\be \label{eq:BulkBoundaryCorrelatorwithBH}
\CA(y, z, \bar z) = \left\langle \CO_H(\infty) \CO_H(1) \CO_L(z, \bar z) \phi_L(y,0,0) \right\rangle
\ee
which can be used to explore the limits of gravitational effective field theory, including in the near horizon region of the black hole microstate created by $\CO_H$.  We will primarily focus on the  pure graviton contributions to this observable.

In the remainder of this introduction we will discuss an aspect of the information paradox associated with Euclidean correlators.  Then we will provide a physical interpretation for the bulk field $\phi$ and  a summary of the technology developed to compute universal contributions to $\CA$.  In this paper we will largely focus on technical machinery, while in future work we hope to use these methods to study infalling observers.

\begin{figure}[th!]
\begin{center}
\includegraphics[width=0.48\textwidth]{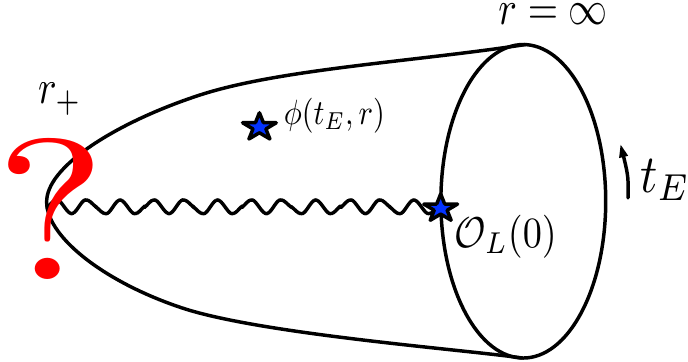}
\caption{This figure depicts a Euclidean bulk-boundary correlator in a black hole microstate.   Although we have forced the correlator to live on the Euclidean BTZ geometry, due to  violations of the KMS condition  the correlator will be multivalued on the Euclidean time circle, and so must have a branch cut.  Thus semiclassical predictions for bulk correlators must breakdown.  In particular, as the Euclidean time circle shrinks to vanishing size at the  horizon, it would seem that exact bulk correlators must differ signficantly from their semiclassical limits at the Euclidean horizon.}
\label{fig:MultisheetedCigarwithPhi}
\end{center}
\end{figure}

\subsection{A Problem at the Euclidean Horizon}

Black hole microstates can be sharply differentiated from the canonical ensemble using Euclidean correlators \cite{Balasubramanian:2007qv, Fitzpatrick:2016ive, Chen:2017yze}.  In the canonical ensemble, correlators are subject to the KMS condition, which means that they must be periodic in Euclidean time.  Black hole solutions such as BTZ reflect this periodicity directly in their Euclidean geometry. 

In contrast, microstate correlators cannot exhibit this periodicity \cite{Fitzpatrick:2016ive}.  If we attempt to parameterize them using BTZ Schwarzschild coordinates, then they must be multivalued on the Euclidean time circle, as pictured in figure \ref{fig:MultisheetedCigarwithPhi}.  This suggests that bulk-boundary correlators will be singular at the horizon, where the size of the Euclidean time circle shrinks to zero.   One of our  goals will be to study bulk-boundary correlators near the Euclidean horizon.

\subsection{Quantum Gravitational Propagation}

We recently derived a prescription \cite{Anand:2017dav} for an exact AdS$_3$ proto-field $\phi$ in Fefferman-Graham gauge.  Instead of recapitulating the formal definition of $\phi$ (see section \ref{app:TechnicalReview} for those details), let us consider some physical scenarios where $\phi$ plays a natural role.  These include first quantized propagation in a quantum gravitational background, and a universe including only a free-field coupled to gravity at low energies.

We can view $\phi$ as a short-hand for an operator sourcing first-quantized propagation in a quantum gravitational background.  That is, to all orders in gravitational perturbation theory about a background created by distant sources, in the vacuum sector we have an operator relation \cite{Anand:2017dav, Maxfield:2017rkn}
\be
\phi(X_1) \phi(X_2) = \exp \left[ - m \int_{X_1}^{X_2} ds \sqrt{g_{\mu \nu} \dot Y^\mu \dot Y^\nu}  \right]
\ee 
This formula includes both quantum gravitational interactions with external sources, such as CFT operators, as well as gravitational self-interactions.

However, $\phi$ does not include loops of matter fields, including itself.  To clarify this, consider a complete AdS$_3$ theory whose sub-Planckian spectrum consists of a single species of scalar particles with purely gravitational interactions.  That is, a theory with a low-energy effective action 
\be
\label{eq:FreeTheoryLagrangian}
S_{\mathrm{universe}} = \int d^3 x \sqrt{-g} \left( \frac{1}{2} (\nabla \varphi)^2 - \frac{m^2}{2} \varphi^2 + \frac{1}{16 \pi G_N} R - 2 \Lambda \right)
\ee
Above the  Planck scale, we do not have any particular requirements for the universe other than those imposed upon us by symmetry, unitarity, and crossing.  

This universe will be a large $c$ CFT$_2$ whose spectrum between the vacuum and the Planck scale\footnote{By the Planck scale we mean an energy scale  $\lesssim \frac{c}{24}$; the details won't be important for this informal discussion.  We do not know if a CFT like this actually exists, nor do we know of any bottom-up constraints that make the existence of such a CFT appear problematic.} consists entirely of a Fock space of states generated by the single-trace operator $\CO$ dual to $\varphi$, with generalized free theory OPE coefficients \cite{Liu, Heemskerk:2009pn, Unitarity} modified only by gravitational effects.  In this universe, the single-particle component\footnote{Graviton exchanges in $n+1$-pt correlators induce mixing between $\varphi$ and $n$-particle states, so that $\< \varphi \CO^n \> \neq 0$, whereas $\phi$ has a vanishing 2-pt function with multi-trace operators. We describe this in more detail in appendix \ref{app:GravMultiTrace}.}
 of the effective field $\varphi$  will correspond  with the proto-field operator $\phi$ constructed from $\CO$.  This follows because $\varphi$ has only gravitational interactions, which are encoded in the Virasoro algebra and were incorporated into the definition of $\phi$.  This universe must contain a Cardy spectrum of black holes at energies  $E > \frac{c}{6}$, so it provides a very convenient laboratory to explore the interactions of particles with black holes, including near horizons.    But the reconstructed proto-field $\phi$ still differs  from the field $\varphi$, as $\phi$ only incorporates gravitational loops, and not loops of itself.  
 
Although we have used the language of perturbation theory to describe $\phi$, as we review in section \ref{sec:DefinitionField}, $\phi$ is  defined using  symmetry considerations at finite $c$.

\subsection*{Universal Contributions to $\CA$}

In this work we will mostly focus on the pure graviton contributions to $\CA$.  But our techniques can be used to compute more general `bulk-boundary Virasoro blocks', where full Virasoro representations are exchanged between a pair of boundary operators and a bulk-boundary pair.    
So it is natural to ask to what extent the behavior of the full $\CA$ correlator will differ in more general holographic CFT$_2$s.

One way to partially address this question is by adapting OPE convergence analyses and large $c$ asymptotics \cite{Pappadopulo:2012jk, Hartman:2014oaa, Cardy:2017qhl, Kraus:2017kyl} to estimate the effect of new interactions and high-energy states on $\CA$.  That is, the correlator can be expanded as
\be
\label{eq:SumOverBlocks}
\CA(y,z, \bar z) = \sum_{h, \bar h} C_{HH; h, \bar h} C_{LL; h, \bar h} \CV_{h, \bar h}(y,z, \bar z)
\ee
where $C$ are conventional OPE coefficients and $\CV$ are new bulk-boundary conformal blocks involving primaries $\CO_{h, \bar h}$ exchanged between the heavy and light operators.  

The convergence rate of this expansion will depend on the kinematic configuration defined by $\CO(z, \bar z ) \phi(y,0,0)$, providing information about the sensitivity of $\CA$ to high-energy (or spin) states and OPE coefficients.  Near the breakdown of convergence, the correlator $\CA$ will be UV sensitive, but in regions where the convergence is rapid, the correlator will be dominated by the exchange of low-dimension primaries,\footnote{In the free field $+$ gravity universe at infinite $c$, the vacuum Virasoro block and its images under crossing will dominate, as discussed in  section \ref{sec:ReviewSemiclassiclProbeCorrelators}.  In a more general holographic CFT$_2$ the correlator will be dominated by the exchange of  low-dimension primaries associated with light bulk fields. \cite{JP, AdSfromCFT}} 
leading to a universal gravitational prediction.  Thus the vacuum or pure gravity contribution\footnote{For simplicity, we only wrote down the holomorphic descendant states in equation (\ref{eq:AVacuumBlockDefinition}), but since $\CV_0(y,z,\bar z)$ does not  factorize, we also need to include the anti-holomorphic descendant states. We will denote a projection operator like that in equation (\ref{eq:AVacuumBlockDefinition}) as $\mathcal{P}_h^{\text{holo}}$ and a full projection operator that also includes the anti-holomorphic contributions as $\mathcal{P}_{h, \bar h}$. We mostly consider scalar exchanged states ($\bar h=h$) and in particular the vacuum ($h=\bar h = 0$) in this paper so we will often omit $\bar h$ in the subscript.}
\begin{smaller}
\begin{equation}
\label{eq:AVacuumBlockDefinition}
\CV_0(y, z, \bar z) = \left\langle \CO_H(\infty) \CO_H(1)  \left( \sum_{\{m_i\},\{ n_j\}}  \frac{L_{-m_1} \cdots L_{-m_i} | 0 \> \< 0 | L_{n_{j}} \cdots L_{n_1} }{\CN_{\{m_i\}, \{n_j\}} }\right) \CO_L(z, \bar z) \phi_L(y,0,0) \right\rangle
\end{equation}
\end{smaller}will be a major focus of study in this work, though the techniques we develop are also applicable to the calculation of  $\CV_{h, \bar h}$ associated with the exchange of any state. 

The full bulk operator $\varphi$ will receive other important corrections, as full bulk fields involve sums of proto-fields.  In perturbation theory, this means that $\varphi$ will contain small admixtures of multi-trace operators \cite{Hamilton:2006az, Kabat:2011rz, Kabat:2016zzr}.  Instead of the sum in equation (\ref{eq:SumOverBlocks}), these effects will appear as sums over the external operators $\CO$ contained in $\varphi$.  We will not explore these effects here, but understanding or constraining their contributions in detail is an important problem as it would shed light on the difference between correlators of proto-fields and full bulk fields.

\subsection{Summary} 

This work largely consists of technical developments to compute the bulk-boundary Virasoro blocks $\CV_{h, \bar h}$ contributing to $\< \CO_H \CO_H \CO_L \phi_L \>$, with $\phi_L$ the Fefferman-Graham gauge proto-field \cite{Anand:2017dav} defined by the bulk primary condition.  
We mostly focus on the vacuum block contribution $\CV_0(y,z, \bar z)$ of equation (\ref{eq:AVacuumBlockDefinition}), though all our methods can be applied to general blocks.  

We review the fact that $\CV_0$  determines the physics of propagation in a semiclassical gravitational background in section \ref{app:TechnicalReview}.  We also briefly review the bulk primary condition and the definition of $\phi$.  Then, in the remaining sections,  
technical developments include:
\begin{itemize}
\item  We compute the semiclassical limit  $\CV_{0}^{\mathrm{semi}}$  (section \ref{sec:SemiclassicalandSymmetry}) and show explicitly that it agrees with BTZ correlators. We develop a monodromy method \cite{ZamolodchikovRecursion,Zamolodchikovq} for computing bulk-boundary blocks.  We also define their symmetry transformations precisely, and show that these greatly constrain their form.  
\item We develop three methods (section \ref{sec:ExactCoorelators}) to compute the bulk-boundary blocks in either a $y$ (radial direction) or $z, \bar z$ expansion, but exactly in $h_H, h_L, c$, and  attach Mathematica implementations.  These methods match the semiclassical BTZ correlators in appropriate limits, as shown in figure \ref{fig:SemiMatchesExact}. In Appendix \ref{app:OPEBlocks}, we  used the OPE block method \cite{Czech:2016xec, Fitzpatrick:2016mtp, Anand:2017dav, Besken:2018zro} to compute $\CV_{0}$ perturbatively at order $1/c^2$. 
\end{itemize}

On a more conceptual level, in section \ref{sec:ExploringHorizon} we demonstrate that the semiclassical approximation fails if we interpret $\CV_0$ as a correlator on the Euclidean BTZ solution.  For explicit results, see figures \ref{fig:ExactVsSemiclassical2} and \ref{fig:ExactSemiDisk}.  In this regard the Euclidean horizon is a special place where derivatives of the correlator become singular.  But in the most conservative interpretation, these singularities may have a non-perturbatively small coefficient.

\section{Brief Technical Review} 
\label{app:TechnicalReview}

In this section we provide a very brief review.  In section \ref{sec:ReviewSemiclassiclProbeCorrelators} we discuss BTZ correlators, emphasizing that in the semiclassical limit, they are entirely determined by summing the vacuum Virasoro block over all possible OPE channels \cite{KeskiVakkuri:1998nw}.  Then in section \ref{sec:DefinitionField} we review our bulk reconstruction prescription, and the relation between BTZ Schwarzschild coordinates and other coordinate systems. 

\subsection{Semiclassical Probe Correlators in a BTZ Black Hole Background}
\label{sec:ReviewSemiclassiclProbeCorrelators}

The spherically symmetric BTZ black hole background has a Euclidean metric
\be \label{eq:UsualBTZMetric}
ds^2 = (r^2 - r_+^2) dt_E^2 + \frac{dr^2}{r^2 - r_+^2} + r^2 d \theta^2
\ee
with the Lorentzian metric related by $t_E \to i t$.  Note that the horizon radius
\be
r_+ = 2 \pi T_H = \sqrt{\frac{24 h_H}{c} - 1}
\ee
where $T_H$ is the Hawking temperature, $h_H$ is the (holomorphic) heavy operator dimension, and $c = \frac{3}{2G_N}$ is the central charge of the CFT$_2$. The full semiclassical bulk-boundary correlator for a free field in this geometry\footnote{By this we mean the limit $c \to \infty$ with $h_L$ and $\frac{h_H}{c}$ fixed, so that the light free field acts as a probe.}  is given by the image sum \cite{KeskiVakkuri:1998nw}
\begin{equation}
\label{eq:semiclassicalbulkboundary}
\CA_{\mathrm{semi}}= \<\phi\CO\>_{\text{BTZ}}= \left( \frac{r_+}{2} \right)^{2h_L} \sum_{n= -\infty}^\infty \frac{1}{\left[ \frac{r}{r_+} \cosh (r_+(\delta \theta + 2 \pi n)) -\frac{\sqrt{ r^2 - r_+^2}}{r_+ } \cos(r_+ \delta t_E)  \right]^{2 h_L}}
\end{equation}
where $\delta \theta$ and $\delta t_E$ are differences between the bulk and boundary values of the cylindrical coordinates $t_E$ and $\theta$, and $r$ is the location of $\phi$ in the radial direction.
The sum guarantees periodicity under $\theta \to \theta + 2 \pi$ for the angular coordinate.  The geometry and the correlator are periodic under $t_E \to t_E + \beta$, enforcing the KMS condition geometrically, and avoiding a conical singularity at the horizon $r = r_+$.

If we take the limit $r \to \infty$ and rescale the bulk-boundary correlator by $r^{2h_L}$, we obtain a probe CFT 2-pt correlator in the BTZ geometry.  This is a semiclassical approximation to a heavy-light CFT 4-pt correlator.  In the OPE limit where the light probe operators collide, this 4-pt function has a Virasoro block decomposition.  The only Virasoro primary states that propagate in this light-light OPE channel are the vacuum and double-trace operators.  

The semiclassical vacuum Virasoro block contribution is simply the $n=0$ term of the sum in equation (\ref{eq:semiclassicalbulkboundary}).  In other words, in the semiclassical limit 
\be
\label{eq:semiclassicalbulkboundaryvacuumblock}
\CV_0^{\mathrm{semi}} &=& \left( \frac{r_+}{2} \right)^{2h_L} \frac{1}{\left[ \frac{r}{r_+} \cosh (r_+ \delta \theta )  -\frac{\sqrt{ r^2 - r_+^2}}{r_+ } \cos (r_+ \delta t_E ) \right]^{2h_L}}
\ee
is the bulk-boundary vacuum block, generalizing the semiclassical heavy-light vacuum block \cite{Fitzpatrick:2014vua}. We'll show how to obtain this semiclassical result in Section \ref{sec:SemiclassicalandSymmetry}.

Clearly the $n \neq 0$ terms in  equation (\ref{eq:semiclassicalbulkboundary}) must also be intimately connected to the Virasoro vacuum block, since all of the terms in the summation have its functional form.  From the point of view of the bootstrap, the image sum simply satisfies crossing symmetry in the simplest possible way,  as it sums the inherently crossing asymmetric Virasoro vacuum block over all possible OPE channels.   This means that in the semiclassical limit, bulk-boundary correlators in a black hole background are fully determined by the vacuum block, suggesting that  universal features of AdS$_3$ quantum gravity can be understood by computing $\CV_0$ of equation (\ref{eq:AVacuumBlockDefinition}) exactly.

\subsection{CFT Definition of the Bulk Proto-Field}
\label{sec:DefinitionField}

For completeness we will now summarize the definition of the bulk proto-field operator $\phi$; for derivations and explanations see \cite{Anand:2017dav}.
In Fefferman-Graham gauge, where the vacuum AdS$_3$ metric takes the form 
\be
\label{eq:FGMetric}
ds^2 = \frac{dy^2 + dz d \bar z}{y^2} -  \frac{1}{2}S(z) dz^2 -  \frac{1}{2} \bar S(\bar z) d \bar z^2
+ y^2 \frac{ S(z) \bar S(\bar z)}{4 } dz d \bar z
\ee
for general holomorphic and anti-holomorphic functions $S, \bar S$, a bulk scalar proto-field must satisfy the bulk primary conditions \cite{Anand:2017dav}
\be \label{eq:BulkPrimaryCondition}
L_{n \geq 2} \phi(y,0,0) | 0 \> = 0, \ \ \ \bar L_{n \geq 2} \phi(y,0,0) | 0 \> = 0
\ee
along with the condition that in the vacuum, the bulk-boundary propagator is
\be \label{eq:NormalizationCondition}
\< \CO(z, \bar z) \phi(y,0,0) \> = \frac{y^{2 h_L}}{(y^2 + z \bar z)^{2 h_L}}
\ee
These conditions uniquely and exactly determine $\phi(y,0,0)$ as a CFT operator defined by its series expansion in the radial $y$ coordinate:
\be
\label{eq:phiasseriesiny}
\phi(y,0,0) = y^{2h_L} \sum_{N=0}^\infty \frac{(-1)^N y^{2N}}{N! (2h_L)_N } \CL_{-N} \bar \CL_{-N} \CO(0)
\ee
The $\CL_{-N}$ are polynomials in the Virasoro generators at level $n$, with coefficients that are rational functions of the dimension $h_L$ of the scalar operator $\CO$ and of the central charge $c$.  
For example
 \be
 \label{eq:MathematicalLMinus2}
\mathcal{L}_{-2}=\frac{(2 h+1) (c+8 h)}{\left(2h+1\right)c+2 h (8 h-5)}\left(L_{-1}^2-\frac{12h}{c+8h}L_{-2}\right)
 \ee
Note that in the limit $c \to \infty$, we have $\CL_{-N} \to L_{-1}^N$ and $\bar \CL_{-N} \to \bar L_{-1}^N$, and our $\phi$ matches known results \cite{Banks:1998dd, Bena:1999jv, Hamilton:2006az} for bulk reconstruction in the absence of gravity.  In some situations it is convenient to compute the properties of a simpler object, which we refer to as the `holomorphic’ part of $\phi$ \cite{Chen:2017dnl}; it is defined by replacing the anti-holomorphic $\bar \CL_{-N} \to \bar L_{-1}^N$, so that anti-holomorphic gravitons are neglected.

This CFT operator $\phi$, inserted in correlation functions such as $\<\phi O T\>$ and $\<\phi\phi\>$ correctly reproduces the result of Witten diagram calculations\footnote{These Witten diagram calculations were performed in the Fefferman-Grahm gauge to facilitate the comparison. In \cite{Verlinde:2015qfa, Lewkowycz:2016ukf} another construction for $\phi$ was proposed, which differs perturbatively from the bulk reconstruction adopted in this paper.} in the bulk~\cite{Anand:2017dav, Chen:2017dnl}. We'll show explicitly in this paper that $\phi$ inserted in states generated by heavy operators correctly reproduces the correlator of a scalar field on the corresponding non-trivial background geometry.

The function $S(z), \bar S(\bar z)$ in the metric (\ref{eq:FGMetric}) are related to expectation values of the boundary stress-energy tensor $T(z), \bar T(\bar z)$ by 
\begin{equation}
S(z) = \frac{12}{c}T(z), \qquad \bar S(\bar z)=\frac{12}{c}\bar T(\bar z).
\end{equation}
Throughout this paper we will work with $\phi$ defined in Fefferman-Graham gauge, which is natural in the coordinates $(y, z, \bar z)$, and in virtually all cases of interest we will have 
\begin{equation}
	T(z)=\frac{h_H}{ z^2},\qquad \bar T (\bar z)= \frac{h_H}{ \bar z^2}
\end{equation}
due to the presence of heavy operators. The semiclassical metric (\ref{eq:FGMetric}) is then describing a BTZ black hole in the coordinate system $(y, z, \bar z)$.  However, for clarity,  we will almost always express correlators of $\phi$ using the BTZ coordinates $(r, t_E, \theta)$.  This is simply a re-labeling of spacetime points, and not a gauge transformation.  The relations between the $(y, z, \bar z)$ coordinates in equation (\ref{eq:FGMetric}) and BTZ coordinates are a bit subtle, and are worked out in appendix \ref{app:CoordinateRelations}.  The result for spherically symmetric black holes is
\be
\label{eq:FGandBTZCoordRelation}
y &=& \frac{2}{\tilde{r}} \left( \frac{r - \sqrt{r^2 - r_+^2 -1}}{r_+^2 + 1} \right)e^{t_E}
\nn \\
z &=& \frac{1}{\tilde{r}} e^{t_E + i \theta}
 \\
\bar z &=& \frac{1}{\tilde{r}}e^{t_E - i \theta} \nn
\ee
where 
\begin{equation}
	\tilde{r}\equiv \left( \frac{r + i r_+ \sqrt{r^2 - r_+^2 - 1}}{(1 + i r_+) \sqrt{r^2 - r_+^2} } \right)^{\frac{i}{r_+}}
\end{equation}
and  $r_+ = \sqrt{\frac{24 h_H}{c} -1}$ is the horizon radius. Notice that for $r^2 < r_+^2 + 1$ the $y$ coordinate must be analytically continued into the complex plane, and that in this range the magnitude of $\frac{y^2}{z \bar z}$ remains constant, with only its phase changing with $r$.  

For the configuration $\left\langle \CO_H(\infty) \CO_H(0)\mathcal{O}_{L}\left(1,1\right)\phi_{L}(y, z, \bar{z}) \right\rangle$ that'll be used in Section \ref{sec:Uniformizing}, we can map to the BTZ coordinates $(r,t_E,\theta)$  via the transformation (\ref{eq:FGandBTZCoordRelation}), since the operator $\CO_L$ at $z=\bar z=1$ has $t_E=\theta=0$. This configuration is intuitive and has the nice interpretation of the correlator as a function of the location of $\phi_L$ with fixed $\CO_L$. However, in Section \ref{sec:ExactCoorelators} (and also parts of Section \ref{sec:SemiclassicalandSymmetry}), in order to take advantage of the bulk primary condition for computation, we'll compute $\mathcal{V}_0$ in  the kinematic configuration  $\left\langle \CO_H(\infty) \CO_H(1) \mathcal{O}_{L}\left(z, \bar{z}\right) \phi_{L}(y, 0,0)\right\rangle$. To map this configuration to the BTZ coordinates $(r,t_E,\theta)$, we first perform a conformal transformation to the new configuration $\left\langle \CO_H(\infty) \CO_H(0) \mathcal{O}_{L}\left(1,1\right)\phi_{L}(y', z', \bar{z}') \right\rangle$ with 
\begin{equation}
	y'=\frac{y}{\sqrt{(1-z)(1-\bar z)}},\quad z'=\frac{1}{1-z}, \quad \bar z'=\frac{1}{1-\bar z}
\end{equation}
and then relate the coordinates $(y', z', \bar z')$ to $(r, t_E, \theta)$.  We obtain the transformation from $\left\langle \CO_H(\infty) \CO_H(1) \mathcal{O}_{L}\left(z, \bar{z}\right) \phi_{L}(y, 0,0)\right\rangle$ to the BTZ coordinates $(r, t_E, \theta)$
\begin{align}
\label{eq:yintermsofr}
y  &= 2 \left( \frac{r-\sqrt{r^{2}-r_{+}^{2}-1}}{r_{+}^{2}+1} \right)\nn\\
z  & = 1- \tilde{r}e^{-t_E-i\theta}\\
\bar z&=1- \tilde{r}e^{-t_E+i\theta}\nn
\end{align}
We explain more details of this relation in appendix \ref{app:FGtoBTZ}. We will be using these  relations implicitly when we probe the Euclidean horizon in section \ref{sec:ExploringHorizon}.

 Ultimately, all of these coordinates and their relations are merely labels for the non-local CFT operator $\phi$, which was precisely defined  by the bulk primary conditions and equation (\ref{eq:phiasseriesiny}).  From these algebraic conditions, it might not be obvious that $\phi$ can be interpreted as a field in a dynamical spacetime, nor do these conditions explicitly encode any information about the black hole geometries we will study.  The bulk dynamics are entirely emergent.

\section{Semiclassical Analyses and Symmetry}
\label{sec:SemiclassicalandSymmetry}

The purpose of this section is to connect the bulk primary condition reviewed in section \ref{sec:DefinitionField} to semiclassical correlation functions.  It was implicit in \cite{Anand:2017dav} that correlators of the bulk proto-field $\phi$ automatically reconstruct the leading semiclassical free-field correlators in any vacuum AdS backgrounds, including BTZ black holes; in section \ref{sec:Uniformizing} we will make this explicit.  In section \ref{sec:Monodromy} we  explain how the monodromy method can be used to compute semiclassical $\phi$ (bulk) conformal blocks.  Finally in section \ref{sec:Symmetries} we will use the symmetry transformation properties of $\phi$ to constrain the coordinate dependence of bulk-boundary Virasoro blocks.  We address both $\< \phi \CO \CO_H \CO_H \>$ and a previously unexplained simplification \cite{Chen:2017dnl} in $\< \phi \phi \>$.

\subsection{Semiclassical Bulk Correlators from Uniformizing Coordinates}
\label{sec:Uniformizing}

In this section, we will show that in the background of a heavy state $|B\>$, vacuum block exchange for the correlator $\< B | \phi_L \CO_L | B\>$ automatically reconstructs the leading semiclassical bulk-to-boundary propagator in the bulk vacuum geometry corresponding to $|B\>$.\footnote{By `bulk vacuum geometry', we mean that the bulk stress tensor vanishes, aside from localized sources. For CFT states $|B\>$ created by a product of local operators $\CO_i$ with large scaling dimensions $\Delta_i$, their corresponding bulk stress tensor will be localized to geodesics in the large $\Delta_i$ limit and therefore produce a bulk vacuum geometry.  More generally, the bulk vacuum geometry can be viewed as an approximation where bulk sources are treated as localized.}  This treatment generalizes an argument from \cite{Fitzpatrick:2015zha} to bulk conformal blocks. 

We restrict to states $|B\>$ created by the product of a finite number of local operators $\CO_i$, so that the sources $\CO_i$ can be separated by a ball from the boundary points of the probes,\footnote{For instance, map to the cylinder, with the light boundary operator $\CO_L$ at $\infty$ and the boundary point corresponding to the proto-field at $-\infty$, so they are separated from the finite region containing the sources.  } and the boundary stress tensor $T(z)$ in the state $|B\>$ is holomorphic outside this ball, where we can define the local operator $B(x)$ that corresponds to the state $|B\>$  The bulk conformal block is the contribution to $\< B B \phi_L \CO_L \>$ from the exchange of the vacuum and its Virasoro descendants between $\phi_L \CO_L$ and $BB$: 
\be
\CV_0 \equiv \left\< B(\infty) B(0)\CP_0 \phi_L(y,z, \bar{z}) \CO_L(1) \right\>,
\label{eq:bulkproj}
\ee
where $\CP_0$ is the projection operator onto the vacuum irrep. The background stress tensor is its expectation value in the state $|B\>$:
\be
T_B(z) \equiv \< B | T(z) | B\>.
\ee
We are interested in the limit of infinite $c$ with $\frac{1}{c} T_B(z)$ fixed.  In this case, one can define  uniformizing coordinates $f(z)$, such that they satisfy
\be
\frac{12 T_B(z)}{c} =  S(f,z),
\ee
where $S(f,z)$ is the Schwarzian derivative\footnote{The Schwarzian derivative is defined to be
\begin{equation}\label{eq:Schwarzian}
S\left(f,z\right)=\left\{ f\left(z\right),z\right\} \equiv\frac{f'''\left(z\right)}{f'\left(z\right)}-\frac{3}{2}\left(\frac{f''\left(z\right)}{f'\left(z\right)}\right)^{2}.
\end{equation}
}, so that $\<B| T(f(z))| B\>=0$ in the uniformizing coordinates. In other words, the OPE coefficient vanishes for $T(f(z))$ in the operator product $B\times B$, and  straightforward power-counting of factors of $c$ shows that at infinite $c$, the OPE coefficients for all powers of $T(f(z))$ (normalized by their two-point functions) vanish as well. This is equivalent to the statement that if $\phi_L$ and $\CO_L$ are conformally mapped to the uniformizing coordinates, then at infinite $c$ the only state that contributes in the projection onto the vacuum irrep in (\ref{eq:bulkproj}) is the vacuum state itself.  Therefore in these coordinates, $\< B B\CP_0 \phi_L \CO_L\>$ is just the usual $\< \phi_L \CO_L\>$ bulk-to-boundary propagator in pure AdS. 

The transformation of $\CO_L$ under $z \rightarrow f(z)$ is simply the usual local scalar operator transformation $\CO_L(f(z)) = (f'(z)\bar{f}'(\bar{z}))^{-h_L}\CO_L(z)$. For $\phi_L$, the transformation must be extended into the bulk; by definition, $\phi_L$ transforms by extending $z\rightarrow f(z)$ into the bulk such that Fefferman-Graham gauge is preserved.  This extension is given  \cite{Roberts:2012aq} by $(y, z,\bar{z}) \rightarrow (u, x,\bar{x})$ with
\begin{align}\label{eq:CoordinateTransformationToPoincare}
u & =  y \frac{4 (f'(z) \bar f'(\bar z))^{\frac{3}{2}} }{ 4 f'(z) \bar f'(\bar z) + y^2 f''(z) \bar f''(\bar z) }\\
x &=  f(z) - \frac{2 y^2 (f'(z))^2 \bar f''(\bar z)}{4 f'(z) \bar f'(\bar z) + y^2 f''(z) \bar f''(\bar z)}
\nn \\
\bar x &= \bar f(\bar z) - \frac{2 y^2 ( \bar f'(\bar z))^2  f''( z)}{4 f'(z) \bar f'(\bar z) + y^2 f''(z) \bar f''(\bar z)}\nn
\end{align}
Under this transformation, $\phi_L$ transforms like a bulk scalar, $\phi_L(y, z,\bar{z}) \rightarrow \phi_L(y, z,\bar{z}) = \phi_L(u, x,\bar{x})$.  So, we have
\begin{small}
\begin{align}\label{eq:BBPnewcoords}
\< B(\infty) B(0)\CP_0 \phi_L(y, z,\bar{z}) \CO_L(1) \> &=(f'(1)\bar{f}'(\bar{1}))^{h_L} \left\langle \phi_L(u, x,\bar{x}) \CO_L(f(1),\bar{f}(1)) \right\rangle \nn\\
 &= (f'(1)\bar{f}'(1))^{h_L} \left( \frac{u}{u^2 + (x-f(1)) (\bar{x}-\bar f(1))} \right)^{2h_L},
 \end{align}
 \end{small}where $u, x,\bar{x}$ should be understood to be the functions of $(y, z,\bar{z})$ in (\ref{eq:CoordinateTransformationToPoincare}). This  result reproduces the leading semiclassical contribution to the bulk-to-boundary propagator in a general vacuum metric, which we can write in Fefferman-Graham gauge (\ref{eq:FGMetric}).  This follows first of all from the fact that the coordinate transformation (\ref{eq:CoordinateTransformationToPoincare}) is also the transformation that takes the Fefferman-Graham gauge metric (\ref{eq:FGMetric}) to be the pure AdS metric
 \be\label{eq:PurePoincareMetric}
 ds^2 = \frac{du^2 + dx d \bar{x}}{u^2}.
 \ee
 The semiclassical bulk-to-boundary propagator is therefore given by the pure AdS bulk-to-boundary propagator in the new coordinates, which is just (\ref{eq:BBPnewcoords}), plus a sum over images arising from the fact that the coordinate transformation is typically not single-valued. The result (\ref{eq:BBPnewcoords}) is just one of these images, but each image can be thought of as just the vacuum block in a particular channel \cite{Anous:2016kss}.  Moreover, if $h_L \gg 1$, then there is a sharp transition between regions where one image dominates and the others are subleading. In this case, one can cleanly think of one image as being the dominant semiclassical contribution, which is reproduced by the bulk vacuum block in the corresponding channel.
 
 In the specific case where the heavy state $|B\>$ is created by  a single primary operator $\CO_H$ of weight $h_H$, we can be more explicit.  Using the coordinate transformation (\ref{eq:CoordinateTransformationToPoincare})
with $f\left(z\right)=z^{\alpha}, \bar f(\bar z)=\bar{z}^{\bar \alpha}$, we find that the bulk-to-boundary propagator transformed to the Fefferman-Graham coordinates is
\begin{align}
 & \alpha^{h_L}\bar{\alpha}^{h_L}\left\langle \phi_{L}\left(y,z,\bar{z}\right)\mathcal{O}_{L}\left(1,1\right)\right\rangle _{\text{FG}}\\
= & \left[\frac{4y\alpha\bar{\alpha}z^{\frac{\alpha+1}{2}}\bar{z}^{\frac{\bar{\alpha}+1}{2}}}{4z\bar{z}\left(z^{\alpha}-1\right)\left(\bar{z}^{\bar{\alpha}}-1\right)+y^{2}\left((\alpha+1)z^{\alpha}+\alpha-1\right)\left(\bar{z}^{\bar{\alpha}}\left(\bar{\alpha}+1\right)+\bar{\alpha}-1\right)}\right]^{2h_{L}}\nonumber 
\end{align}
By the above argument,  this is also the semiclassical limit $\mathcal{V}_{0}^{\text{semi}}$
of the bulk-boundary vacuum block $\<\CO_H(\infty)\CO_H(0)\CP_0\CO_L\left(y,z,\bar{z}\right)\CO_L(1)\>$, i.e.
\begin{equation}
\mathcal{V}_{0}^{\text{semi}}=\alpha^{h_L}\bar{\alpha}^{h_L}\left\langle \phi_{L}\left(y,z,\bar{z}\right)\mathcal{O}_{L}\left(1,1\right)\right\rangle _{\text{FG}}.
\end{equation}
To obtain the result in the usual BTZ coordinates $(r,t_E, \theta)$, we can use the coordinate transformations (\ref{eq:FGandBTZCoordRelation}), and the result is exactly the same as equation (\ref{eq:semiclassicalbulkboundaryvacuumblock}). We have also checked this semi-classical result with the result of $\mathcal{V}_{0}$
from the recursion relation (to be introduced in next section) analytically
at low orders and numerically up to order $z^{10}\bar{z}^{10}$
in the limit where $\frac{h_{H}}{c}$ is fixed, and $h_{L}\ll c$.

\subsection{Monodromy Method}
\label{sec:Monodromy}

Our goal in this subsection is to extend Zamolodchikov's monodromy method\footnote{For a nice pedagogical introduction to the monodromy method, see appendix D of \cite{HarlowLiouville}.} \cite{ZamolodchikovRecursion,Zamolodchikovq} for Virasoro conformal blocks to bulk-boundary blocks with three boundary and one bulk proto-field operator.    Although boundary blocks factorize into holomorphic and anti-holomorphic pieces, once a bulk field enters the correlator this does not occur. In \cite{Chen:2017dnl}, we developed the monodromy method for the two-point function $\< \phi \phi\>$ of two bulk proto-fields in a ``holomorphic'' version where only the holomorphic stress tensors are included (all  global descendants, under either $L_{-1}$ or $\bar{L}_{-1}$, are also included).\footnote{This holomorphic bulk block can be obtained by taking a chiral limit where $c_R \gg c_L$ and in particular $c_R$ is infinitely larger than all the other parameters that determine the correlator, so that the right-moving stress tensors decouple; it can therefore be thought of as a chiral gravity limit. }  In this subsection, we will continue to work in this limit for the sake of simplicity, and will relegate some discussion of how to apply the monodromy method to the full block to appendix \ref{app:nonchiralMonoBulk}. 
 
 As usual, the monodromy method begins by considering the wavefunction $\psi$ for a degenerate light operator $\hat{\psi}$ acting on the correlator in the large $c$ limit, where it exponentiates to the form
 \be
 \<  \CO_H(z_1) \CO_H(z_2) \phi_L(y_3, z_3, \bar{z}_3) \CO_L(z_4,\bar{z}_4)\> = e^{\frac{c}{6} g},
 \ee
 with $g \sim \CO(c^0)$ at large $c$.   The wavefunction $\psi$ satisfies the degenerate equation of motion
 \be
 \psi''(z) + \frac{6}{c} T(z) \psi(z) = 0,
 \label{eq:psiwfeq}
 \ee
 where the potential $T(z)$ is the stress tensor acting on the bulk correlator.  Because the bulk field necessarily involves both $z$ and $\bar{z}$ dependence, we will also need to consider the analogous anti-holomorphic degenerate wavefunction $\bar{\psi}$, which satisfies the conjugate of (\ref{eq:psiwfeq}). 

The action of the stress tensors $T(z),\bar{T}(\bar{z})$ on the correlator are determined by the singular parts of their OPE with the bulk and boundary operators.   For the boundary operators $\CO_L, \CO_H$, these singular terms are the standard ones for primary operators and simply depend on the primary operator weights as well as their derivatives, which bring down derivatives of the exponent $g$.  For the bulk operator $\phi$, however, the OPE is more complicated:
\begin{equation}
T(z) \phi(y,w,\bar{w}) \sim - y^2 \frac{ \frac{\partial_{\bar{w}} + y^2 \frac{6}{c} \bar{T}(\bar{w}) \partial_w}{1- y^4 \frac{36}{c^2} T(w) \bar{T}(\bar{w})}}{(z-w)^3} \phi(y,w,\bar{w}) + \frac{1}{2} \frac{y \partial_y \phi(y,w,\bar{w})}{(z-w)^2} + \frac{\partial_w \phi(y,w,\bar{w})}{z-w}.
\label{eq:TphiOPE}
\end{equation}
The origin of the complicated cubic term is the fact that $\phi$ transforms under special conformal transformation $L_1$ by moving around in the bulk in a way that depends on the background geometry.  A similar formula holds for the $\bar{T}(\bar{z}) \phi(y,w,\bar{w})$ OPE, related to the above one by conjugation.  These expressions require some care because, as we will discuss in more detail, the $T,\bar{T}$s that appear on the RHS have singularities that must be regulated appropriately.  We will begin by considering the limit where $h_L/c$ is small, so to leading order $T$ and $\bar{T}$ are just given by their behavior in the heavy state background.  For holomorphic backgrounds, i.e. $\bar{h}_H=0$, we therefore have at leading order in $h_L/c$ that
\be
T(z) \phi(y,w,\bar{w}) &\sim& - y^2 \frac{\partial_{\bar{w}}\phi(y,w,\bar{w}) }{(z-w)^3}  + \frac{1}{2} \frac{y \partial_y \phi(y,w,\bar{w})}{(z-w)^2} + \frac{\partial_w \phi(y,w,\bar{w})}{z-w}, \\
\bar{T}(\bar{z}) \phi(y,w,\bar{w}) &\sim& - y^2 \frac{ \partial_{w} + y^2 \frac{6}{c} T_H(w) \partial_{\bar{w}}}{(\bar{z}-\bar{w})^3} \phi(y,w,\bar{w}) + \frac{1}{2} \frac{y \partial_y \phi(y,w,\bar{w})}{(\bar{z}-\bar{w})^2} + \frac{\partial_{\bar{w}} \phi(y,w,\bar{w})}{\bar{z}-\bar{w}}, \nn
 \ee
 where $T_H$ includes only the contribution from the heavy boundary operators $\CO_H$,
 \be
 T(z) = T_H(z) + T_L(z), \qquad T_H(z) \equiv \frac{\< T(z) \CO_H(z_1) \CO_H(z_2)\>}{\< \CO_H(z_1) \CO_H(z_2)\>},
 \ee
 and therefore $T_H(w)$ is regular when $\phi$ is separated from the $z$ positions of the heavy operators. 
 
 Using the bulk OPE (\ref{eq:TphiOPE}) and the standard boundary OPEs, the potentials for the correlator $\< \CO_H(z_1) \CO_H(z_2) \phi(y_3, z_3, \bar{z}_3) \CO_L(z_4)\>$ are easily seen to be
 \be
 \frac{6}{c} T(z) &=& -\frac{y_3^2 c_{\bar{z}_3}}{\left(z-z_3\right){}^3}+\frac{y_3 c_{y_3}}{2
   \left(z-z_3\right){}^2}+\frac{c_{z_1}}{z-z_1}+\frac{c_{z_2}}{z-z_2}+\frac{c_{z_3}}{z-z
   _3}+\frac{c_{z_4}}{z-z_4} \nn\\
   &&+\frac{h_H}{\left(z-z_1\right){}^2}+\frac{h_H}{\left(z-z_2\right){}^2}+\frac{h_L}{\left(z-z_4\right){}^2}
   \ee
   for the holomorphic potential and 
   \begin{align}
  \frac{6}{c}  \bar{T}(z) =& -\frac{y_3^2 \left(c_{z_3}+y_3^2 \frac{6}{c} T_H\left(z_3\right) c_{\bar{z}_3}\right)}{\left(\bar{z}-\bar{z}_3\right){}^3}+\frac{y_3 c_{y_3}}{2
   \left(\bar{z}-\bar{z}_3\right){}^2}+\frac{c_{\bar{z}_1}}{\bar{z}-\bar{z}_1}+\frac{c_{\bar{z}_2}}{\bar{z}-\bar{z}_2}+\frac{c_{\bar{z}_3}}{\bar{z}-\bar{z}_3}+\frac{c_{\bar{z}_4}}{\bar
   {z}-\bar{z}_4} \nn\\
    & +\frac{\bar{h}_L}{\left(\bar{z}-\bar{z}_4\right){}^2}
   \end{align}
for the anti-holomorphic one, where the $c_i$s are the derivatives of the semiclassical function $g$:
\be
c_X \equiv \frac{\partial}{\partial X} g .
\ee
The dependence of the function $g$ on the positions of the operators must be invariant under global coordinate transformations.  An efficient way to impose this constraint is that the potentials $T(z)$ and $\bar{T}(\bar{z})$ must decay at large $z,\bar{z}$ like $z^{-4}, \bar{z}^{-4}$, respectively.  This constraint imposes six conditions (the first three inverse powers of $z$ and $\bar{z}$), so we are able to eliminate the derivatives with respect to all coordinates except for three, which we will choose to be $y_3, z_4, \bar{z}_4$.  We set the other six coordinates to  
\be
z_1=\bar{z}_1 = \infty, z_2 = \bar{z}_2 = 1, z_3 = \bar{z}_3 = 0.
\ee
The remaining derivatives $c_X$ fixed indirectly by the two Schrodinger equations for $\psi$ and $\bar{\psi}$, by demanding that the monodromy of the solutions to these Schrodinger equations along cycles in the complex $z$ and $\bar{z}$ plane correspond to the weights of the operators contained within those cycles.  Setting the heavy operator to be purely holomorphic, i.e. $\bar{h}_H=0$, makes the anti-holomorphic condition particularly useful, since it means that the monodromy of the $\bar{\psi}$ solutions around a cycle containing only the points $\bar{z}_1$ and/or $\bar{z}_2$ must vanish.  First of all, this condition immediately implies that the coefficient $c_{\bar{z}_2}$ of the $\bar{z}=\bar{z}_2$ pole in $\bar{T}$ must vanish;  we then obtain the following condition when we eliminate $c_{\bar{z}_2}$ in terms of the $y_3, z_4, \bar{z}_4$ derivatives:
\be
0 &=& -\bar{h}_L - \frac{1}{2} y_3 c_{y_3} - \bar{z}_4 c_{\bar{z}_4} .
\ee
This condition is equivalent to the statement that the  correlator depends on $\bar{z}_4$ and $y_3$ only in the combination 
\be
x \equiv \frac{y_3^2}{z_4 \bar{z}_4}
\ee
after we factor out an overall $y^{-2 \bar{h}_L}$ from the correlator.  In other words, the function $g$ must be of the form
\begin{equation}\label{eq:SemiFReduction1}
g(y_3, z_4, \bar{z}_4) = g(z_4, x) -2 \bar{h}_L \log y_3.
\end{equation}

Next, we consider the monodromy of the $\bar{\psi}$ solutions around the point $\bar{z}_1$. This monodromy must also be trivial. In the limit $\bar{z}_1 \rightarrow \infty$ that we have taken, this condition implies that $\lim_{\bar{z}_1 \rightarrow \infty} c_{\bar{z}_1} =0$.\footnote{This is probably most explicitly seen by changing variables of the Schrodinger equation from $\bar{z}$ to $t=\frac{1}{\bar{z}}$, in which case the condition $\lim_{\bar{z}_1 \rightarrow \infty} c_{\bar{z}_1} =0$ is simply that the coefficient of the pole of $\bar{T}$ at $t=0$ must vanish.  Since the map $\bar{z}=1/t$ maps the point $\bar{z}_1=\infty$ to 0, a small cycle around $t=0$ contains only the heavy operator $\CO_H(z_1)$.}  We can then use our solution for $c_{\bar{z}_1}$ in terms of $c_{y_3}, c_{z_4}, c_{\bar{z}_4}$ together with the constraint (\ref{eq:SemiFReduction1}) on $g$ to write this condition on $c_{\bar{z}_1}$ in terms of derivatives of $g(z_4, x)$:
\begin{small}
\begin{equation}
0 = -\left(x z_4+2\right) \bar{h}_L+x g^{(0,1)}\left(z_4,x\right) \left(x^2 z_4^2
  \frac{6}{c}  T_H(0)+x+1\right)+x \left(z_4-1\right) z_4 g^{(1,0)}\left(z_4,x\right)+x z_4 h_L .
\end{equation}
\end{small}
The general solution to this equation is of the form
\be
g(z_4, x) &=& g\left( z_{\rm eff}(z_4, x) \right) - h_L \log(1-z_4) - \bar{h}_L \log \left( \frac{1-\frac{(1+ \frac{2}{x z_4})^2}{\alpha_H^2} }{4\left(1+\frac{1}{x}\right)^2} \right) , 
\label{eq:genholbulk}
\ee
where we have defined the combination
\be
z_{\rm eff}(z_4, x) &\equiv&  1+ (z_4 -1) \left( \frac{2-x z_4 (\alpha_H-1)}{2+x z_4 (\alpha_H+1)} \right)^{\frac{1}{\alpha_H}} 
\ee
so that it reduces to $z_4$ at the boundary $y_3=0$ ($x = 0$).  This parameterization also depends on the stress tensor in the heavy operator background, through the parameter $\alpha_H\equiv \sqrt{1- \frac{24 T_H(0)}{c}}$.   Remarkably, the dependence on all bulk coordinates has been reduced to the dependence on a single coordinate! 

 In the limit that $\phi$ approaches the boundary, the bulk block reduces to the boundary block, so the problem is reduced to the previously solved problem of the boundary block behavior.
Note that we did not need to use the holomorphic Schrodinger equation monodromy condition to accomplish this reduction.
So far, this result holds only to leading order in the small $h_L/c$ limit, where we can neglect the subleading pieces of $T$  that depend on the light operator. It would be interesting to extend this analysis to higher orders, where additional conceptual issues arise due to the necessity of regulating the singularities in $T(z)$ at $z=0$.

\subsection{Constraining Bulk Correlators Using Symmetries}
\label{sec:Symmetries}

In this section we will discuss the semiclassical and quantum symmetries of various correlators involving the bulk proto-field $\phi$.  Our main focus is on the heavy-light bulk-boundary propagator, discussed in section \ref{sec:BulkBoundarySymmetries}, but we also discuss the bulk-to-bulk propagator in section \ref{sec:SymmetryPropagator}, and the discrete inversion symmetry in section \ref{sec:Inversions}.

\subsubsection{Heavy-Light Bulk-Boundary Correlator}
\label{sec:BulkBoundarySymmetries}

Because the result (\ref{eq:genholbulk}) at the end of section \ref{sec:Monodromy} followed essentially from demanding certain residues of $\bar{T}(\bar{z})$ vanished, it should be equivalent to demanding that the corresponding conformal symmetries are satisfied.  
In this subsection, we will go through this explicitly, though here we will specialize to the case $h_L= \bar{h}_L$ for simplicity. 

We will apply the method to holomorphic heavy operators with $\bar{h}_{H}=0$ and
that therefore $\langle \CO_{L}\left(z,\bar{z}\right)\phi_{L}\left(y,z_{3},\bar{z}_{3}\right)\CO_{H}\left(z_{1}\right)\CO_{H}\left(z_{2}\right)\rangle$
 have no dependence on $\bar{z}_1, \bar{z}_2$. Now this four-point function depends on seven coordinates and we can fix five of them using  the symmetry transformations $L_{-1,0,1}$ and $\bar{L}_{-1,0}$, and we get
\begin{equation}
\CA=\langle \CO_{L}\left(z,1\right)\phi_{L}\left(y,0,0\right)\CO_{H}\left(1\right)\CO_{H}\left(\infty\right)\rangle 
\end{equation}
The remaining generator $\bar{L}_{1}$ acts on a bulk point as the  vector field \cite{Anand:2017dav}
\begin{equation}\label{eq:L1barSymmetry}
\bar{L}_{1}\left(y',z',\bar{z}'\right)=\left(y'\bar{z}',\frac{4y'^{2}}{-4+y'^{4}\CS\bar{\CS}},\frac{2 y'^{4}\CS}{-4+\CS\bar{\CS}y'^{4}}+\bar{z}'^{2}\right)
\end{equation}
interpreted as a differential operator $\bar L_1^A \partial_A$ in the bulk (with $A$ running over ($y',z',\bar z'$)).  Here $\CS$ is defined as 
\begin{equation}\label{eq:CSdefinition}
\CS\left(z'\right)=\frac{12}{c}\frac{\langle\left[\CO_{L}(z, 1)\phi_{L}(y,0,0)T\left(z'\right)\right]\left[\CO_{H}(1)\CO_{H}(\infty)\right]\rangle}{\langle\left[\CO_{L}(z, 1)\phi_{L}(y,0,0)\right]\left[\CO_{H}(1)\CO_{H}(\infty)\right]\rangle}
\end{equation}
 where the brackets represent the normal ordering defined in \cite{Fitzpatrick:2016mtp}. In the semiclassical limit  subtleties concerning normal ordering are irrelevant. $\bar{\CS}$ would be defined in a similar way, but it vanishes since we are considering the case that $\bar{h}_{H}=0$.  

We can identify a certain linear combination of $\bar{L}_{1}$ with other global
conformal generators that will move $z$ and
$y$ while keeping the other coordinate fixed.  We will denote this linear combination by  $\tilde{L}$. 
We find that $\tilde{L}$ acts on a bulk point as the vector field: 
\begin{smaller}
\begin{equation}
\tilde{L}\left(y',z',\bar{z}'\right)=\left(\frac{1}{4}y'\left(4\bar{z}'-\CS(0)y^{4}-2y^{2}-2\right),-y'^{2}-y^{2}(z'-1),\frac{1}{2}\left(\left(\bar{z}'-1\right)\left(2\bar{z}'-y^{4}\CS(0)\right)-y'^{4}\CS(z')\right)\right)
\end{equation}
\end{smaller}
This transformation is a global conformal symmetry which leaves the vacuum invariant, 
\begin{equation}
\left\langle\left[\tilde{L},\CO_{L}\left(z,1\right)\phi_{L}\left(y,0,0\right)\CO_{H}(1)\CO_{H}(\infty)\right] \right\rangle=0
\end{equation}
Therefore, the correlator $e^{\mathcal{I}}\equiv\CA$ must be a solution to the differential equation
\begin{small}
\begin{equation}
-h_{L}y^{2}+h_{L}-y^{2}(z-1)\partial_{z}\mathcal{I}-\frac{1}{2}\left(1+y^{2}\right)y\partial_{y}\mathcal{I}-\frac{1}{2}y^{4}\left(h_{L}\CS(0)+\CS(0)\frac{1}{2}y\partial_{y}\mathcal{I}+\frac{1}{2}y\partial_{y}\CS(0)\right)=0\label{eq:DiffEqFromSymmetry}
\end{equation}
\end{small}
In the semiclassical limit of $c\rightarrow\infty$ with $\frac{h_{H}}{c}$
fixed, we simply have 
\begin{equation}
\CS\left(0\right)=\frac{12h_{H}}{c}+\mathcal{O}\left(\frac{1}{c}\right)
\end{equation}
Solving this equation while requiring the $y\rightarrow0$ limit to
match the boundary heavy-light Virasoro vacuum block, we find 
\begin{equation}
\CV_0^{\text{semi}}=y^{-2h_{L}}\left(\frac{\alpha\left(1-z\right)^{\frac{\alpha-1}{2}}}{\alpha+\left(\left(1-z\right)^{\alpha}-1\right)\left(\frac{\left(\alpha-1\right)}{2}-\frac{1}{y^{2}}\right)}\right)^{2h_{L}}
\end{equation}
which agrees with the bulk-boundary vacuum block obtained using the uniformizing coordinates (with $f(z)=z^\alpha$ and $\bar f(\bar z)=\bar z$, since we are setting $\bar h_H$=0) and the semiclassical monodromy method in previous subsections.

In the large $c$ limit with $h_L, h_H$ fixed, using the OPE block method developed in \cite{Fitzpatrick:2016mtp, Anand:2017dav}, we can compute the next to leading order correction to $\CS\left(0\right)$, which is given by
\begin{small}
\begin{align}
\CS(0)=& \frac{12h_H}{c}+\frac{24h_{H}h_{L}}{c^{2}}\frac{1}{\left(y^2+z\bar{z}\right)z^{3}}\left[z\left(2z((z-12)z+12)\bar{z}-y^{2}(z(z(z+2)+6)-12)\right)\right.\nn\\
&\left. -12(z-1)\left(y^{2}-(z-2)z\bar{z}\right)\log(1-z)\right]+\CO(1/c^3)
\label{eq:S0}
\end{align}
\end{small}with $\bar z=1$ for $\CS(0)$ defined in (\ref{eq:CSdefinition}). Inserted into (\ref{eq:DiffEqFromSymmetry}), this gives a differential
equation satisfied by the vacuum block $\CV_0$ up to order $\CO(1/c^2)$. In Appendix \ref{app:OPEBlocks}, we used the OPE block method to compute $\CV_0$ up to order $\CO(1/c^2)$ and checked that the result (with $\bar h_H=0$) does satisfy this differential equation.

\subsubsection{Symmetry Analysis of the Propagator $\< \phi \phi\>$}
\label{sec:SymmetryPropagator}

We can perform a similar analysis of the bulk-bulk propagator in the vacuum.  In recent work \cite{Chen:2017dnl} we found that when $\< \phi(X) \phi(Y) \>$ is computed while incorporating only holomorphic gravitons (we denote this as $\<\phi\phi\>_\text{holo}$), it depends only the the geodesic separation between $X$ and $Y$.  We will now explain this fact using symmetry.  

We can immediately use the translations $L_{-1}$ and $\bar L_{-1}$ to write the propagator as
\be
G(y_1, y_2, z, \bar z) = \< \phi(y_1, z, \bar z)  \phi(y_2, 0, 0) \>_{\text{holo}}
\ee
The transformations $L_{0}$ and $\bar L_{0}$  also do not depend on $\CS$ or $\bar \CS$, and so they act simply, giving the differential equations
\be
0 &=& \left(y_1 \partial_{y_1} + y_2 \partial_{y_2} + 2 z \partial_z \right) G 
\nn \\
0 &=& \left(y_1 \partial_{y_1} + y_2 \partial_{y_2} + 2 \bar z \partial_{\bar z} \right) G 
\ee
These require $G$ to depend on only the quantities $\frac{y_1^2}{z \bar z}$ and $\frac{y_2^2}{z \bar z}$.  This is as far as we can go in general, as the action of $L_{1}$ and $\bar L_{1}$ depend on $\CS$ and $\bar \CS$, which themselves will depend on the bulk fields $\phi$.

However, if we are only computing the holomorphic propagator \cite{Chen:2017dnl}, then we can ignore anti-holomorphic gravitons, and so $\bar \CS = 0$.  In that case $L_1$ acts simply, so that $G$ must satisfy the addition differential equation
\be
\left(y_1 z \partial_{y_1} + z^2 \partial_z - y_1^2 \partial_{\bar z} + y_2^2 \partial_{\bar z} \right) G = 0
\ee
This then implies that
\be
\< \phi \phi \>_{\mathrm{holo}} = G \left( \frac{2 y_1 y_2}{y_1^2 + y_2^2 + z \bar z} \right)
\ee
or in words, that the holomorphic propagator can only depend on the geodesic separation (in the AdS$_3$ vacuum) between the bulk points.  It would be interesting to study this method at higher orders in $1/c$ using the additional $\bar L_1$ generator and the $\CS$ determined by gravitational back-reaction.

\subsubsection{A Note on Inversion Symmetry}
\label{sec:Inversions}

CFTs may have a discrete symmetry under inversions in the plane, which take 
\be
(z, \bar z) \to \left(\frac{1}{\bar z}, \frac{1}{z} \right) 
\ee
After transforming to the cylinder, inversions correspond to the $t \to -t$ time reversal symmetry.  The vacuum conformal block of CFT$_2$ possesses these symmetries in both the $1/c$ expansion and also at finite central charge. Correlation functions in vacuum AdS and probe correlators in classical BTZ black hole backgrounds also inherit this inversion symmetry.  For example, the semiclassical bulk-boundary conformal block in equation (\ref{eq:semiclassicalbulkboundaryvacuumblock}) is manifestly symmetric under $\delta t_E \to - \delta t_E$.  

However, complications arise when extending this symmetry to bulk proto-fields at the quantum level.  First, we must extend inversions into the bulk in the $(y, z, \bar z)$ coordinate system in the chosen Fefferman-Graham gauge.  Formally, this is fairly simple.  If we obtain the vacuum AdS metric of equation \ref{eq:FGMetric} via maps $f(z), \bar f (\bar z)$ from the pure AdS metric 
\be
ds^2 = \frac{du^2 + dx  d \bar x}{u^2}
\ee
by the coordinate transformation (\ref{eq:CoordinateTransformationToPoincare}) \cite{Anand:2017dav, Roberts:2012aq}, then inversions correspond to the identification between unprimed and primed coordinates through the relations
\be
\label{eq:Inversion}
u\left(y,f\left(z\right),\bar{f}\left(\bar{z}\right)\right)&=u\left(y',f\left(\frac{1}{\bar z'}\right),\bar{f}\left(\frac{1}{{z}'}\right)\right)\nn\\
x\left(y,f\left(z\right),\bar{f}\left(\bar{z}\right)\right)&=x\left(y',f\left(\frac{1}{\bar z'}\right),\bar{f}\left(\frac{1}{z'}\right)\right)  \\ 
\bar{x}\left(y,f\left(z\right),\bar{f}\left(\bar{z}\right)\right)&=\bar{x}\left(y',f\left(\frac{1}{\bar z'}\right),\bar{f}\left(\frac{1}{{z}'}\right)\right) \nn
\ee
Note that because $S(z)$ in equation (\ref{eq:FGMetric}) is determined by the Schwarzian derivative of $f(z)$, it is automatic that equation (\ref{eq:Inversion}) is a discrete symmetry of the spacetime.  We provide a few examples and details in appendix \ref{app:InversionSymmetry}, but although equation (\ref{eq:Inversion}) is simple, the relation between the original and primed coordinates may be rather involved.

Beyond heavy-light semiclassical limit, to determine the inversion symmetry transformations explicitly we must incorporate the backreaction on the geometry from $\phi$ itself.  This echoes complications  encountered when extending Virasoro transformations, such as equation (\ref{eq:L1barSymmetry}), to the quantum level in the bulk. To extend the inversion symmetry into the bulk, the coordinates $(y,z,\bar z)$ must transform in a way that depends on $\CS(z)$ and $\bar \CS(\bar z)$.

A further issue arises when interpreting inversion symmetry in F-G coordinates as time reversal in the BTZ coordinate system.  The connection between F-G coordinates $(r, t_E, \theta)$ and the BTZ Schwarzschild coordinates $(y, z, \bar z)$ obtained in section \ref{sec:DefinitionField} was semiclassical, and did not account for the backreaction of $\phi$ or quantum corrections.  In other words, the Schwarzschild coordinates were introduced as a re-labeling of the F-G coordinates, and it's challenging to extend this re-labeling beyond the semiclassical probe limit.

We demonstrate some of these points  in appendix \ref{app:InversionSymmetry}, where we show explicitly how bulk-boundary correlators transform under the inversion symmetry, including quantum effects in $1/c$ perturbation theory.   As a consequence of such effects, when the exact correlators are plotted using the semiclassical BTZ coordinates $(r, t_E, \theta)$, \emph{they are not manifestly symmetric under a $t_E \to -t_E$ reflection}.  Violations of this symmetry are very small, but become noticeable for BTZ $r$ coordinates very near the horizon.  We emphasize that this apparent asymmetry comes from the application of the (merely) semiclassical coordinate transformations from section \ref{sec:DefinitionField}.

\section{Exact Correlators}
\label{sec:ExactCoorelators}

In this section we discuss two different methods that can be used to automate the calculation of the bulk-boundary conformal blocks $\CV_{h, \bar h}(y, z, \bar z)$, where its most convenient to use  the kinematic configuration
\begin{equation}
\left\langle \CO_H(\infty) \CO_H(1)\CO_L(z, \bar z) \phi_L(y,0,0) \right\rangle.
\end{equation}
  The two direct methods of section \ref{sec:DirectCalculation} are based on a brute force sum over Virasoro descendants.  These methods have the advantage of providing either exact $y$-dependence to some order in $z$, or (nearly) exact $z$-dependence to fixed order in $y$.  Then in section \ref{sec:ZRR} we discuss a generalization of the Zamolodchikov recursion relations; this enables a higher order numerical evaluations of $\CV_{h, \bar h}(z,y)$.  The direct methods are most useful for computing correlators in the Lorentzian regime, as they permit extremely high accuracy in the boundary coordinate and Lorentzian time.  The recursion relation is more efficacious in the Euclidean regime, where it's possible to obtain $\CV_0$ as an expansion in $z,\bar z$ with coefficients exact in $y$. The plots in this paper are made with results from the recursion relation up to order $z^{60}\bar z^{60}$.

\begin{figure}
\begin{center}
\includegraphics[width=0.98\textwidth]{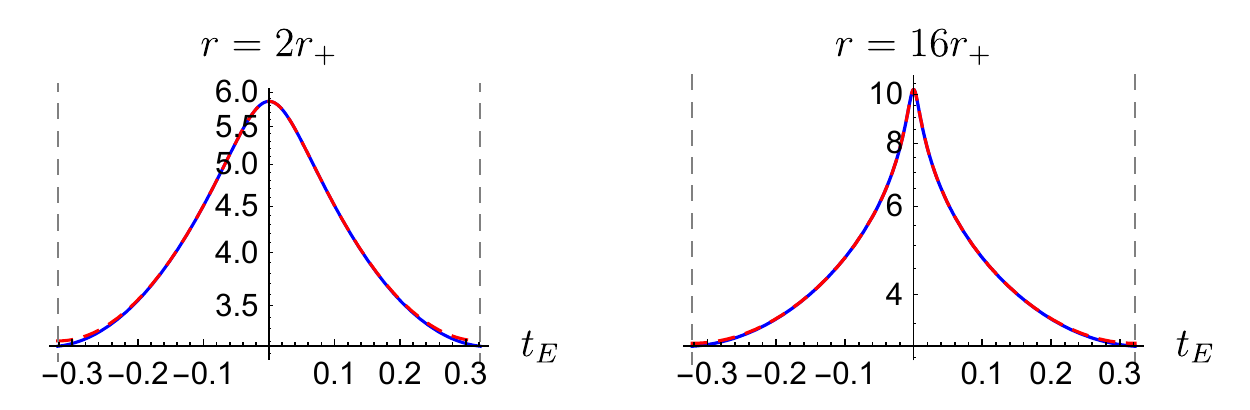}
\caption[Caption for LOF]{ These plots compare the exact  (blue, $\log(|\mathcal{V}_0^{\mathrm{exact}}|)$) and semiclassical (pink, $\log(|\mathcal{V}_0^{\mathrm{semi}}|)$) correlators for different values of $r$.  The parameters for these plots are $c=30.1, h_L=0.505, \frac{h_H}{c}=4$, so that $r_+ \approx 9.7$.  The semiclassical approximation is  excellent for these values of $t_E$ and $r$.  
The gray dashed lines are $\pm \beta/2$. We used the exact result from  recursion up to order $z^{60}\bar z^{60}$, with convergence $\left|\frac{\mathcal{V}_0^{\mathrm{exact}}(\text{60 orders})-\mathcal{V}_0^{\mathrm{exact}}(\text{59 orders})}{\mathcal{V}_0^{\mathrm{exact}}(\text{60 orders})}\right|<10^{-12}$.}\label{fig:SemiMatchesExact}
\end{center}
\end{figure}

We have attached Mathematica code implementing these three  methods.
Figure   \ref{fig:SemiMatchesExact} provides visual confirmation that the bulk primary reproduces  semiclassical physics in black hole backgrounds at large $c$.

\subsection{Direct Calculations}
\label{sec:DirectCalculation}

The bulk-boundary blocks can be directly evaluated in two ways.  The first leverages the simplicity of the bulk primary condition, while the second attempts to exploit the availability of high-precision information \cite{Chen:2017yze} on the boundary blocks.  Thus the first method computes $\CV_h(y, z)$ exactly in $y$ but only to low-order in $z$ (practically up to order $\sim z^{14}$), while the second method computes the blocks only to low order in $y$, but to extremely high precision in the boundary coordinates (so the result can be written in terms of the $q$ coordinate \cite{Zamolodchikovq, Maldacena:2015iua}, which provides far better convergence, along with the ability to analytically continue deep into the Lorentzian regime).  

\subsubsection{Using the Bulk Primary Condition}
\label{sec:BlockFromBulkPrimary}

Consider the direct evaluation of the general bulk-boundary conformal block
\begin{small}
\begin{equation}
\label{eq:ExplicitBlock}
\CV_h(y, z, \bar z) = \left\langle \CO_H(\infty) \CO_H(1)  \left( \sum_{\{m_i\},\{ n_j\}}  \frac{L_{-m_1} \cdots L_{-m_i} | h\> \< h | L_{n_{j}} \cdots L_{n_1} }{\CN_{\{m_i\}, \{n_j\}} }\right) \CO_L(z, \bar z) \phi_L(y,0,0) \right\rangle
\end{equation}
\end{small}For simplicity we have only explicitly included a holomorphic intermediate primary $|h\>$ along with a sum over holomorphic Virasoro descendants, but in general we would also simultaneously include an anti-holomorphic intermediate state and a sum over anti-holomorphic Virasoro descendants. Due to the presence of $\phi_L(y,0,0)$ this block will not factor into a product of holomorphic and anti-holomorphic contributions, although the coefficients of any given power $y^{2h_L + 2N}$ do factorize in this way.

We can compute using equation (\ref{eq:ExplicitBlock}) almost as efficiently as in the pure boundary case of $\< \CO_H \CO_H \CO_L \CO_L \>$.  This follows because the bulk primary condition
\be
L_{m \geq 2 } \phi(y,0,0) | 0 \> = 0
\ee
implies that almost all Virasoro generators act trivially on $\phi$, meaning that
\begin{smaller}
\begin{align}
\< h | (L_{n_k} \cdots L_{n_2}) L_{n_1} \CO_L(z, \bar z) \phi_L(y,0,0)  \>
 &= \< h | (L_{n_k} \cdots L_{n_2}) [L_{n_1}, \CO_L(z)] \phi_L(y,0,0) \>
 \\
&=z^{n_1} (h_L (1+n_1) + z \partial_z) \< h | (L_{n_k} \cdots L_{n_2}) \CO_L(z, \bar z)\phi_L(y,0,0) \>
\nn
\end{align}
\end{smaller}whenever $n_1 \geq 2$.  Thus we can simply extract any string of Virasoro generators.  When computing  the vacuum block, we have $\< \CO_L(z, \bar z ) \phi_L(y,0,0) \>=\left(\frac{y}{y^2+z \bar z}\right)^{2h_L}$ and we can choose a basis where all $n_i \geq 2$, so that all calculations can be performed in this way.

The calculation of the other factors in equation (\ref{eq:ExplicitBlock}) are just a standard application of the Virasoro algebra, and are easily automated.  This makes it possible to compute $\CV_0(y, z, \bar z)$ to reasonably high order order (e.g. at least $z^{14}$ for the holomorphic $\phi$) with exact, algebraic coefficients, including the exact $y$ dependence.  For example, up to order $z^4$ we find that the contributions from the exchanged vacuum state and its holomorphic descendants are
\begin{align}
\frac{\CV_0 (y, z, \bar{z})}{\< \CO_L(z, \bar z ) \phi_L(y,0,0) \>} = & 1+\frac{2 h_L h_H (1+3 x) z^2 }{c (1+x)}+\frac{2  h_L h_H (1+2 x) z^3}{c
   (1+x)}
    \\
&   + \frac{h_L h_H z^4 }{c (5 c+22) (x+1)^2} \left(   12 x \left(9+2 c\right) +   (2 + 12x)(h_L+h_H+5 h_L h_H) \right.
\nn \\ 
&	+ \left.  3 x^2 \left(24+5 c+6 h_L+10 h_H +30 h_L h_H \right) +9 c+ 40\right)  + \cdots\nn
\end{align}
where we define \footnote{We apologize for the usage of $x$ in several different places in this paper (e.g. $x$ is also used in equation (\ref{eq:PurePoincareMetric}) as the coordinate in the pure Poincare metric). But its meaning should be clear from the context. }
 $x \equiv \frac{y^2}{z \bar z}$ and note that when $x \to 0$ this reduces to the usual boundary Virasoro block.  We have also verified that these results agree with those of  section \ref{sec:ZRR}, which are based on an adaptation of the Zamolodchikov recursion relations \cite{ZamolodchikovRecursion}.  At large $c$ with $h_H/c$ and $h_L$ fixed, these results match the semiclassical correlators reviewed in section \ref{sec:ReviewSemiclassiclProbeCorrelators}.

These methods imply that terms of order $z^{2n}$ or $z^{2n+1}$ are always given by polynomials of degree $n$ in $x$ times a factor of $\frac{1}{(1+x)^n}$.  This  follows  because each $L_m$ includes only a single $\partial_z$ derivative acting on $\< \CO_L(z, \bar z) \phi_L(y,0,0) \>$, and since $m \geq 2$ we have at most $n$ such derivatives producing the $z^{2n}$ or $z^{2n+1}$ terms.  This insight makes it possible to extract the exact $x$ dependence from the methods of section \ref{sec:ZRR}, which formally only produce a series expansion in the variables $x, z, \bar z$.  In practice, this is how we study bulk-boundary correlators in the Euclidean region.

\subsubsection{Using Knowledge of the Boundary Correlators}

As our starting point, we can instead use the expression
\be
\CV_h(y, z, \bar z) = \left\< \CO_H(\infty) \CO_H(1)  \CP_h  \CO_L(z, \bar z) \sum_{n=0}^{\infty} \frac{y^{2h_L + 2n}}{n! (2h_L)_n} \CL_{-n} \bar \CL_{-n} \CO_L(0) \right \>
\ee
for the bulk-boundary block.
The $\CL_{-n}$ are linear combinations of products of Virasoro generators at level $n$, determined by the bulk primary condition from section \ref{sec:DefinitionField}, and $\CP_h$ is the Virasoro projector onto the block with primary dimension $h$.  All Virasoro generators $L_m$ commute with $\CP_h$, so we can compute $\CV_h$ by commuting the individual Virasoro generators in $\CL_{-n}$ to the left, where they act on $\CO_L(z, \bar z)$ and $\CO_H(1)$ before annihilating the $\< 0|\CO_H(\infty)$ state.  

This method outputs the coefficient of $y^{2h_L + 2n}$ in $\CV_h$ as a differential operator acting on the boundary Virasoro block
\be
V_h(y, z, \bar z) = \left\< \CO_H(\infty) \CO_H(1)  \CP_h  \CO_L(z, \bar z) \CO_L(0) \right \>
\ee
As a concrete example, in the kinematic configuration $z = \bar z$, the first three terms are 
\be
\label{eq:exactextrap}
\CV_h &=& \frac{y^{2 h_L} (1-z)^{2 h_L}}{ z^{4 h_L}} \left( V_h(z)^2 -  y^{2} \frac{  \left(2 h_L V_h(z)-(1-z) z V_h'(z)\right){}^2}{2 h_L z^2} \right.
\\ 
&& \left. +\resizebox{0.83\textwidth}{!}{\text{\ensuremath{ y^{4} \frac{   \left(1+2 h_L\right) \left(2 h_L \left(c-6 z^2 h_H+2 h_L \left(c+8 h_L-5\right)\right) V_h(z)-2 (1-z) z \left(c z+2 h_L \left(-3+c+z+8 h_L\right)\right)
   V_h'(z)+(-1+z)^2 z^2 \left(c+8 h_L\right) V_h''(z)\right)^2}{4 h_L z^4 \left(c+2 h_L \left(-5+c+8 h_L\right)\right)^2}
} }}  + \cdots   \right)
\nn
\ee
The boundary blocks $V_h(z)$ can be computed to extremely high precision \cite{Chen:2017yze} using the Zamolodchikov recursion relations.  In particular, $\CV_h$ can be computed in the $q$-expansion, which remains convergent after arbitrary analytic continuation into the Lorentzian regime.  This last property will make this method very useful for studying Lorentzian bulk-boundary correlators.  We have attached Mathematica code implementing this computation.

We can also use this method to compute $\CA$ directly from the boundary correlator $\< \CO_H \CO_H \CO_L \CO_L\>$.  In particular, in regimes where the boundary correlator is extremely well-approximated by its semiclassical limit, we can simply feed the semiclassical $V_h(z)$ into this algorithm.  When our goal is to uncover new effects from bulk reconstruction (rather than from deviations between the exact and semiclassical boundary correlators), this is a useful trick: any deviations between the result and the semiclassical bulk correlator will be due to the difference between extrapolating boundary operators into the bulk via classical bulk wave equations vs via the protofield construction.\footnote{To be more precise, for any heavy-heavy-light-light boundary correlator we can compare a `semiclassical' and an `exact' extrapolation of one of the boundary operators into the bulk. The `semiclassical' extrapolation is defined as using the bulk wave equation for the classical geometry corresponding to the heavy state, whereas the `exact' extrapolation is defined as using the protofield, as in (\ref{eq:exactextrap}). }

\subsection{Recursion Relations}
\label{sec:ZRR}

The Zamolodchikov recursion relations \cite{ZamolodchikovRecursion, Zamolodchikovq, Perlmutter:2015iya} can be adapated to compute the bulk-boundary block $\CV_h$.  This requires a sum over holomorphic and anti-holomorphic Virasoro descendants from both the Virasoro projector $\CP_h$ and from the definition of $\phi$.  Thus the bulk-boundary correlator $\CV_h$ has the complexity of two coupled 5-pt Virasoro blocks \cite{Cho:2017oxl}.  In this section we will present the $c$-recursion relations for computing $\CV_h$.

\subsubsection{Order by Order Factorization of the Bulk-boundary Blocks}
At each order of $y$, the proto-field 
\begin{equation}
\phi =y^{2h}\sum_{n=0}^{\infty}\left(-1\right)^{n}y^{2n}\lambda_{n}\mathcal{L}_{-n}\bar{\mathcal{L}}_{-n}\mathcal{O}\left(z,\bar{z}\right), \qquad \lambda_n=\frac{1}{n!(2h_L)_n}
\end{equation} 
factorize in to the
product of holomorphic and antiholomorphic parts. This will lead to
the factorization of the bulk-boundary blocks at each order of $y$.
Thus we can compute the ``holomorphic'' part of the bulk-boundary
block first and recover the full block at the end. We define the holomorphic
part of the proto-field to be \footnote{Note that the definition of the holomorphic part of the proto-field
$\phi$ is different the definition of that in \cite{Chen:2017dnl}. The definition
here is simply for computational convenience. } 
\begin{equation}\label{eq:HoloField}
\tilde{\phi}_{h}^{\text{holo}}\left(y,z,\bar{z}\right)\equiv y^{2h}\sum_{n=0}^{\infty}\lambda_n y^{2n}\mathcal{L}_{-n}\mathcal{O}_{h,h}\left(z,\bar{z}\right).
\end{equation}
Then the holomorphic bulk-boundary block is given by\footnote{For the convenience of discussing the recursion relation later on,
here we are being more general by setting the dimensions of the intermediate
state and the proto-field to be arbitrary $h_{1}$ and $h_{2}$. Eventually,
we are interested in the case that $h_{1}=0$ and $h_{2}=h_{L}$. } 
\begin{equation}
\mathcal{V}_{\text{holo}}\left(h_{1},h_{2},c\right)\equiv\left\langle \mathcal{O}_{H}(\infty)\mathcal{O}_{H}(1)\mathcal{P}_{h_{1}}^{\text{holo}}\mathcal{O}_{L}(z,\bar{z})\tilde{\phi}_{h_{2}}^{\text{holo}}\left(y,0,0\right)\right\rangle ,
\end{equation}
where the holomorphic projection operator $\mathcal{P}_{h_{1}}^{\text{holo}}$
only includes the holomorphic descendants of the $\mathcal{O}_{h_{1}}$.
We'll introduce a recursion relation to compute $\mathcal{V}_{\text{holo}}\left(h_{1},h_{2},c\right)$
in next sub section. Eventually, we are interested in $\mathcal{V}_{\text{holo}}\left(0,h_{L},c\right)$,
which will be given as an expansion in terms of $y^{2}$, that is
\begin{equation}
\mathcal{V}_{\text{holo}}\left(0,h_{L},c\right)=\left(\frac{y}{z}\right)^{2h_{L}}\sum_{n=0}^{\infty}\left(\frac{y^2}{z}\right)^nF_{n}\left(z\right).
\end{equation}
where $F_{n}\left(z\right)$ is an expansion in terms of $z$ (starting
from $z^{0}$). And we can obtain the full bulk-boundary vacuum block
via 
\begin{equation}
\mathcal{V}_{0}\equiv\mathcal{V}\left(0,h_{L},c\right)=\left(\frac{y}{z\bar{z}}\right)^{2h_{L}}\sum_{n=0}^{\infty}\frac{\left(-1\right)^{n}}{\lambda_{n}}x^{n}F_{n}\left(z\right)F_{n}\left(\bar{z}\right)
\end{equation}
where $F_{n}\left(\bar{z}\right)$ is defined to be $F_{n}\left(z\right)$
with $z$ replaced by $\bar{z}$ and $x\equiv\frac{y^{2}}{z\bar{z}}$. 

The above result is an expansion of $\mathcal{V}_{0}$ in terms of
$x,z,\bar{z}$. On the other hand, as explained at the end of section \ref{sec:BlockFromBulkPrimary},  we know that the
vacuum block is of the form
\begin{equation}
\mathcal{V}_{0}=\left(\frac{y}{y^{2}+z\bar{z}}\right)^{2h_{L}}\tilde{\mathcal{V}}_{0}.
\end{equation}
Here $\tilde{\mathcal{V}}_{0}=1+\cdots$ is an expansion of $z,\bar{z}$ with the coefficient of $z^{n}\bar{z}^{m}$  being a product of
$\frac{1}{\left(1+x\right)^{\lfloor m/2\rfloor+\lfloor n/2\rfloor}}$
and a polynomial of degree $\lfloor m/2\rfloor+\lfloor n/2\rfloor$
in $x$, where $\lfloor k\rfloor$ means the maximum integer that's
small or equal to $k$. So we can use the coefficients of $z^{n}\bar{z}^{m}$
in $\tilde{\mathcal{V}}_{0}$ up to $x^{\lfloor m/2\rfloor+\lfloor n/2\rfloor}$
and extract its exact dependence on $x$. Eventually, the result we
obtain for the vacuum block $\mathcal{V}_{0}$ is an expansion in
terms of $z$ and $\bar{z}$, with coefficients exact in $x$. 

\subsubsection{Recursion relation}

Now our task is to compute $\mathcal{V}_{\text{holo}}\left(h_{1},h_{2},c\right)$.
We'll show that $\mathcal{V}_{\text{holo}}\left(h_{1},h_{2},c\right)$
can be computed via the following recursion relation
\begin{align}
\mathcal{V}_{\text{holo}}\left(h_{1},h_{2},c\right)= & \mathcal{V}_{\text{holo}}\left(h_{1},h_{2},c\rightarrow\infty\right)\label{eq:Recursion}\\
 & +\sum_{m\ge2,n\ge1}\frac{R_{m,n}\left(h_{1},h_{2}\right)}{c-c_{m,n}\left(h_{1}\right)}\mathcal{V}_{\text{holo}}\left(h_{1}\rightarrow h_{1}+mn,h_{2},c\rightarrow c_{mn}\left(h_{1}\right)\right)\nonumber \\
 & +\sum_{m\ge2,n\ge1}\frac{S_{m,n}\left(h_{1},h_{2}\right)}{c-c_{m,n}\left(h_{2}\right)}\mathcal{V}_{\text{holo}}\left(h_{1},h_{2}\rightarrow h_{2}+mn,c\rightarrow c_{mn}\left(h_{2}\right)\right),\nonumber 
\end{align}
with
\begin{align}
R_{m,n}\left(h_{1},h_{2}\right) & =-\frac{\partial c_{m,n}\left(h_{1}\right)}{\partial h_{1}}A_{m,n}^{c_{m,n}\left(h_{1}\right)}P_{m,n}^{c_{m,n}\left(h_{1}\right)}\left[\begin{array}{c}
h_{H}\\
h_{H}
\end{array}\right]P_{m,n}^{c_{m,n}\left(h_{1}\right)}\left[\begin{array}{c}
h_{L}\\
h_{2}
\end{array}\right].\nonumber \\
S_{m,n}\left(h_{1},h_{2}\right) & =-\frac{\partial c_{m,n}\left(h_{2}\right)}{\partial h_{2}}A_{m,n}^{c_{m,n}\left(h_{2}\right)}P_{m,n}^{c_{m,n}\left(h_{2}\right)}\left[\begin{array}{c}
h_{1}\\
h_{L}
\end{array}\right].
\end{align}
We'll parametrize the central charge $c$ in terms of $b$ as $c=13+6\left(b^{2}+b^{-2}\right)$.
The poles $c_{m,n}\left(h\right)$ are given by 
\begin{equation}
c_{m,n}\left(h\right)=13+6\left[\left(b_{m,n}\left(h\right)\right)^{2}+\left(b_{m,n}\left(h\right)\right)^{-2}\right]
\end{equation}
 with 
 \begin{small}
\begin{equation}
\left(b_{m,n}\left(h\right)\right)^{2}=\frac{2h+mn-1+\sqrt{\left(m-n\right)^{2}+4\left(mn-1\right)h+4h^{2}}}{1-m^{2}},m=2,3,\cdots,n=1,2,\cdots.
\end{equation}
\end{small}
The functions $A_{m,n}^{c}$ and $P_{m,n}^{c}\left[\begin{array}{c}
h_{1}\\
h_{2}
\end{array}\right]$ are given by
\begin{equation}\label{eq:Amnc}
A_{m,n}^{c}=\frac{1}{2}\prod_{k=1-m}^{m}\prod_{l=1-n}^{n}\frac{1}{kb+\frac{l}{b}},\qquad\left(k,l\right)\ne\left(0,0\right),\left(m,n\right),
\end{equation}
and 
\begin{equation}\label{eq:Pmnc}
P_{m,n}^{c}\left[\begin{array}{c}
h_{1}\\
h_{2}
\end{array}\right]=\prod_{p,q}\frac{\lambda_{1}+\lambda_{2}+pb+qb^{-1}}{2}\frac{\lambda_{1}-\lambda_{2}+pb+qb^{-1}}{2}
\end{equation}
with $\lambda_{i}^{2}=b^{2}+b^{-2}+2-4h_{i}$. The ranges of $p$
and $q$ in the above product are 
\begin{align*}
p & =-m+1,-m+3,\cdots,m-3,m-1,\\
q&=  -n+1,-n+3,\cdots,n-3,n-1.
\end{align*}
Note that in $R_{m,n}\left(h_{1},h_{2}\right)$, $A_{m,n}^{c_{m,n}\left(h_{1}\right)}$
means that the $b$ in $A_{m,n}^{c}$ should be replaced by $b_{m,n}\left(h_{1}\right)$,
and similarly for other terms in $R_{m,n}\left(h_{1},h_{2}\right)$
and $S_{m,n}\left(h_{1},h_{2}\right)$. 

The last piece of information we need for the recursion (\ref{eq:Recursion})
is the bulk-boundary global blocks
\begin{equation}
G\left(h_{1},h_{2}\right)\equiv\mathcal{V}_{\text{holo}}\left(h_{1},h_{2},c\rightarrow\infty\right).
\end{equation}
In the limit that $c\rightarrow0$, all the Virasoro generators will
be suppressed, therefore in the projection operator $\mathcal{P}_{h_{1}}$
and the holomorphic proto-field $\phi_{h_{2}}^{\text{holo}}$, all
that left are the global descendants. Thus we have 
\begin{equation}
G\left(h_{1},h_{2}\right)=\sum_{m_{1},m_{2}=0}^{\infty}y^{2h_{2}+2m_{2}}\frac{\left\langle \mathcal{O}_{H}\mathcal{O}_{H}L_{-1}^{m_{1}}|h_{1}\right\rangle \left\langle h_{1}|L_{1}^{m_{1}}\mathcal{O}_{L}\left(z\right)L_{-1}^{m_{2}}|h_{2}\right\rangle }{\left|L_{-1}^{m_{1}}\left|h_{1}\right\rangle \right|^{2}\left|L_{-1}^{m_{2}}\left|h_{2}\right\rangle \right|^{2}}.
\end{equation}
The details for computing $G\left(h_{1},h_{2}\right)$ is provided
in Appendix \ref{app:RRdetails}, and the result is given by 
\begin{equation}
G\left(h_{1},h_{2}\right)=z^{h_{1}}\left(\frac{y^{2}}{z}\right)^{h_{2}}\sum_{m_{1},m_{2}=0}^{\infty}\frac{\left(h_{1}\right)_{m_{1}}s_{m_{1},m_{2}}\left(h_{1},h_{L},h_{2}\right)}{\left(2h_{1}\right)_{m1}m_{1}!\left(2h_{2}\right)_{m_{2}}m_{2}!}z^{m_{1}}\left(\frac{y^{2}}{z}\right)^{m_{2}}
\end{equation}
with \cite{Alkalaev:2015fbw}
\begin{align}
s_{k,m}\left(h_{1},h_{2},h_{3}\right)\equiv & \left\langle h_{1}|L_{1}^{m_{1}}\mathcal{O}_{h_{2}}\left(1\right)L_{-1}^{m_{2}}|h_{3}\right\rangle \label{eq:skm}\\
 =& \sum_{p=0}^{\text{min\ensuremath{\left(k,m\right)}}}\frac{k!}{p!\left(k-p\right)!}\left(2h_{3}+m-p\right)_{p}\left(m-p+1\right)_{p}\nonumber \\
 & \times\left(h_{3}+h_{2}-h_{1}\right)_{m-p}\left(h_{1}+h_{2}-h_{3}+p-m\right)_{k-p}.\nonumber 
\end{align}

Solving the recursion (\ref{eq:Recursion}) will give $\mathcal{V}_{\text{holo}}\left(h_{1},h_{2},c\right)$
as a sum over global blocks
\begin{equation}
\mathcal{V}_{\text{holo}}\left(h_{1},h_{2},c\right)=\sum_{m,n=0}^{\infty}C_{m,n}G\left(h_{1}+m,h_{2}+n\right).
\end{equation}
The global block $G\left(h_{1}+m,h_{2}+n\right)$ is the contribution
to  $\mathcal{V}_{\text{holo}}$ from a level-$m$ quasi-primary in
$\mathcal{P}^{\text{holo}}_{h_{1}}$ and a level-$n$ quasi-primary in $\phi_{h_{2}}^{\text{holo}}$.
The coefficients $C_{m,n}$ are functions of the operators dimensions
and the central charge $c$. As shown in equation (\ref{eq:Cmn}),
they are related to three point
functions of primaries with one or two quasi-primaries and the norms of the quasi-primaries. Specifically, $C_{m,n}G(h_1+m,h_2+n)$ computes the total contribution to $\mathcal{V}_{\text{holo}}$ from all the level-$m$ quasi-primaries in
$\mathcal{P}^{\text{holo}}_{h_{1}}$ and  level-$n$ quasi-primaries in $\phi_{h_{2}}^{\text{holo}}$.
 One way of
understanding the recursion (\ref{eq:Recursion}) is that it provides
an efficient way of computing these coefficients. More details about the recursion relation and the algorithm for implementing it in Mathematica can be found in Appendix \ref{app:RRdetails}.

After obtaining $\mathcal{V}_{\text{holo}}\left(0,h_{L},c\right)$, we can use the method discussed in last subsection to compute $\tilde{\mathcal{V}}_0$. Concretely, the first several terms of $\tilde{\mathcal{V}}_0$ are given by 
\begin{smaller}
\begin{align}\label{eq:tildeV}
\tilde{\mathcal{V}}_0 & =1+\frac{2(3x+1)h_{H}h_{L}}{c(x+1)}\left(\bar{z}^{2}+z^{2}\right)\\
 & +\frac{4h_{H}^{2}h_{L}\left(x(5x-2)+\left(1+2x-3x^{2}\right)h_{L}+\left(17x^{2}+12x+2\right)h_{L}^{2}+12x^{2}h_{L}^{3}-4x^{2}h_{L}^{4}\right)}{c^{2}(x+1)^{2}\left(2h_{L}+1\right)}z^{2}\bar{z}^{2}+\cdots\nonumber
 \end{align}
\end{smaller}

We've checked that all the three methods discussed in this section for computing $\mathcal{V}_0$ give the same result, which also agrees with the large $c$ expansion of $\CV_0$ (Appendix \ref{app:OPEBlocks}) and the semiclassical result $\CV_0^{\mathrm{semi}}$ (Section \ref{sec:Uniformizing}) in the appropriate limits.

In next section, we'll compare the result from the recursion with the semiclassical result. For clarity, we'll convert all results to the usual BTZ coordinates $(r,t_E, \theta)$, where the semiclassical result is given by 
\begin{equation}
\mathcal{V}_{0}^{\text{semi}}\left(r,t_E,\theta\right) =\left(\frac{r_{+}}{2}\right)^{2h_{L}}\frac{1}{\left[\frac{r}{r_{+}}\cosh\left(r_{+}\theta\right)-\sqrt{\frac{r^{2}}{r_{+}^{2}}-1}\cos\left(r_{+}t_E\right)\right]^{2h_{L}}}.\label{eq:SemiClassicalResultInrtautheta}
\end{equation} 
As discussed in section \ref{sec:DefinitionField} and appendix \ref{app:FGtoBTZ}, the right object to compare with $\mathcal{V}_{0}^{\text{semi}}$ is the following
\begin{equation}
	\mathcal{V}_{0}^{\text{exact}}\left(r,t_E,\theta\right)\equiv\left(1-z\right)^{h_{L}}\left(1-\bar{z}\right)^{h_{L}}\left(\frac{y}{y^2+ z \bar z}\right)^{2h_L} \tilde{\CV}_0
\end{equation}
with the coordinate transformation from $(y, z, \bar z)$ to $(r,t_E,\theta)$ via (\ref{eq:yintermsofr}) and $\tilde{\CV}_0$ as given in (\ref{eq:tildeV}). For better visibility of the plots, we'll actually divide both $\mathcal{V}_{0}^{\text{semi}}$ and $\mathcal{V}_{0}^{\text{exact}}$ by $y^{2h_L}$ (which is not singular in the region we are interested in).

\section{Exploring the Euclidean Horizon}
\label{sec:ExploringHorizon}

Now we will explore the behavior of the correlator when the bulk operator $\phi$ approaches the Euclidean horizon\footnote{The bulk field operator $\phi(y,z, \bar z)$ was defined  in terms of a local CFT$_2$ primary and its descendants via the bulk primary conditions of section \ref{sec:DefinitionField}.    So when we discuss the `horizon', we are referring to certain values of the $(y,z,\bar z)$ coordinate labels determined mathematically in terms of the BTZ black hole coordinates $(t,r,\theta)$ through equation (\ref{eq:FGandBTZCoordRelation}).  Bulk interpretations of these labels are emergent. } of a black hole microstate.   For simplicity we study spherically symmetric black holes with $h_H = \bar h_H$, and since $\phi$ is a scalar we have $h_L = \bar h_L$.  Our plots always indicate bulk-boundary correlators with no angular separation, so that the correlators depend only on $(r, t_E)$.

\begin{figure}[th!]
\begin{center}
\includegraphics[width=0.35\textwidth]{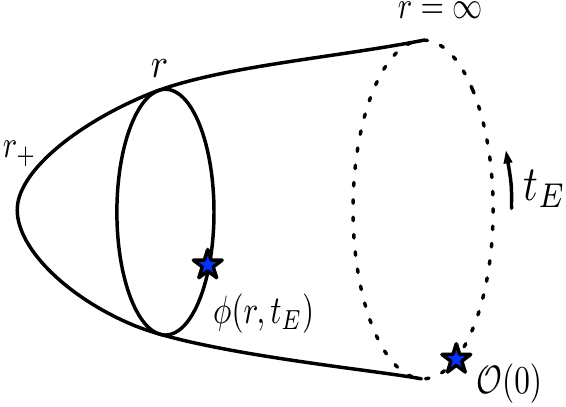} \ \ \ \ \  \includegraphics[width=0.55\textwidth]{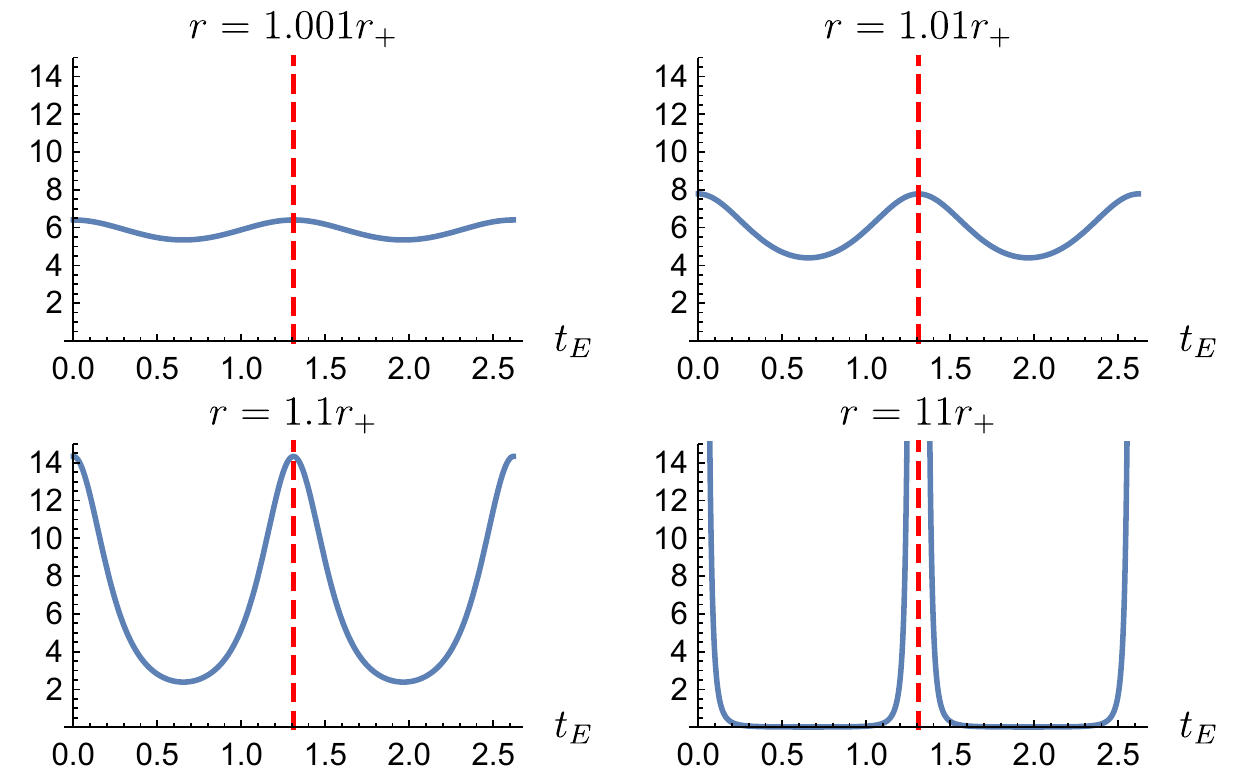}
\caption{{\bf Left}:  This figure depicts a Euclidean bulk-boundary correlator $|\mathcal{V}_0^\mathrm{semi}|$ on the BTZ `cigar' geometry, focusing on slices at fixed $r$, where we can easily study Euclidean time periodicity. 
{\bf Right}:  These plots display the semiclassical bulk-boundary correlator $\mathcal{V}_0^\mathrm{semi}$ on constant-$r$ slices. The semiclassical correlator is periodic in $t_E$, and its range of variation becomes smaller as we approach the horizon $r=r_+$, where it is constant in $t_E$. The red dashed line is $t_E=\beta$ and the parameters are $\frac{h_H}{c}=1, h_L=1$.}
\label{fig:CigarBulkBoundary}
\end{center}
\end{figure}

The Euclidean horizon is the region where $r \gtrsim r_+$ with purely Euclidean BTZ time  coordinate $t_E$.  We have reason to expect a sharp, order-one deviation between the semiclassical and exact correlators in this region.   As one can see from figure \ref{fig:CigarBulkBoundary}, the classical BTZ geometry and the semiclassical correlators are periodic in Euclidean time.  But exact CFT correlators in a pure state (or even in the microcanonical ensemble) cannot be periodic \cite{Balasubramanian:2007qv, Fitzpatrick:2016ive}.  As illustrated in figure \ref{fig:MultisheetedCigarwithPhi},  the exact CFT correlators must lift to multivalued functions on the `cigar' geometry.  This  suggests that the correlators will be badly behaved at the Euclidean horizon where the $t_E$ circle shrinks to zero size.  We will confirm this expectation with an explicit numerical computation using the exact correlators.  We will also see that the region where the exact and semiclassical correlators differ shrinks as we increase $c$.

Near the Euclidean horizon, the corresponding Fefferman-Graham coordinates $z, \bar z$ remain in the Euclidean region with $\bar z = z^*$, and thus the correlator can be best approximated using the algorithm of section \ref{sec:ZRR}.  With it we can compute the correlator to order $z^{60} \bar z^{60}$ with coefficients that capture the exact dependence on $h_H, h_L, c$ and the kinematic $y$-coordinate.   For clarity, we will convert
all results into the usual BTZ coordinates $\left(r,t_E,\theta\right)$ as discussed at the end of section \ref{sec:ZRR}.

To any finite order in $y$, these results should converge for all $|z| < 1$.  However, since we are only computing to finite order in the $z$ expansion, the radius of convergence will be smaller, and must be estimated empirically based on the growth of terms in the series expansion.  We find that the recursion relations of section \ref{sec:ZRR} converge best when $24 h_H / c \gg 1$, $h_L \ll 1$, and $c > 1$ is relatively small.   For the most part we will focus on this regime, as our goal is to compare the exact and semiclassical correlators as precisely as possible.  Note that in this regime there are two relevant length scales in the bulk, the AdS scale $R_{AdS} =1$ in our conventions, and the larger  horizon scale $r_+ = \sqrt{ \frac{24h_H}{c} - 1} \gg 1$.  Typically with our chosen parameters $r_+ \sim 10-100$.  As explained in section \ref{sec:Inversions}, the exact results are not exactly symmetric under $t_E \to -t_E$ in the BTZ coordinates.  

\begin{figure}
\begin{center}
\includegraphics[width=0.99\textwidth]{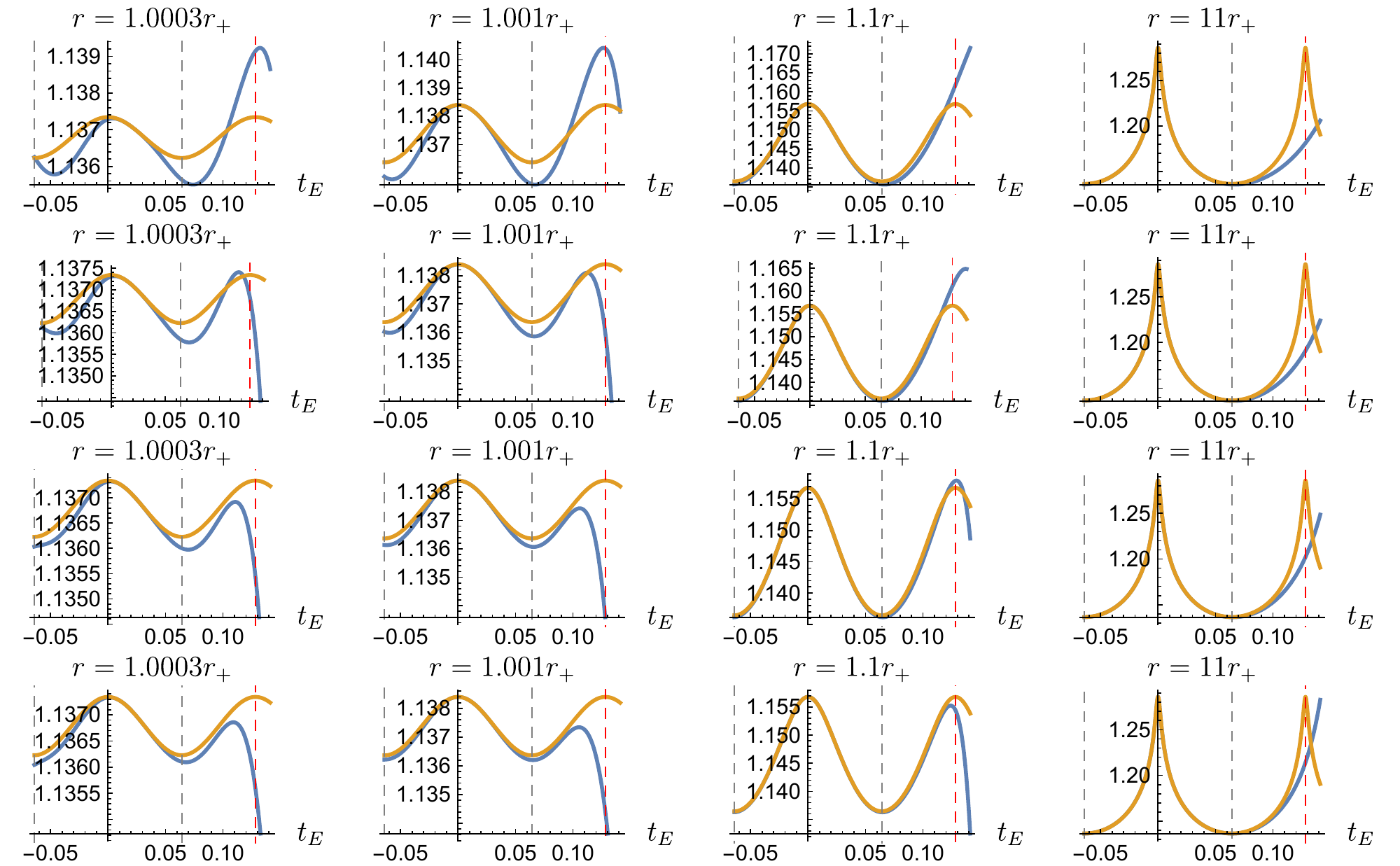}
\caption{The blue lines are the exact result $|\mathcal{V}_0^{\text{exact}}|$ and the yellow lines are the semiclassical  $|\mathcal{V}_0^{\text{semi}}|$. From top  to bottom the rows of plots correspond to $c=8.1, 16.1, 32.1, 64.1$, respectively. Other parameters for these plots are $h_L=0.01, \frac{h_H}{c}=100$, and $r_+ \approx 50$.  The first two plots in each row  are in the region whose distance from the horizon is much smaller than the AdS radius. The red dashed line is $t_E=\beta$ and the gray dashed line is $t_E=\pm\beta/2$. The exact results in the visible plot range have converged to better than $10^{-13}$ precision (the precision of convergence is defined as in figure \ref{fig:SemiMatchesExact}).  \label{fig:ExactVsSemiclassical2}}
\end{center}
\end{figure}

We compared the exact and semiclassical results for small $t_E$ and large $r$ in Figure \ref{fig:SemiMatchesExact} and we found excellent agreement.  Now let us investigate $r \approx r_+$, larger $t_E$, and small $c$.  In figure \ref{fig:ExactVsSemiclassical2} we have compared the exact and semiclassical correlators as functions of the Euclidean time $t_E$ for various fixed values of the radius $r$.  We see that the exact and semiclassical correlators are very similar for $t_E < \beta$ when  $r \gg r_+$, though the correlators deviate significantly for $t_E \approx \beta$, as expected based on the boundary behavior \cite{Chen:2017yze}. But as we approach the horizon, the correlators disagree for a greater and greater range of $t_E$ values, such that for $r \approx r_+$ the exact and semiclassical correlators are significantly different for all $t_E$.  

\begin{figure}
\begin{center}
\includegraphics[width=0.4\textwidth]{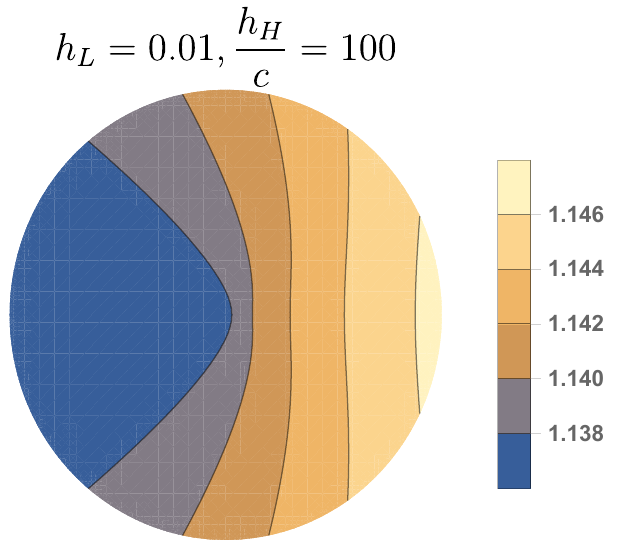}
\caption{This is a plot of $|\mathcal{V}_0^{\text{semi}}|$ zoomed in to the tip of the Euclidean `cigar', with $r_+<r<1.025r_+$ and $0<t_E<\beta$. The radial coordinate of the disk is $r - r_+$ and the angular direction is $\frac{2 \pi}{\beta} t_E$; the BTZ angular coordinate $\theta = 0$. The center of the plot is the position of the Euclidean horizon and $r_+ \approx 49$.} 
\label{fig:SemiDisk}
\end{center}
\end{figure}

\begin{figure}
\begin{center}
\includegraphics[width=0.98\textwidth]{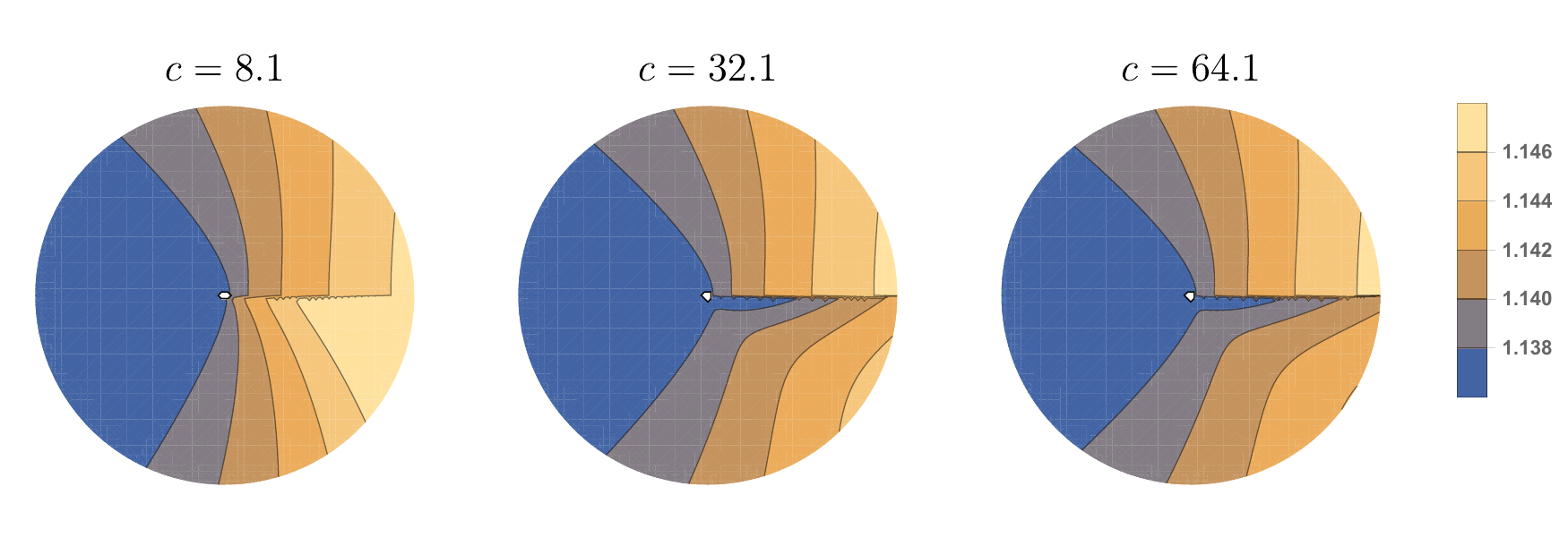}
\caption{ These are plots of $|\mathcal{V}_0^{\text{exact}}|$ for $h_L=0.01, h_H/c=100$ but with different values of $c$. These are plotted in the same region and use the same range as figure \ref{fig:SemiDisk} for ease of comparison. These results have converged to better than $10^{-10}$ accuracy except for a tiny region at the origin of the disk (i.e. the white point at the center).}\label{fig:ExactDisk}
\end{center}
\end{figure}

\begin{figure}
\begin{center}
\includegraphics[width=0.98\textwidth]{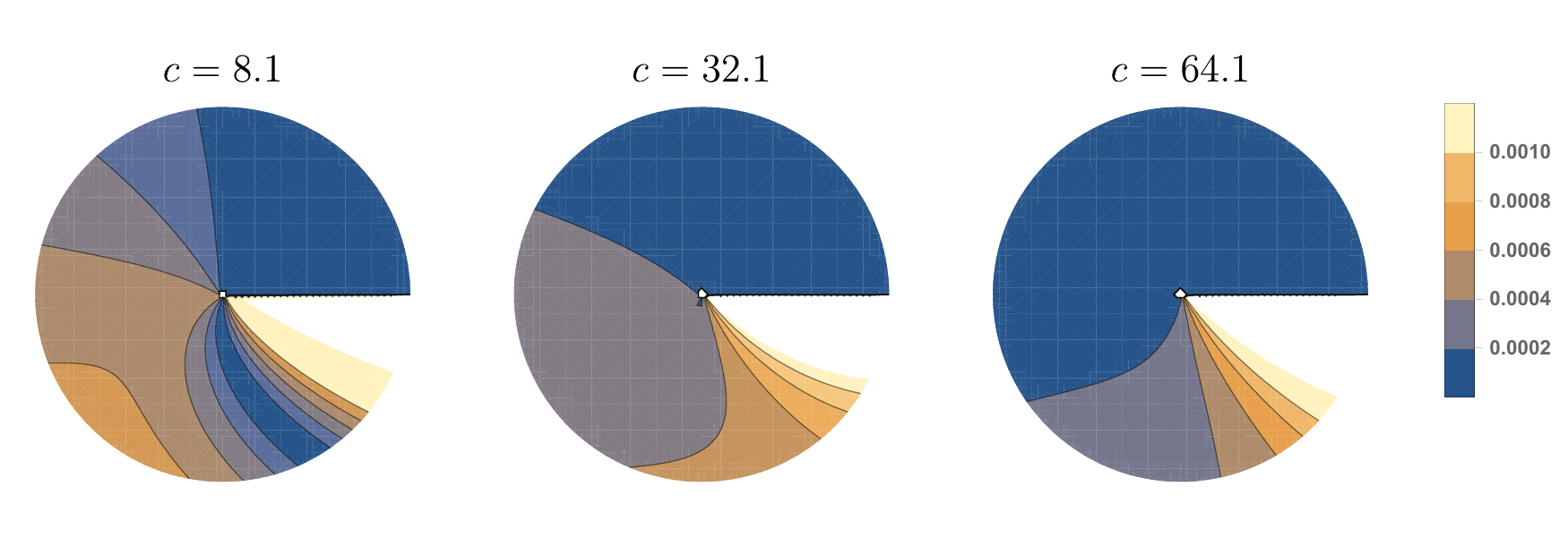}
\caption{These plots show the difference between the exact and semiclassical results: $\left|\frac{\mathcal{V}_0^{\mathrm{exact}}-\mathcal{V}_0^{\mathrm{semi}}}{\mathcal{V}_0^{\mathrm{semi}}}\right|$ in the same region as figure \ref{fig:SemiDisk} and \ref{fig:ExactDisk}. They have the same parameters as figure \ref{fig:ExactDisk}: $h_L=0.01,h_H/c=100$.  The difference between exact result and semiclassical result is numerically small because we've chosen very small $h_L = \frac{1}{100}$ for better convergence, and this means that both the exact result and the semiclassical result are very close to 1. Qualitatively, we can see that as we increase $c$, the agreement between the exact result and the semiclassical result improves. The exact results have converged to better than $10^{-10}$ accuracy. } \label{fig:ExactSemiDisk}
\end{center}
\end{figure}

We compare the exact and semiclassical correlators on the full Euclidean `cigar' geometry in figures \ref{fig:SemiDisk}, \ref{fig:ExactDisk}, and \ref{fig:ExactSemiDisk}.  These plots indicate the full dependence on $r$ and $t_E$, and give some idea of the way the results change with $c$.  However the `migration' of the discrepancy from $t_E \approx \beta$ to the full range of $t_E$ is easier to see in figure \ref{fig:ExactVsSemiclassical2}.

\subsection*{Minimizing Violations of Bulk Effective Field Theory}

Since these results are somewhat preliminary, we would like to interpret them as conservatively as possible.  So its natural to ask how to minimize the discrepancy between a naive bulk effective field theory description -- i.e. the semiclassical correlator -- and the exact correlator.

The discrepancy between the exact and semiclassical correlators becomes unavoidable once we approach $t_E - \beta \sim O \left(\frac{1}{\sqrt{c}} \right)$.  And for $|t_E|$ larger than $\beta$ the semiclassical description completely fails.   We have now seen that this applies both on the boundary and in the bulk.  This unsuppressed effect is due to non-perturbative corrections in the large $c$ limit, though surprisingly, there are already hints of this phenomenon in $1/c$ perturbation theory  \cite{Fitzpatrick:2015dlt, Fitzpatrick:2016ive}.

However, one can brush this problem under the rug by defining the correlator on the Euclidean cigar using the exact correlator evaluated in the range $t_E \in \left[-\frac{\beta}{2}, \frac{\beta}{2} \right]$.  On the boundary, the disagreement between the exact and perturbative correlators will be extremely (non-perturbatively) small for this range of $t_E$.  This fact has been discussed previously \cite{Chen:2017yze}, as analytic continuation in $t_E$ to $\frac{\beta}{2}$ can also be used to mimic the correlators in the double-sided eternal black hole geometry.  

Even if the bulk correlators can be smoothly connected between $t_E = \pm \frac{\beta}{2}$ up to $\sim e^{-c}$ corrections, then at exponentially small values of $|r - r_+|$ we might nevertheless see a large deviation from naive effective field theory predictions.  This follows because the derivative of the correlator will grow as $\frac{1}{r - r_+}$, and so eventually even a tiny effect may become significant.    But this will only occur at a distance exponentially close to the horizon, and so it's unclear if it would affect observers. 

We also see indications in figure \ref{fig:ExactVsSemiclassical2} that the exact and semiclassical correlators disagree for a greater range of $t_E$ as $r \to r_+$.  We have confirmed this phenomena for some other choices of parameters.  Unfortunately, due to the limitations of numerical convergence we do not have the dynamic range to determine if this effect is perturbative or non-perturbative in nature, or to work out its empirical dependence on $h_L, r_+,$ and $c$.  Hopefully some of these issues can be clarified through a more detailed analysis, or by directly studying the Lorentzian regime in future work.

\section{Discussion}
\label{sec:Discussion}

The primary purpose of this paper was to develop methods for computing the  gravitational contributions to the bulk-boundary propagator in a black hole microstate at finite $G_N$.    In CFT$_2$ terminology, we  studied the conformal block decomposition of a 4-pt correlator involving three CFT primaries and a single bulk proto-field $\phi$, which has been defined as a specific infinite sum of Virasoro descendants \cite{Anand:2017dav} parameterized by the bulk coordinate $y$.  We  explored the semiclassical limit of these correlators, and demonstrated that they reduce to known results in the probe or heavy-light limit. 

It would be interesting to better understand the convergence of the bulk reconstruction algorithm and of the associated conformal blocks.  It would be especially useful to develop an analog of the $q$ variable \cite{Zamolodchikovq, Maldacena:2015iua} that can achieve a maximum radius of convergence for these objects.  To reach the interior of a microstate black hole, it seems that one must analytically continue through a \emph{bulk-boundary} light-cone OPE \cite{Fitzpatrick:2012yx, KomargodskiZhiboedov}, as the bulk field must cross the past lightcone of $\CO_H$ in the bulk.  The $q$ variable allows analytic continuation through infinitely many boundary light-cone limits, so an analog in the bulk might clarify the definition of correlators in the black hole interior.  

We performed a preliminary comparison of exact vs semiclassical Euclidean bulk-boundary correlators.  Our  goal was to understand the bulk implications of the fact that correlators in black hole microstate backgrounds violate the Euclidean periodicity manifest in the classical black hole geometry.  The result was that Euclidean bulk correlators  deviate from their semiclassical limit in a way that appears to be unsuppressed at $t_E \gtrsim \beta$.  The effect appears increasingly impactful as $r \to r_+$, since the naive Euclidean-time circle contracts to zero size.  We also found  evidence that the effect spreads to a much greater range of $t_E$ as one approaches very near to the horizon.  

The most conservative interpretation still allows for an exponentially suppressed deviation for physical observables.  Furthermore, violations of Euclidean-time periodicity in perturbative CFT$_2$ computations \cite{Fitzpatrick:2015dlt} are an important case where even for boundary correlators, the distinction between effects that are and are not visible in bulk effective field theory remains to be understood.  Note that even if our results have implications for `drama' at the horizon \cite{Mathur:2009hf, Almheiri:2012rt}, they would not immediately apply  to eternal black holes or the canonical ensemble, which  satisfy the KMS condition exactly.

Do physical observers  see violations of bulk effective field theory outside the horizon, and are there relatively unambiguous predictions for what observers might see inside a black hole?  To  address these questions, we must investigate the behavior of the Lorentzian correlators pertaining to physical observers.  It will also be important to differentiate between corrections to  CFT correlators and qualitatively new effects due to the bulk reconstruction process itself.  Non-perturbative corrections to reconstruction can dramatically alter the bulk equations of motion and invalidate bulk locality \cite{Chen:2017dnl}; it is the investigation of such effects in black hole backgrounds that necessitates exact bulk reconstruction.

\section*{Acknowledgments}

We would like to thank Tarek Anous, Ibou Bah, Xi Dong, Ethan Dyer, Diego Hofman, Tom Hartman,  Shamit Kachru, David E. Kaplan, Zuhair Khandker, Nima Lashkari,  Joao Penedones, Suvrat Raju, Mukund Rangamani, Matt Walters, and Junpu Wang for discussions. ALF was supported in part by the US Department of Energy Office of Science under Award Number DE-SC-DE-SC0015845, and by a Sloan Foundation fellowship.   JK and HC  have been supported in part by NSF grant PHY-1454083.  ALF, JK and DL were also supported in part by the Simons Collaboration Grant on the Non-Perturbative Bootstrap. HC was also supported by a KITP Graduate Fellowship and in part by the NSF grant No. NSF PHY-1748958.

\appendix
 
\addtocontents{toc}{\protect\setcounter{tocdepth}{1}}

\section{Coordinate Systems}
\label{app:CoordinateRelations}

The purpose of this appendix is to explain the relationship between the BTZ black hole in its standard form and in the coordinate system that we use in this paper.  We will see that the relation has a surprising feature:  real values of the standard AdS-Schwarzschild coordinates $(t_E, r, \theta)$ correspond with complex values for the Fefferman-Graham radial coordinate $y$.  As far as we are aware, this feature has not been noted in the literature.  For completeness and perhaps for pedagogical value, we will also make some elementary comments concerning the connection between diffeomorphisms and conformal transformations.

\subsection{Various Coordinate Relations}

The Euclidean BTZ black hole metric is typically written using Schwarzchild coordinates 
\be
\label{eq:MetricSpinningBTZ}
ds^2 = \frac{(r^2-r_+^2)(r^2 + r_-^2)}{r^2} dt_E^2 + \frac{r^2 dr^2}{(r^2-r_+^2)(r^2  + r_-^2)} + r^2 \left( d \theta + \frac{r_+ r_-}{r^2} dt_E \right)^2
\ee
where we note that to avoid a conical singularity at the horizon, we must identify $t_E \sim t_E + \frac{2 \pi}{r_+}$, and by definition we identify $\theta \sim \theta + 2 \pi$.    As we take $r \to \infty$ with fixed $t_E, \theta$ we approach the boundary cylinder, with metric $ds^2 = dt_E^2 + d \theta^2$.  We can easily obtain the Lorentzian BTZ metric via the simultaneous analytic continuations $t_E \to i t$ and $r_- \to i r_-$.  

Our exact results are based in a Fefferman-Graham coordinate system, where in the presence of a heavy source the Euclidean metric takes the form
\be
\label{eq:HeavySource}
ds^2 = \frac{dy^2 + dz d \bar z}{y^2} -  \frac{6h_H}{c z^2} dz^2 -  \frac{6\bar h_H}{c \bar z^2} d \bar z^2
+ y^2 \frac{36  h_H \bar h_H}{c^2 z^2 \bar z^2} dz d \bar z
\ee
The boundary corresponds to $y \to 0$, and if we take this limit uniformly (without scaling by any function of $z, \bar z$) then we obtain a flat boundary metric $ds^2 = dz d \bar z$.  The heavy sources with conformal weights $(h_H, \bar h_H)$ are located at $z = 0$ and $z = \infty$ on the boundary.  When the sources are absent, this metric reduces to that of the standard Euclidean Poincar\'e patch for AdS$_3$.

 Throughout, we'll use the relation $\alpha = \sqrt{1 - \frac{24 h_H}{c}}$, and $\bar{\alpha} = \sqrt{1 - \frac{24 \bar{h}_H}{c}}$, and by convention when $\alpha,\bar{\alpha}$ are imaginary we take them to have opposite signs.  These parameters are related to the outer and inner horizon radii of the Euclidean black hole  via $\alpha = i r_+ - r_-$ and $\bar \alpha = -i r_+ - r_-$. 
We mostly focus on the spherically symmetric case with $r_- = 0$.

Now let us discuss the coordinate relations.  First, let us note that equation (\ref{eq:MetricSpinningBTZ}) does have a simple relationship with a metric that looks superficially like our Fefferman-Graham coordinate system.  This is a third distinct form of the metric 
\be
\label{eq:GlobalCoordCylinder}
ds^2 = \frac{dn^2}{n^2} + \frac{\alpha^2}{4 } \frac{d \xi^2}{\xi^2} + \frac{\bar \alpha^2}{4 } \frac{d \bar \xi^2}{\bar \xi^2} + \left(\frac{1}{n^2} + n^2 \frac{\alpha^2 \bar \alpha^2}{16} \right) \frac{d \xi d \bar \xi}{\xi \bar \xi}
\ee
Notice that in the absence of sources, when $\alpha = \bar \alpha = 1$, this metric does not reduce to the Poincar\'e patch form of AdS$_3$.  Relatedly, when we approach the boundary by taking the limit $n \to 0$, the term $\frac{\alpha^2}{4 } \frac{d \xi^2}{\xi^2}$ in the metric has an interpretation as an expectation value for the CFT stress tensor $\< T \> = \frac{\alpha^2}{4 \xi^2}$, and it is non-zero even in the vacuum.  Both of these facts follow because in equation (\ref{eq:GlobalCoordCylinder}), the coordinates $\xi, \bar \xi$ parameterize the surface of a cylinder, rather than a flat plane when we take $n \to 0$.  This is manifest with $\xi = e^{t_E + i \theta}$ and $\bar \xi = e^{t_E - i \theta}$.

We can relate the metric (\ref{eq:GlobalCoordCylinder}) and the standard form of BTZ (\ref{eq:MetricSpinningBTZ}) straightforwardly; we take $\xi = e^{t_E + i \theta}$ and $\bar \xi = e^{t_E - i \theta}$ as above, while
\be
r^2 = \frac{\left(\alpha ^2 n^2-4\right) \left(\bar \alpha^2 n^2-4\right)}{16 n^2}
\ee
This means that the horizon is located at\footnote{Recall that in the Euclidean region this actually represents a line, rather than a 2d surface, because the thermal Euclidean time circle shrinks to a point at the horizon.}
\be
n_* = \frac{2}{\sqrt{\alpha \bar \alpha}} = \frac{2}{\sqrt{r_+^2 + r_-^2}}
\ee
in the coordinate system of equation (\ref{eq:GlobalCoordCylinder}).

Now let us identify a relation between the metric (\ref{eq:GlobalCoordCylinder}) and the Fefferman-Graham metric (\ref{eq:HeavySource}) that we are using in this paper.  This is more complicated, but it can be achieved by obtaining both metrics as sub-regions of empty Poincar\'e patch AdS$_3$.  Starting with 
\be
ds^2 = \frac{d u^2  + d x d \bar x}{u^2}
\ee
we can obtain any vacuum metric by identifying \cite{Roberts:2012aq}
\be \label{eq:OurFavoriteDiff}
u & = & y \frac{4 (f'(z) \bar f'(\bar z))^{\frac{3}{2}} }{ 4 f'(z) \bar f'(\bar z) + y^2 f''(z) \bar f''(\bar z) }\\
x &= & f(z) - \frac{2 y^2 (f'(z))^2 \bar f''(\bar z)}{4 f'(z) \bar f'(\bar z) + y^2 f''(z) \bar f''(\bar z)}
\nn \\
\bar x &= & \bar f(\bar z) - \frac{2 y^2 ( \bar f'(\bar z))^2  f''( z)}{4 f'(z) \bar f'(\bar z) + y^2 f''(z) \bar f''(\bar z)}\nn
\ee
for suitable $f, \bar f$.  To obtain the metric of equation (\ref{eq:GlobalCoordCylinder}) we use $f(z) = e^{\alpha z}$ followed by $z = \log \xi$, whereas to obtain equation (\ref{eq:HeavySource}) we directly use $f(z) = z^\alpha$.  These two transformations are subtly different because the derivatives of $f$ are respect to different variables.  The end result is a dictionary between coordinate systems
\be
y \frac{4 z \bar z \sqrt{\alpha \bar \alpha z^{\alpha -1} \bar z^{\bar \alpha-1}}}{4 z \bar z + (\alpha-1)(\bar \alpha - 1) y^2} = u &=&  \frac{4 n \sqrt{\alpha \bar \alpha \xi^\alpha \bar \xi^{\bar \alpha}}}{4 + \alpha \bar \alpha n^2}\nn\\
z^\alpha \frac{4 z \bar z - (\alpha+1)(\bar \alpha - 1) y^2}{4 z \bar z + (\alpha-1)(\bar \alpha - 1) y^2} = x &=& \frac{\xi^\alpha (4 - \alpha \bar \alpha n^2)}{4 + \alpha \bar \alpha n^2}\\
\bar z^{\bar \alpha} \frac{4 z \bar z - (\alpha-1)(\bar \alpha + 1) y^2}{4 z \bar z + (\alpha-1)(\bar \alpha - 1) y^2}  = \bar x &=&  \frac{\bar \xi^{\bar \alpha} (4 - \alpha \bar \alpha n^2)}{4 + \alpha \bar \alpha n^2}\nn
\ee
Notice that at small $y$ and $n$, we have $z \approx \xi$, $\bar z \approx \bar \xi$, and $\frac{y^2}{z \bar z} \approx n$.  This means that taking the limit $y \to 0$  results in a different boundary metric from $n \to 0$; in the former case we obtain a CFT in flat space, whereas in the latter case we obtain the CFT on a cylinder.

One can solve the relation between coordinates explicitly.  Defining a discriminant 
\be
D^2 \equiv \alpha^4 \bar \alpha^4 n^8 - 16 \alpha^2 \bar \alpha^2 n^6 + 32(2 \alpha^2 + 2 \bar \alpha^2 - \alpha^2 \bar \alpha^2)n^4 - 256 n^2 + 256
\ee
we find  the results
\be
\frac{y^2}{z \bar z} &=&  \frac{\alpha^2 \bar \alpha^2 n^4 - 8 n^2 + 16 -  D} {2 n^2 (1 - \alpha^2)(1- \bar \alpha^2)} 
\nn \\
z^\alpha &=& \xi^\alpha \left( \frac{ \alpha^2 \bar \alpha^2 n^4 - 8 \alpha^2 n^2 + 16 + \alpha D}{(1 + \alpha)(16 - \alpha^2 \bar \alpha^2 n^4)} \right)
\nn \\
\bar z^{\bar \alpha} &=& \bar \xi^{\bar \alpha} \left( \frac{ \alpha^2 \bar \alpha^2 n^4 - 8 \bar \alpha^2 n^2 + 16 + \bar \alpha D}{(1 + \alpha)(16 - \alpha^2 \bar \alpha^2 n^4)} \right)
\ee
This makes it possible to connect the standard BTZ metric and our Fefferman-Graham coordinate system; for completeness note that
\be
n^2 =  \frac{2 \left(\alpha ^2+\bar \alpha^2+4 r^2-\sqrt{\left(\alpha ^2 - \bar \alpha^2\right)^2+16 r^4+8 r^2 \left(\alpha ^2+\bar \alpha^2\right)}\right)}{\alpha ^2 \bar \alpha^2}
\ee
which allows us to write $y, z, \bar z$ explicitly in terms of $r, t_E, \theta$.  The results simplify somewhat in the spherically symmetric case $r_ - = 0$ when we connect directly to the BTZ coordinates.  In that case we find
\be
\label{eq:FromFGtoBTZ}
y &=& \frac{2\sqrt{\xi \bar \xi}}{\tilde{r}} \left( \frac{r - \sqrt{r^2 - r_+^2 -1}}{r_+^2 + 1} \right)\nn \\
z &=& \frac{1}{\tilde{r}}\xi
 \\
\bar z &=& \frac{1}{\tilde{r}}\bar \xi \nn
\ee
with $\tilde{r}\equiv \left( \frac{r + i r_+ \sqrt{r^2 - r_+^2 - 1}}{(1 + i r_+) \sqrt{r^2 - r_+^2} } \right)^{\frac{i}{r_+}}$. 
Recall that we can rewrite these results in terms of $t_E, \theta$ of the BTZ metric via $\xi = e^{t_E + i \theta}$ and $\bar \xi = e^{t_E - i \theta}$; note that the BTZ and Fefferman-Graham time coordinates are only identical at the boundary.

However, these expressions imply something unexpected about the 3d real manifold in the $(y, z, \bar z)$ coordinate systems associated with real $ r, t_E, \theta$ in the standard BTZ metric -- the $y$ coordinate takes complex values when $(r, t_E, \theta)$ are real.  This occurs whenever $r^2 < 1 + r_+^2$.  In particular, the horizon corresponds with
\be
\frac{y^2}{z \bar z} = \frac{4}{(r_+ \pm i)^2 - r_-^2} 
\ee
Despite these complex values for $y$, by definition the line element $ds^2$ from (\ref{eq:HeavySource}) will be real when evaluated as a function of real $t, r, \theta$ (and also after a Lorentzian continuation via $t_E \to it$ and $r_- \to i r_-$).  Nevertheless, these complex values for $y$ are a feature of the relationships between these coordinate systems.

\subsubsection*{Eddington-Finkelstein}

To study the horizon of a BTZ black hole, it is useful to use coordinates that are well-behaved in its vicinity.  Thus we can use the Eddington-Finkelstein coordinate
\be
v = t - \frac{1}{r_+} \tanh^{-1} \left( \frac{r}{r_+} \right) + i\frac{\pi}{2 r_+}
\ee
which we have written in terms of the Lorentzian BTZ time coordinate ($t=-it_E$) and radius.  In the spherically symmetric case, this produces a metric
\be
ds^2 = - (r^2 - r_+^2) dv^2 + 2 dv dr + r^2 d \theta^2
\ee
which is non-singular through the horizon.

\subsubsection*{Holomorphic Limit}

The relationship between the $\frac{y^2}{z \bar z}$ and $r$ coordinates simplifies in the holomorphic limit, where $\bar h_H = 0$ and $\bar \alpha = 1$.  In that case we simply find that
\be
\frac{y^2}{z \bar z} = \frac{4}{4 r^2 + \alpha^2 - 1}
\ee
where the Fefferman-Graham coordinates are on the left hand side.  We see that even in the case of deficit angles (with real $\alpha$), when $r$ becomes sufficiently small we must analytically continue to complex values of $y$.  However, the relationship between $z, \bar z$ and $r$ remains quite complicated.

\subsection{Bulk-boundary Vacuum Block in BTZ Coordinates}
\label{app:FGtoBTZ}

In Section \ref{sec:ExactCoorelators}, we've developed several methods to compute the bulk-boundary
vacuum block $\mathcal{V}_{0}\left(y,z,\bar{z}\right)$ in the
following configuration: $\left\langle \mathcal{O}_{H}\left(\infty\right)\mathcal{O}_{H}\left(1\right)\mathcal{O}_{L}\left(z,\bar{z}\right)\phi\left(y,0,0\right)\right\rangle $,
where the heavy sources $\mathcal{O}_{H}$ are at $z=1$ and $z=\infty$.
However, the Euclidean BTZ metric (\ref{eq:HeavySource}) in the Fefferman-Graham coordinate system
has heavy sources located at $z=0$ and $z=\infty$ on the boundary.
To make the physics more transparent in that metric, we can move
the heavy operator $\mathcal{O}_{H}$ at $z=1$ to $z=0$, by using
a conformal transformation that takes $\infty\rightarrow\infty,1\rightarrow0,z\rightarrow1$.
This uniquely fixes the conformal transformation to be
\begin{equation}
x\rightarrow\frac{1-x}{1-z}.
\end{equation}
Under this transformation, the bulk position $\left(y,0,0\right)$
transforms as
\begin{equation}\label{eq:ypzpzbpToyzzb}
\left(y,0,0\right)\rightarrow\left(\frac{y}{\sqrt{\left(1-z\right)\left(1-\bar{z}\right)}},\frac{1}{1-z},\frac{1}{1-\bar{z}}\right)\equiv\left(y',z',\bar{z'}\right),\end{equation}
and we find
\begin{align}
 & \left\langle \mathcal{O}_{H}\left(\infty\right)\mathcal{O}_{H}\left(0\right)\mathcal{O}_{L}\left(1\right)\phi\left(y',z',\bar{z'}\right)\right\rangle \\
= & \left(1-z\right)^{h_{L}}\left(1-\bar{z}\right)^{h_{L}}\left\langle \mathcal{O}_{H}\left(\infty\right)\mathcal{O}_{H}\left(1\right)\mathcal{O}_{L}\left(z,\bar{z}\right)\phi\left(y,0,0\right)\right\rangle .\nonumber 
\end{align}
Now we can use equation (\ref{eq:FromFGtoBTZ}) to map $\left(y',z',\bar{z'}\right)$
to the usual BTZ coordinates $\left(r,t_{E},\theta\right)$, i.e
\begin{equation}
y'=\frac{2}{\tilde{r}}\left(\frac{r-\sqrt{r^{2}-r_{+}^{2}-1}}{r_{+}^{2}+1}\right)e^{t_E},\quad z'=\frac{1}{\tilde{r}}e^{t_{E}+i\theta},\quad\bar{z'}=\frac{1}{\tilde{r}}e^{t_{E}-i\theta},
\end{equation}
with $\tilde{r}\equiv\left(\frac{r+ir_{+}\sqrt{r^{2}-r_{+}^{2}-1}}{\left(1+ir_{+}\right)\sqrt{r^{2}-r_{+}^{2}}}\right)^{\frac{i}{r_{+}}}.$
Using the relationship between $\left(y',z',\bar{z'}\right)$
and $\left(y,z,\bar{z}\right)$, that is, equation (\ref{eq:ypzpzbpToyzzb}),
we find that to map the bulk-boundary vacuum block $\mathcal{V}_{0}\left(y,z,\bar{z}\right)$
of Section \ref{sec:ExactCoorelators} from $\left(y,z,\bar{z}\right)$ to the BTZ coordinates
$\left(r,t_{E},\theta\right)$, we need to use
\begin{align}\label{eq:yzzbToBTZCoordinates}
y & =2\frac{r-\sqrt{r^{2}-r_{+}^{2}-1}}{r_{+}^{2}+1},\nonumber \\
z & =1-\tilde{r}e^{-t_{E}-i\theta},\\
\bar{z} & =1-\tilde{r}e^{-t_{E}+i\theta}.\nonumber 
\end{align}

\subsection{Inversion Symmetry}
\label{app:InversionSymmetry}

In this section, we give two examples of inversion symmetry discussed in section \ref{sec:Inversions}. In Feffereman-Graham gauge, the AdS$_{3}$ metric 
\begin{equation}
ds^{2}=\frac{dy^{2}+dzd\bar{z}}{y^{2}}-\frac{S\left(z\right)}{2}dz^{2}-\frac{\bar{S}\left(\bar{z}\right)}{2}d\bar{z}^{2}+y^{2}\frac{S\left(z\right)\bar{S}\left(\bar{z}\right)}{4}dzd\bar{z}
\end{equation}
can be obtained from the pure Poincare metric $ds^{2}=\frac{du^{2}+dxd\bar{x}}{u^{2}}$
with transformations (\ref{eq:OurFavoriteDiff}), where $S\left(z\right)$ and $\bar{S}\left(\bar{z}\right)$
are given by Schwarzian derivatives $S\left(z\right)=\left\{ f\left(z\right),z\right\} $, $\bar{S}\left(\bar{z}\right)=\left\{ \bar{f}\left(\bar{z}\right),\bar{z}\right\} $.
This metric has an inversion symmetry, because the same metric can
be obtained by the same functions $f,\bar{f}$, but with inverse
arguments, i.e. $f\left(\frac{1}{\bar{z}}\right),\bar{f}\left(\frac{1}{z}\right)$.
Specifically, the inversion corresponds to the identification between
unprimed and primed coordinates through the relations
\begin{align}
\bar{u}\left(y,f\left(z\right),\bar{f}\left(\bar{z}\right)\right) & =u\left(y',f\left(\frac{1}{\bar{z}'}\right),\bar{f}\left(\frac{1}{z'}\right)\right)\nonumber \\
x\left(y,f\left(z\right),\bar{f}\left(\bar{z}\right)\right) & =x\left(y',f\left(\frac{1}{\bar{z}'}\right),\bar{f}\left(\frac{1}{z'}\right)\right)\label{eq:InversionEquations}\\
\bar{x}\left(y,f\left(z\right),\bar{f}\left(\bar{z}\right)\right) & =\bar{x}\left(y',f\left(\frac{1}{\bar{z}'}\right),\bar{f}\left(\frac{1}{z'}\right)\right)\nonumber 
\end{align}
The solutions to these equations (i.e. $\left(y,z,\bar{z}\right)$
in terms of $\left(y',z',\bar{z}'\right)$) are often rather
complicated. Here, we give two examples: the Poincare AdS$_{3}$ and
BTZ black holes.

\subsubsection*{Poincare AdS$_{3}$ }
The Poincare metric $ds^2=\frac{dy^2+dz d\bar z}{y^2}$ can be simply obtained by $f\left(z\right)=z,\bar{f}\left(\bar{z}\right)=\bar{z}$.
So we have $f\left(\frac{1}{\bar{z}'}\right)=\frac{1}{\bar{z}'}$
and $\bar{f}\left(\frac{1}{z'}\right)=\frac{1}{z'}$. And equations
(\ref{eq:InversionEquations}) become
\begin{align}
\bar{u}\left(y,z,\bar{z}\right) & =y=\bar{u}\left(y',\frac{1}{z'},\frac{1}{\bar{z}'}\right)=\frac{y'}{y'^{2}+z'\bar{z}'}\nonumber \\
x\left(y,z,\bar{z}\right) & =z=x\left(y',\frac{1}{z'},\frac{1}{\bar{z}'}\right)=\frac{z'}{y'^{2}+z'\bar{z}'}\label{eq:InversionVaccumSolution}\\
\bar{x}\left(y,z,\bar{z}\right) & =\bar{z}=\bar{x}\left(y',\frac{1}{z'},\frac{1}{\bar{z}'}\right)=\frac{\bar{z}'}{y'^{2}+z'\bar{z}'}\nonumber 
\end{align}
where the relations between $\left(y,z,\bar{z}\right)$ and $\left(y',z',\bar{z}'\right)$ are manifest.

\subsubsection*{BTZ black holes}

The BTZ black hole case is more relevant to this work; it is also more complicated. To obtain the BTZ black hole metric (\ref{eq:HeavySource})
in terms of $\left(y,z,\bar{z}\right)$, we used $f\left(z\right)=z^{ir_{+}},\bar{f}\left(\bar{z}\right)=\bar{z}^{-ir_{+}}$,
with $r_{+}=\sqrt{\frac{24h_{H}}{c}-1}$. So we have $f\left(\frac{1}{\bar{z}'}\right)=\bar{z}^{'-ir_{+}},\bar{f}\left(\frac{1}{z'}\right)=z^{'ir_{+}}$.
Then equations (\ref{eq:InversionEquations}) become
\begin{smaller}
\begin{align}
u\left(y,z^{ir_{+}},\bar{z}^{-ir_{+}}\right) & =\frac{4r_{+}yz^{\frac{1}{2}+\frac{ir_{+}}{2}}\bar{z}^{\frac{1}{2}-\frac{ir_{+}}{2}}}{4z\bar{z}+\left(r_{+}^{2}+1\right)y^{2}}=u\left(y',\bar{z}^{'-ir_{+}},z^{'ir_{+}}\right)=\frac{4r_{+}y'z'^{\frac{1}{2}+\frac{ir_{+}}{2}}\bar{z}'^{\frac{1}{2}-\frac{ir_{+}}{2}}}{4z'\bar{z}'+\left(r_{+}^{2}+1\right)y'^{2}} \\
x\left(y,z^{ir_{+}},\bar{z}^{-ir_{+}}\right) & =\frac{z^{ir_{+}}\left(4z\bar{z}-\left(r_{+}-i\right){}^{2}y^{2}\right)}{4z\bar{z}+\left(r_{+}^{2}+1\right)y^{2}}=x\left(y',\bar{z}^{'-ir_{+}},z^{'ir_{+}}\right)=\frac{\bar{z}'^{-ir_{+}}\left(4z'\bar{z}'-\left(r_{+}+i\right){}^{2}y'^{2}\right)}{4z'\bar{z}'+\left(r_{+}^{2}+1\right)y'^{2}}\nn\\
\bar{x}\left(y,z^{ir_{+}},\bar{z}^{-ir_{+}}\right) & =\frac{\bar{z}^{-ir_{+}}\left(4z\bar{z}-\left(r_{+}+i\right){}^{2}y^{2}\right)}{4z\bar{z}+\left(r_{+}^{2}+1\right)y^{2}}=\bar{x}\left(y',\bar{z}^{'-ir_{+}},z^{'ir_{+}}\right)=\frac{z'^{ir_{+}}\left(4z'\bar{z}'-\left(r_{+}-i\right){}^{2}y'^{2}\right)}{4z'\bar{z}'+\left(r_{+}^{2}+1\right)y'^{2}}\nn
\end{align}
\end{smaller}
where the solution gives the coordinate relations after an inversion:
\begin{align}
y & =\frac{2}{\sqrt{z'\bar{z}'}\tilde{r}^2}\left(\frac{r-\sqrt{r^{2}-r_{+}^{2}-1}}{r_{+}^{2}+1}\right)\nonumber \\
z & =\frac{1}{\bar{z}'\tilde{r}^2}\\
\bar{z} & =\frac{1}{z'\tilde{r}^2}\nonumber 
\end{align}
with $\tilde{r}=\left(\frac{r+ir_{+}\sqrt{r^{2}-r_{+}^{2}-1}}{(1+ir_{+})\sqrt{r^{2}-r_{+}^{2}}}\right)^{\frac{i}{r_{+}}}$
and $r=\frac{\left(r_{+}^{2}+1\right)y'^{2}+4z'\bar{z}'}{4y'\sqrt{z'\bar{z}'}}$.
We emphasize that although the above solution looks  complicated,
in terms of the BTZ metric (\ref{eq:UsualBTZMetric}) in coordinates $\left(r,t_{E},\phi\right)$,
this just corresponds to the time reversal symmetry $t_{E}\rightarrow-t_{E}$.
One can also  check that expanding the above solution in small $h_{H}$, the leading term are indeed given by the inversion solution
(\ref{eq:InversionVaccumSolution}) for the pure Poincare metric.

\subsection{Elementary Note on Diffeomorphisms and Conformal Symmetries}

Here we will make some very elementary comments about bulk diffeomorphisms and boundary conformal transformations.  These ideas are probably well-known among experts, but they are rarely stated explicitly, so for completeness we will briefly review them.  The ultimate point is to contrast the diffeomorphism (\ref{eq:OurFavoriteDiff}) with a different and more naive procedure for implementing conformal transformations in AdS/CFT.  First, let us remind ourselves of a trivial point concerning the definition of conformal transformations.

Consider a CFT$_2$ in the metric $ds ^2= dz d \bar z$.  If we introduce new coordinate labels via $z \equiv f(\zeta)$, then we obtain a new expression for the metric, so $ds ^2= dz d \bar z = f' \bar f' d\zeta d \bar \zeta$.  This is the same physical metric; we have just re-written it using a different set of labels for the points.  However, if we now perform a Weyl transformation and multiply our metric by $\frac{1}{f' \bar f'}$, then we obtain a physically distinct metric $ds'^2 = d \zeta d \bar \zeta$.  This metric once again appears flat, but distances between points have clearly changed as a consequence of the Weyl factor.  The key point is that the metrics are physically different because we have fixed the relation $z \equiv f(\zeta)$.

Now let us consider the transformation rule for a primary operator.  When we transform from $\CO(z)$ to $\CO(\xi)$, what we really mean is that we define $z \equiv f(\xi)$ \emph{and} we change the metric from $dz d \bar z \to d \xi d \bar \xi$.  Then the transformation rule is
\be
(d \xi)^h \CO(\xi) = (d z)^h \CO(z)
\ee
Since we have that $z = f(\xi)$ this means that as usual
\be
\CO(\xi) = (f'(\xi))^h \CO(z).
\ee
For example, we can verify the standard result for $f(\xi) = e^\xi$ that
\be
\< \CO(\xi_1) \CO(\xi_2) \> = \left( e^{\xi_1} e^{\xi_2} \right)^h \left( \frac{1}{\left(e^{\xi_1} - e^{\xi_2} \right)^2 } \right)^h = \left( \frac{1}{2 \sinh \left(\frac{\xi_1 - \xi_2}{2} \right)} \right)^{2 h}
\ee
providing  a quick check of the logic.

Now we can see why the diffeomorphism of equation (\ref{eq:OurFavoriteDiff}) implements a general conformal transformation in the CFT$_2$.  Under this transformation, the boundary metric $ds^2 = dx d \bar x$ corresponding to the limit $n \to 0$ becomes a new boundary metric $ds^2 = dz d \bar z$ when we take the (different) limit $y \to 0$.  Though these boundary metrics appear identical, they are physically distinct, since by definition $x = f(z)$.  

We have utilized the bulk diffeomorphism (\ref{eq:OurFavoriteDiff}) to move the CFT from one spacetime metric to another via a (Virasoro) conformal transformation.  We can distinguish this operation from another kinematical procedure, which appears to function in any number of spacetime dimensions, and is often discussed in the context of the null cone embedding of AdS/CFT.  In this procedure we write
\be
ds^2 = \frac{dy^2 + dx_i^2}{y^2}
\ee
and then take $y = \epsilon F(x_i)$ followed by $\epsilon \to 0$, resulting in a boundary metric
\be
\label{eq:GeneralWeyl}
ds^2 = \frac{1}{F^2(x_i)} dx_i^2
\ee
that differs from the flat metric by a completely general Weyl factor.  

While this procedure appears to correctly implement the transformation rule for primary operators, it is purely kinematical.  In this sense it is somewhat misleading, as knowledge of CFT$_d$ correlators in flat spacetime does not determine the correlators in the general metric of equation (\ref{eq:GeneralWeyl}).  For example, this procedure does not account for effects such as the expectation value of the stress tensor in the new background metric, which arises automatically (as a Schwarzian derivative of the conformal transformation) when we use the diffeomorphism (\ref{eq:OurFavoriteDiff}) in the context of AdS$_3$/CFT$_2$.  Thus the diffeomorphism (\ref{eq:OurFavoriteDiff})  correctly implements conformal transformations in CFT$_2$; the fact that no equivalent diffeomorphism exists in higher dimensions reflects the physical fact that the conformal group is finite dimensional, and cannot be used to implement non-constant Weyl transformations.

\section{Bulk-boundary Vacuum Block via OPE Blocks}
\label{app:OPEBlocks}

In this section, we'll use the OPE block formalism developed in \cite{Anand:2017dav, Fitzpatrick:2016mtp} to compute the vacuum bulk-boundary block $\mathcal{V}_{0}=\left\langle \mathcal{O}_{H}\left(\infty\right)\mathcal{O}_{H}\left(1\right)\mathcal{P}_{0}\mathcal{O}_{L}\left(z,\bar{z}\right)\phi_{L}\left(y,0,0\right)\right\rangle $
up to order $1/c^{2}$. Here, we are considering the large $c$ limit, with $h_L$ and $h_H$ fixed.

The vacuum bulk-boundary OPE block for $\mathcal{O}_{L}\phi_{L}$
is given by 
\begin{equation}
\frac{\phi_{L}\left(y,0,0\right)\mathcal{O}_{L}\left(z,\bar{z}\right)}{\langle \phi_{L}\left(y,0,0\right)\mathcal{O}_{L}\left(z,\bar{z}\right)\rangle}=e^{K_{T}^{\text{bulk}}+K_{\bar{T}}^{\text{bulk}}+K_{T\bar{T}}^{\text{bulk}}+K_{TT}^{\text{bulk}}+K_{\bar{T}\bar{T}}^{\text{bulk}}+\cdots}
\end{equation}
 with
\begin{align}
K_{T}^{\text{bulk}} & =\frac{12h_L}{c}\int_{0}^{z}dz'\frac{\left(y^{2}+z'\bar{z}\right)\left(z-z'\right)}{z\bar{z}+y^{2}}T\left(z'\right),\nonumber \\
K_{TT}^{\text{bulk}} & =\frac{72h_L}{c^{2}\left(y^{2}+z\bar{z}\right)^{2}}\int_{0}^{z}dz'\int_{0}^{z'}dz''\left(z-z'\right)^{2}\left(y^{2}+\bar{z}z''\right)^{2}T\left(z'\right)T\left(z''\right),\\
K_{T\bar{T}}^{\text{bulk}} & =-\frac{72y^{2}h_L}{c^{2}\left(y^{2}+z\bar{z}\right){}^{2}}\int_{0}^{z}dz'\left(z-z'\right)^{2}\int_{0}^{\bar{z}}d\bar{z}'\left(\bar{z}-\bar{z}'\right)^{2}T\left(z'\right)\bar{T}\left(\bar{z}'\right).\nonumber 
\end{align}
$K_{\bar{T}}^{\text{bulk}}$ is anti-holomorphic version of $K_{T}^{\text{bulk}}$
, that is, $K_{T}^{\text{bulk}}$ with $T\rightarrow\bar{T}$
and $z\leftrightarrow\bar{z}$, and similarly $K_{\bar{T}\bar{T}}^{\text{bulk}}$
is the anti-holomorphic version of $K_{TT}^{\text{bulk}}$ . The OPE
block of $\mathcal{O}_{H}\mathcal{O}_{H}$ factorizes and we have
\[
\mathcal{O}_{H}\left(\infty\right)\mathcal{O}_{H}\left(1\right)=e^{K_{T}^{\text{bdy}}+K_{TT}^{\text{bdy}}+\cdots}e^{K_{\bar{T}}^{\text{bdy}}+K_{\bar{T}\bar{T}}^{\text{bdy}}+\cdots}
\]
with
\begin{align}
K_{T}^{\text{bdy}} & =\frac{12h_{H}}{c}\int_{1}^{\infty}dz'\left(z'-1\right)T\left(z'\right),\\
K_{TT}^{\text{bdy}} & =\frac{72h_{H}}{c^{2}}\int_{1}^{\infty}dz'\int_{1}^{z'}dz''\left(z''-1\right)^{2}T\left(z'\right)T\left(z''\right).\nonumber 
\end{align}
and the anti-holomorphic $K_{\bar{T}}^{\text{bdy}}$ and $K_{\bar{T}\bar{T}}^{\text{bdy}}$.
The superscript ``$\text{bdy}$'' means ``boundary''.

To obtain the vacuum block $\mathcal{V}_{0}$, we need to compute
the correlation functions of OPE blocks of $\phi_{L}\mathcal{O}_{L}$
and $\mathcal{O}_{H}\mathcal{O}_{H}$. The one-holomorphic-graviton-exchange
contribution is 
\begin{align}
\left\langle K_{T}^{\text{bulk}}K_{T}^{\text{bdy}}\right\rangle = & \left\langle \frac{12h_{L}}{c}\int_{0}^{z}dz'\frac{\left(y^{2}+z'\bar{z}\right)\left(z-z'\right)}{z\bar{z}+y^{2}}T\left(z'\right)\frac{12h_{H}}{c}\int_{1}^{\infty}dz''\left(z''-1\right)T\left(z''\right)\right\rangle \nonumber \\
= & -\frac{12h_{L}h_{H}}{c}\frac{\left(z\left(2\bar{z}+y^{2}\right)+\log(1-z)\left(y^{2}-(z-2)\bar{z}\right)\right)}{\left(z\bar{z}+y^{2}\right)},\label{eq:OneGravitonExchangOPEBlocks}
\end{align}
and the one-anti-holomorphic-graviton-exchange contribution $\left\langle K_{\bar{T}}^{\text{bulk}}K_{\bar{T}}^{\text{bdy}}\right\rangle $
is simplify $\left\langle K_{T}^{\text{bulk}}K_{T}^{\text{bdy}}\right\rangle $
with $z,\bar{z}$ exchanged, ie $z\leftrightarrow\bar{z}$.

The one-graviton-exchange contribution computed above is order $1/c$,
and two-graviton-exchanges will contribution at order $1/c^{2}$.
There are three types of two-graviton-exchanges,
\begin{align}
\mathcal{K}_{1}\equiv & \left\langle \left(K_{T\bar{T}}^{\text{bulk}}+K_{T}^{\text{bulk}}K_{\bar{T}}^{\text{bulk}}\right)\left(K_{T}^{\text{bdy}}K_{\bar{T}}^{\text{bdy}}\right)\right\rangle ,\\
\mathcal{K}_{2}\equiv & \left\langle \left(\frac{K_{T}^{\text{bulk}}K_{T}^{\text{bulk}}}{2}+K_{TT}^{\text{bulk}}\right)\left(\frac{K_{T}^{\text{bdy}}K_{T}^{\text{bdy}}}{2}+K_{TT}^{\text{bdy}}\right)\right\rangle ,\\
\mathcal{K}_{3}\equiv & \left\langle \left(\frac{K_{\bar{T}}^{\text{bulk}}K_{\bar{T}}^{\text{bulk}}}{2}+K_{\bar{T}\bar{T}}^{\text{bulk}}\right)\left(\frac{K_{\bar{T}}^{\text{bdy}}K_{\bar{T}}^{\text{bdy}}}{2}+K_{\bar{T}\bar{T}}^{\text{bdy}}\right)\right\rangle ,
\end{align}
where we've grouped them by contributions from different types of
gravitons. $\mathcal{K}_{3}$ is simply the anti-holomorphic version
of $\mathcal{K}_{2}$, ie, $\mathcal{K}_{2}$ with $z$, $\bar{z}$
exchanged, so we'll focus on $\mathcal{K}_{1}$ and $\mathcal{K}_{2}$. 

$\mathcal{K}_{1}$ is contribution from exchanges of one holomorphic
graviton and one anti-holomorphic graviton. The first term in $\mathcal{K}_{1}$
is
\begin{align}
 & \left\langle K_{T\bar{T}}\left(K_{T}^{\text{bdy}}K_{\bar{T}}^{\text{bdy}}\right)\right\rangle \\
= & -\frac{10368h_{H}^{2}h_{L}y^{2}}{c^{4}\left(y^{2}+z\bar{z}\right)^{2}}\int_{0}^{z}dz'\int_{0}^{\bar{z}}d\bar{z}'\int_{1}^{\infty}dz''\int_{1}^{\infty}d\bar{z}''\left(z-z'\right)^{2}\left(\bar{z}-\bar{z}'\right)^{2}\left(z''-1\right)\left(\bar{z}''-1\right)\nonumber \\
 & \times\left\langle T\left(z'\right)\bar{T}\left(\bar{z}'\right)T\left(z''\right)\bar{T}\left(\bar{z}''\right)\right\rangle \nonumber \\
= & -\frac{72h_{L}h_{H}^{2}y^{2}}{c^{2}\left(y^{2}+z\bar{z}\right){}^{2}}\left((z-2)z+2(z-1)\log(1-z)\right)\left(\left(\bar{z}-2\right)\bar{z}+2\left(\bar{z}-1\right)\log\left(1-\bar{z}\right)\right),\nonumber 
\end{align}
where we've used $\left\langle T\left(z'\right)\bar{T}\left(\bar{z}'\right)T\left(z''\right)\bar{T}\left(\bar{z}''\right)\right\rangle =\frac{c^{2}}{4}\frac{1}{\left(z''-z'\right)^{4}}\frac{1}{\left(\bar{z}''-\bar{z}'\right)^{4}}$.
The second term in $\mathcal{K}_{1}$ is
\begin{small}
\begin{align}
 & \left\langle K_{T}K_{\bar{T}}\left(K_{T}^{\text{bdy}}K_{\bar{T}}^{\text{bdy}}\right)\right\rangle \\
= & \frac{20736h_{H}^{2}h_{L}^{2}}{c^{4}\left(y^{2}+z\bar{z}\right)^{2}}\int_{0}^{z}dz'\int_{0}^{\bar{z}}d\bar{z}'\int_{1}^{\infty}dz''\int_{1}^{\infty}d\bar{z}''\left(y^{2}+z'\bar{z}\right)\left(z-z'\right)\left(y^{2}+\bar{z}'z\right)\left(\bar{z}-\bar{z}'\right)\nonumber \\
 & \times\left(z''-1\right)\left(\bar{z}''-1\right)\left\langle T\left(z'\right)\bar{T}\left(\bar{z}'\right)T\left(z''\right)\bar{T}\left(\bar{z}''\right)\right\rangle \nonumber \\
= & \frac{144h_{L}^{2}h_{H}^{2}\left(\bar{z}\left(y^{2}+2z\right)+\left(y^{2}-z\left(\bar{z}-2\right)\right)\log\left(1-\bar{z}\right)\right)\left(z\left(2\bar{z}+y^{2}\right)+\left(y^{2}-(z-2)\bar{z}\right)\log(1-z)\right)}{c^{2}\left(y^{2}+z\bar{z}\right)^{2}}.\nonumber 
\end{align}
\end{small}
And the sum of these two terms gives us
\begin{small}
\begin{align*}
\mathcal{K}_{1}= & \frac{72h_{H}^{2}h_{L}}{c^{2}\left(y^{2}+z\bar{z}\right)^{2}}\left[y^{2}\left((2-z)z+2(1-z)\log(1-z)\right)\left(\left(\bar{z}-2\right)\bar{z}+2\left(\bar{z}-1\right)\log\left(1-\bar{z}\right)\right)\right.\\
 & \left.+2h_{L}\left(z\left(2\bar{z}+y^{2}\right)+\log(1-z)\left(y^{2}-(z-2)\bar{z}\right)\right)\left(\bar{z}\left(y^{2}+2z\right)+\left(y^{2}+2z-z\bar{z}\right)\log\left(1-\bar{z}\right)\right)\right]. 
\end{align*}
\end{small}
 Similarly, $\mathcal{K}_{2}$ is given by
 \begin{smaller}
\begin{align*}
\mathcal{K}_{2}= & \frac{72h_{H}h_{L}}{c^{4}\left(y^{2}+z\bar{z}\right)^{2}}\int_{0}^{z}dz'\int_{0}^{z'}dz''\left[2h_{L}\left(y^{2}+z'\bar{z}\right)\left(z-z'\right)\left(y^{2}+z''\bar{z}\right)\left(z-z''\right)+\left(z-z'\right)^{2}\left(y^{2}+\bar{z}z''\right)^{2}\right]\\
 & \ensuremath{\times\int_{1}^{\infty}dz'''\int_{1}^{z'''}dz''''\left[2h_{H}\left(z'''-1\right)\left(z''''-1\right)+\left(z''''-1\right)^{2}\right]\left\langle \left[T\left(z'\right)T\left(z''\right)\right]\left[T\left(z'''\right)T\left(z''''\right)\right]\right\rangle }\\
= & \resizebox{\textwidth}{!}{\text{\ensuremath{\frac{72h_{H}h_{L}}{c^{2}\left(z\bar{z}+y^{2}\right)^{2}}\left[\log^{2}(1-z)\left(h_{H}\left(h_{L}\left(y^{2}-(z-2)\bar{z}\right)^{2}-2(z-1)\bar{z}\left(\bar{z}+y^{2}\right)\right)+(1-z)\bar{z}\left(2h_{L}\left(\bar{z}+y^{2}\right)-z\bar{z}+\bar{z}\right)\right)-2\text{Li}_{2}(z)\left(2y^{2}\bar{z}-(z-2)z\bar{z}^{2}+y^{4}\right)\right.}}}\\
 & \resizebox{\textwidth}{!}{\text{\ensuremath{+\frac{1}{12}\left(2y^{2}z\bar{z}\left(6h_{H}\left(z\left(4h_{L}-1\right)-6\right)-6(z+6)h_{L}-z+34\right)+16z^{2}\bar{z}^{2}\left(3h_{H}\left(h_{L}-1\right)-3h_{L}+2\right)+y^{4}\left(2h_{L}\left(z^{2}\left(6h_{H}+1\right)+6z-24\right)+z^{2}\left(6h_{H}+1\right)+z\left(10-36h_{H}\right)+24\right)\right)}}}\\
 & \resizebox{\textwidth}{!}{\text{\ensuremath{\left.+\frac{\log(1-z)\left(-2y^{2}z\bar{z}\left(6h_{H}\left(z^{2}h_{L}-4zh_{L}-2z+3\right)-12zh_{L}+18h_{L}+6z-5\right)-(z-2)z^{2}\bar{z}^{2}\left(6h_{H}\left(4h_{L}-1\right)-6h_{L}-1\right)+y^{4}\left(6h_{L}\left(2z^{2}h_{H}+3z-4\right)+12z^{2}h_{H}-18zh_{H}-13z+12\right)\right)}{6z}\right]}}}.
\end{align*}
\end{smaller}The four-point function of $T$ in the second line is regularized
as in \cite{Anand:2017dav, Fitzpatrick:2016mtp} and it's given by 
$\left\langle \left[T\left(z'\right)T\left(z''\right)\right]\left[T\left(z'''\right)T\left(z''''\right)\right]\right\rangle =\frac{c^{2}}{4}\left(\frac{1}{\left(z'-z'''\right)^{4}}\frac{1}{\left(z''-z''''\right)^{4}}+\frac{1}{\left(z''-z'''\right)^{4}}\frac{1}{\left(z'-z''''\right)^{4}}\right) +\mathcal{O}(c).$
Note that the above equation, in the first term of the first line,
we've made use of the symmetry between $z'$ and $z''$ to change
the integration range from $\int_{0}^{z}dz'\int_{0}^{z}dz''$ to $\frac{1}{2}\int_{0}^{z}dz'\int_{0}^{z'}dz''$,
so that both terms of the first line have the same integration ranges.
And similarly for the second line.

Adding up all the above contributions, we obtain the result for the
bulk-boundary vacuum block $\mathcal{V}_{0}$ up to order $1/c^{2}$,
and it's given by
\begin{equation}
\mathcal{V}_{0}=\left(\frac{y}{y^2+ z \bar z}\right)^{2h_L}\left[1+\left\langle K_{T}^{\text{bulk}}K_{T}^{\text{bdy}}\right\rangle +\left\langle K_{\bar{T}}^{\text{bulk}}K_{\bar{T}}^{\text{bdy}}\right\rangle +\mathcal{K}_{1}+\mathcal{K}_{2}+\mathcal{K}_{3}+\mathcal{O}\left(\frac{1}{c^{3}}\right)\right].
\end{equation}

In Section \ref{sec:Inversions} and Appendix \ref{app:InversionSymmetry}, we discussed the inversion symmetry, i.e. the symmetry under $t_E\rightarrow -t_E$. Here, we would to comment that the above large $c$ expansion of $\mathcal{V}_0$ is symmetric under $t_E\rightarrow -t_E$ at order $1/c$ but not $1/c^2$. The order $1/c$ terms of the above result is just the same as the $1/c$ terms of the semiclassical result when expanded at large $c$. But at $1/c^2$, there are quantum-correction terms in the above result that are not included in the semiclassical result, which breaks this symmetry if we use the naive semiclassical transformation (\ref{eq:yzzbToBTZCoordinates}) and expand to order $1/c^2$.

\section{Details of the Recursion Relation and Algorithm}
\label{app:RRdetails}
In this section, we'll analyze the structure of the proto-field $\phi\left(y,z,\bar{z}\right)$
and bulk-boundary blocks in more details and explain why the recursion relation of Section \ref{sec:ZRR} works.

In the main text, we've written the proto-field as a sum over descendant
levels $N$ as follows
\begin{equation}
\phi\left(y,z,\bar{z}\right)=y^{2h}\sum_{n=0}^{\infty}\left(-1\right)^{n}y^{2n}\lambda_{n}^{(h)}\mathcal{L}_{-n}\bar{\mathcal{L}}_{-n}\mathcal{O}\left(z,\bar{z}\right), \qquad \lambda_{m}^{\left(h\right)}\equiv\frac{1}{\left(2h\right)_{m}m!}.
\end{equation}
with $\mathcal{L}_{-n}$ and $\bar{\mathcal{L}}_{-n}$ uniquely
determined by the bulk primary condition (\ref{eq:BulkPrimaryCondition}) and the normalization
condition (\ref{eq:NormalizationCondition}). As shown in \cite{Anand:2017dav}, we can solve these conditions and
write the proto-field $\phi$ as a sum over quasi-primaries and their
global descendants. That is, we can write $\phi$ as
\begin{equation}
\phi\left(y,z,\bar{z}\right)=\sum_{n,\bar{n}}\sum_{i,j}\phi_{i,j}^{n,\bar{n}}
\end{equation}
where $\phi_{i,j}^{n,\bar{n}}$ means the contribution to $\phi$
from the $i$th level $n$ holomorphic quasi-primary and $j$th level
$\bar{n}$ anti-holomorphic quasi-primary and their global descendants.
And in the above sum, we sum over all quasi-primaries. Here we've
assume that all the quasi-primaries are orthogonal. It can be shown
that $\phi_{i,j}^{n,\bar{n}}$ is given by \cite{Anand:2017dav}
\begin{small}
\begin{equation}
\phi_{i,j}^{n,\bar{n}}\left(y,z,\bar{z}\right)\equiv y^{2h+2n}\sum_{m=0}^{\infty}\frac{\left(-1\right)^{n+m}y^{2m}}{\lambda_{n+m}^{\left(h\right)}}\left(\frac{L_{-1}^{m}\mathcal{L}_{-n}^{\text{quasi},i}}{\left|L_{-1}^{m}\mathcal{L}_{-n}^{\text{quasi},i}\mathcal{O}\right|^{2}}\right)\left(\frac{\bar{L}_{-1}^{m+n-\bar{n}}\mathcal{\bar{L}}_{-\bar{n}}^{\text{quasi},j}}{\left|\bar{L}_{-1}^{m+n-\bar{n}}\mathcal{\bar{L}}_{-\bar{n}}^{\text{quasi},j}\mathcal{O}\right|^{2}}\right)\mathcal{O}\left(z,\bar{z}\right).
\end{equation}
\end{small}
In writing down the above equation, we've assumed that $n\ge\bar{n}$,
but the case with $n<\bar{n}$ is similar. One interesting fact about
the above equation is that the contribution to $\phi_{i,j}^{n,\bar{n}}$
from each descendant of $\mathcal{O}$ is normalized by its norm with
other factors independent of the central charge $c$. As we'll see,
this is a feature that also holds for the the Virasoro projection
operator. And this is one of the reason that we can use the $c$-recursion
to compute the bulk-boundary blocks. 

Similar to $\phi$, the holomorphic part of the proto-field $\tilde{\phi}_{h}^{\text{holo}}$
defined in equation (\ref{eq:HoloField}), i.e
\begin{equation}
\tilde{\phi}_{h}^{\text{holo}}\left(y,z,\bar{z}\right)=y^{2h}\sum_{n=0}^{\infty}\lambda_n^{(h)} y^{2n}\mathcal{L}_{-n}\mathcal{O}_{h}\left(z,\bar{z}\right).
\end{equation}
can be written as a sum over contributions from different quasi-primaries
as
\begin{equation}
\tilde{\phi}_{h}^{\text{holo}}\left(y,z,\bar{z}\right)=\sum_{n=0}^{\infty}\phi_{i}^{n}
\end{equation}
with
\begin{equation}
\phi_{i}^{n}\left(y,z,\bar{z}\right)=\frac{y^{2h+2n}}{\left|\mathcal{L}_{-n}^{\text{quasi},i}\mathcal{O}\right|^{2}}\sum_{m=0}^{\infty}\lambda_{m}^{\left(h+n\right)}y^{2m}L_{-1}^{m}\mathcal{L}_{-n}^{\text{quasi},i}\mathcal{O}\left(z,\bar{z}\right).\label{eq:holophiQP}
\end{equation}
In order to make the structure of the holomorphic bulk-boundary block
$\mathcal{V}_{\text{holo}}\left(h_{1},h_{2},c\right)$ more transparent,
we can also write the holomorphic Virasoro projection operator in terms of quasi-primaries
and their global descendants:
\begin{align}
\mathcal{P}^{\text{holo}}_{h_{1}} & =\sum_{m=0}^{\infty}\sum_{j}\sum_{m_{1}=0}^{\infty}\frac{\left|L_{-1}^{m_{1}}\mathcal{L}_{-m}^{\text{quasi},j}\mathcal{O}_{h_{1}}\right\rangle \left\langle L_{-1}^{m_{1}}\mathcal{L}_{-m}^{\text{quasi},j}\mathcal{O}_{h_{1}}\right|}{\left|L_{-1}^{m_{1}}\mathcal{L}_{-m}^{\text{quasi},j}\mathcal{O}_{h_{1}}\right|^{2}}\label{eq:ProjectionQP}\\
 & =\sum_{m=0}^{\infty}\sum_{j}\frac{1}{\left|\mathcal{L}_{-m}^{\text{quasi},j}\mathcal{O}_{h_{1}}\right|^{2}}\left[\sum_{m_{1}=0}^{\infty}\lambda_{m_{1}}^{\left(h_{1}+m\right)}\left|L_{-1}^{m_{1}}\mathcal{L}_{-m}^{\text{quasi},j}\mathcal{O}_{h_{1}}\right\rangle \left\langle L_{-1}^{m_{1}}\mathcal{L}_{-m}^{\text{quasi},j}\mathcal{O}_{h_{1}}\right|\right]\nonumber. 
\end{align}
Plugging equation (\ref{eq:holophiQP}) and (\ref{eq:ProjectionQP}) 
into the definition of $\mathcal{V}_{\text{holo}}\left(h_{1},h_{2},c\right)$,
we obtain
\begin{align}
\mathcal{V}_{\text{holo}}\left(h_{1},h_{2},c\right)\equiv & \left\langle \mathcal{O}_{H}\mathcal{O}_{H}\mathcal{P}^{\text{holo}}_{h_{1}}\mathcal{O}_{L}\phi_{h_{2}}^{\text{holo}}\left(y,0,0\right)\right\rangle \nonumber \\
= & y^{2h_2}\sum_{m,n=0}^{\infty}\sum_{j}\frac{\lambda_{m_1}^{\left(h_{1}+m\right)}}{\left|\mathcal{L}_{-m}^{\text{quasi},i}\mathcal{O}_{h_{1}}\right|^{2}}\frac{y^{2n}}{\left|\mathcal{L}_{-n}^{\text{quasi},j}\mathcal{O}_{h_2}\right|^{2}}\sum_{m_{1},m_{2}=0}^{\infty}\lambda_{m_{2}}^{\left(h_2+n\right)}y^{2m_{2}}\label{eq:VholoExplicit}\\
 & \times\left\langle \mathcal{O}_{H}\mathcal{O}_{H}\left|L_{-1}^{m_{1}}\mathcal{L}_{-m}^{\text{quasi},i}\mathcal{O}_{h_{1}}\right\rangle \left\langle L_{-1}^{m_{1}}\mathcal{L}_{-m}^{\text{quasi},i}\mathcal{O}_{h_{1}}\right|\mathcal{O}_{L}\left(z\right)L_{-1}^{m_{2}}\mathcal{L}_{-n}^{\text{quasi},j}\mathcal{O}_{h_{2}}\right\rangle \nonumber 
\end{align}
The two factors in the last line can be simplified to be
\begin{equation}
\left\langle \mathcal{O}_{H}|\mathcal{O}_{H}\left(1\right)|L_{-1}^{m_{1}}\mathcal{L}_{-m}^{\text{quasi},i}\mathcal{O}_{h_{1}}\right\rangle =\left(h_{1}+m\right)_{m_{1}}\left\langle \mathcal{O}_{H}|\mathcal{O}_{H}\left(1\right)|\mathcal{L}_{-m}^{\text{quasi},i}\mathcal{O}_{h_{1}}\right\rangle ,
\end{equation}
and
\begin{align}
 & \left\langle L_{-1}^{m_{1}}\mathcal{L}_{-m}^{\text{quasi},i}\mathcal{O}_{h_{1}}|\mathcal{O}_{L}\left(z\right)|L_{-1}^{m_{2}}\mathcal{L}_{-n}^{\text{quasi},j}\mathcal{O}_{h_2}\left(0,0\right)\right\rangle \\
= & s_{m_{1},m_{2}}\left(h_{1}+m,h_{L},h_{2}+n\right)z^{h_{1}+m_{1}-h_{2}-m_{2}}\left\langle \mathcal{L}_{-m}^{\text{quasi},i}\mathcal{O}_{h_{1}}|\mathcal{O}_{L}|\mathcal{L}_{-n}^{\text{quasi},j}\mathcal{O}_{h_{2}}\right\rangle ,\nonumber 
\end{align}
with $s_{m_{1},m_{2}}\left(h_{1}+m,h_{L},h_{2}+n\right)$ given in
(\ref{eq:skm}).

Now we can separate the factors in $\mathcal{V}_{\text{holo}}$ that
depend on $c$ (i.e. those terms that involve $\mathcal{L}^{\text{quasi}}$)
from those that don't depend on $c$, and write $\mathcal{V}_{\text{holo}}$
as a sum over global blocks
\begin{equation}
\mathcal{V}_{\text{holo}}\left(h_{1},h_{2},c\right)=\sum_{m,n=0}^{\infty}C_{m,n}G\left(h_{1}+m,h_{2}+n\right).
\end{equation}
with
\begin{equation}\label{eq:Cmn}
C_{m,n}=\sum_{i,j}\frac{\left\langle \mathcal{O}_{H}\mathcal{O}_{H}|\mathcal{L}_{-m}^{\text{quasi},i}\mathcal{O}_{h_{1}}\right\rangle \left\langle \mathcal{L}_{-m}^{\text{quasi},i}\mathcal{O}_{h_{1}}|\mathcal{O}_{L}\left(1\right)|\mathcal{L}_{-n}^{\text{quasi},j}\mathcal{O}_{h_{2}}\right\rangle }{\left|\mathcal{L}_{-m}^{\text{quasi},i}\mathcal{O}_{h_{1}}\right|^{2}\left|\mathcal{L}_{-n}^{\text{quasi},j}\mathcal{O}_{h_2}\right|^{2}}
\end{equation}
and the global blocks are
\begin{equation}
G\left(h_{1},h_{2}\right)=z^{h_{1}}\left(\frac{y^{2}}{z}\right)^{h_{2}}\sum_{m_{1},m_{2}=0}^{\infty}\frac{\left(h_{1}\right)_{m_{1}}s_{m_{1},m_{2}}\left(h_{1},h_{L},h_{2}\right)}{\left(2h_{1}\right)_{m1}m_{1}!\left(2h_{2}\right)_{m_{2}}m_{2}!}z^{m_{1}}\left(\frac{y^{2}}{z}\right)^{m_{2}}.
\end{equation}
It's easily seen from the above derivation that $G\left(h_{1}+m,h_{2}+n\right)$
are the contributions from the global descendants of quasi-primaries
of dimension $h_{1}+m$ and $h_{2}+n$. The sum over $i,j$ in equation (\ref{eq:Cmn}) is summing over the level-$m$ quasi-primaries of $\CO_{h_1}$ and level-$n$ quasi-primaries of $\CO_{h_2}$. So $C_{m,n}$ is the sum of the product of 3-pt functions of quasi-primaries with primaries normalized by the norms of the quasi-primaries, at specific levels.

A detail derivation of the recursion can then be obtained along the line
of \cite{Cho:2017oxl}. Basically, the function $A_{m,n}^{c}$ in (\ref{eq:Amnc})
encode the information about the norms of the states in the denominator
of (\ref{eq:Cmn}), and $P_{m,n}^{c}$ in (\ref{eq:Pmnc}) encodes the 3-pt functions
of one quasi-primaries with two primaries. In (\ref{eq:Cmn}), we
have a 3-pt function with 2 quasi-primaries, but  at the residues
of (\ref{eq:Recursion}), one of the quasi-primaries becomes a primary,
that's why $P_{m,n}^{c}$ can be used to compute this 3-pt function. The reason that the Zamolodchikov recursion relation can be modified to compute the bulk-boundary Virasoro blocks is that the the structure of the proto-field $\phi$ is very similar to the structure of the projection operator, i.e. the proto-field is built up of descendant states of $\CO$ normalized by their norms.

\subsection*{Algorithm for Solving the Recursion}
Solving the recursion (\ref{eq:Recursion}) (reproduced here for convenience)
\begin{align}\label{eq:RecursionApp}
\mathcal{V}_{\text{holo}}\left(h_{1},h_{2},c\right)= & \mathcal{V}_{\text{holo}}\left(h_{1},h_{2},c\rightarrow\infty\right)\\
 & +\sum_{m\ge2,n\ge1}\frac{R_{m,n}\left(h_{1},h_{2}\right)}{c-c_{m,n}\left(h_{1}\right)}\mathcal{V}_{\text{holo}}\left(h_{1}\rightarrow h_{1}+mn,h_{2},c\rightarrow c_{mn}\left(h_{1}\right)\right)\nonumber \\
 & +\sum_{m\ge2,n\ge1}\frac{S_{m,n}\left(h_{1},h_{2}\right)}{c-c_{m,n}\left(h_{2}\right)}\mathcal{V}_{\text{holo}}\left(h_{1},h_{2}\rightarrow h_{2}+mn,c\rightarrow c_{mn}\left(h_{2}\right)\right),\nonumber 
\end{align}
will give us the coefficients $C_{M,N}$ (here we use $M,N$ instead
of $m,n$ for clarity). The basic idea of the algorithm for obtaining $C_{M,N}$ is
similar to that of the algorithm for computing $\left\langle \phi\phi\right\rangle _{\text{holo}}$
in \cite{Chen:2017dnl} using the $c$-recursion relation, as described in detail
in Appendix D of that paper. Here, we briefly describe the algorithm for this more complicated recursion.

From recursion (\ref{eq:RecursionApp}), we know that $C_{M,N}$ get contribution from every
decomposition of $M,N$ in the following forms 
\begin{equation}
M=m_{1}\tilde{m}_{1}+...+m_{k}\tilde{m}_{k}...+m_{i}\tilde{m}_{i},\qquad N=n_{1}\tilde{n}_{1}+...+n_{l}\tilde{n}_{l}...+n_{j}\tilde{n}_{j}
\end{equation}
where $m_{k},\tilde{m}_{k},n_{l},\tilde{n}_{l}$ are integers with
$m_{k},n_{l}\ge2$ and $\tilde{m}_{k},\tilde{n}_l\ge1$ and the orders
of the products in the sums matter. We can imaging obtaining $\left(M,N\right)$
from $\left(0,0\right)$ step by step, where at each step, we either
choose $m_{k}\tilde{m}_{k}$ or $n_{l}\tilde{n}_{l}$. Different ways
of arriving at $\left(M,N\right)$ give different contributions to
$C_{M,N}$. Denoting the contribution to $C_{M,N}$ whose last step
is $m_{i}\tilde{m}_{i}$ as $C_{\left(M,m_{i},\tilde{m}_{i}\right),\left(N,n_{j},\tilde{n}_{j}\right),1}$
and the contribution to $C_{M,N}$ whose last step is $n_{j}\tilde{n}_{j}$
as $C_{\left(M,m_{i},\tilde{m}_{i}\right),\left(N,n_{j},\tilde{n}_{j}\right),2}$.
Then we have
\begin{equation}
C_{M,N}=\sum_{2\le m_{i}\tilde{m}_{i}\le M}\sum_{2\le n_{i}\tilde{n}_{j}\le N}\left[C_{\left(M,m_{i},\tilde{m}_{i}\right),\left(N,n_{j},\tilde{n}_{j}\right),1}+C_{\left(M,m_{i},\tilde{m}_{i}\right),\left(N,n_{j},\tilde{n}_{j}\right),2}\right].
\end{equation}
$C_{\left(M,m_{i},\tilde{m}_{i}\right),\left(N,n_{j},\tilde{n}_{j}\right),1}$
and $C_{\left(M,m_{i},\tilde{m}_{i}\right),\left(N,n_{j},\tilde{n}_{j}\right),2}$
are computed as follows (from special and simple to more general cases):
\begin{enumerate}
\item The simplest case is 
\begin{align}
C_{\left(m_{1}\tilde{m}_{1},m_{1},\tilde{m}_{1}\right),\left(0,0,0\right),1} & =\frac{R_{m_{1},\tilde{m}_{1}}\left(h_{1},h_{2}\right)}{c-c_{m_{1},\tilde{m}_{1}}\left(h_{1}\right)}\\
C_{\left(0,0,0\right),\left(n_{1}\tilde{n}_{1},n_{1},\tilde{n}_{1}\right),2} & =\frac{S_{n_{1},\tilde{n}_{1}}\left(h_{1},h_{2}\right)}{c-c_{n_{1},\tilde{n}_{1}}\left(h_{2}\right)}\nonumber 
\end{align}
where the recursion is only used once.
\item For the case with $N=0$ and $M-m_{i}\tilde{m}_{i}\ge2$,
we have
\begin{equation*}
\resizebox{0.9\textwidth}{!}{\text{\ensuremath{ C_{\left(M,m_{i},\tilde{m}_{i}\right),\left(0,0,0\right),1}=\sum_{2\le m_{k}\tilde{m}_{k}\le M-m_{i}\tilde{m}_{i}}\frac{R_{m_{i},\tilde{m}_{i}}\left(h_{1}+M-m_{i}\tilde{m}_{i},h_{2}\right)C_{(M-m_{i}\tilde{m}_{i},m_{k},\tilde{m}_{k}),\left(0,0,0\right),1}}{c_{m_{k},\tilde{m}_{k}}\left(h_{1}+M-m_{i}\tilde{m}_{i}-m_{k}\tilde{m}_{k}\right)-c_{m_{i},\tilde{m}_{i}}\left(h_{1}+M-m_{i}\tilde{m}_{i}\right)}}}}
\end{equation*}
Similarly for $M=0$ and $N-n_{j}\tilde{n}_{j}\ge2$, we have 
\begin{smaller}
\begin{equation*}
C_{\left(0,0,0\right),\left(N,n_{j},\tilde{n}_{j}\right),2}=\sum_{2\le n_{k}\tilde{n}_{k}\le N-n_{j}\tilde{n}_{j}}\frac{S_{n_{j},\tilde{n}_{j}}\left(h_{1},h_{2}+N-n_{j}\tilde{n}_{j}\right)C_{(0,0,0),\left(N-n_{j}\tilde{n}_{j},n_{k},\tilde{n}_{k}\right),2}}{c_{n_{k},\tilde{n}_{k}}\left(h_{2}+N-n_{j}\tilde{n}_{j}-n_{k}\tilde{n}_{k}\right)-c_{n_{j},\tilde{n}_{j}}\left(h_{2}+N-n_{j}\tilde{n}_{j}\right)}
\end{equation*}
\end{smaller}
\item For terms with $M-m_{i}\tilde{m}_{i}\ge2$ and $N=n_{1}\tilde{n}_{1}$,
we have 
\begin{align}
 & C_{\left(M,m_{i},\tilde{m}_{i}\right),\left(n_{1}\tilde{n}_{1},n_{1},\tilde{n}_{1}\right),1}\nonumber \\
= & \sum_{2\le m_{k}\tilde{m}_{k}\le M-m_{i}\tilde{m}_{i}}R_{m_{i},\tilde{m}_{i}}\left(h_{1}+M-m_{i}\tilde{m}_{i},h_{2}+n_1\tilde{n}_1\right)\\
 &\resizebox{0.9\textwidth}{!}{\text{\ensuremath{ \times\left[\frac{C_{(M-m_{i}\tilde{m}_{i},m_{k},\tilde{m}_{k}),\left(n_{1}\tilde{n}_{1},n_{1},\tilde{n}_{1}\right),1}}{c_{m_{k},\tilde{m}_{k}}\left(h_{1}+M-m_{i}\tilde{m}_{i}-m_{k}\tilde{m}_{k}\right)-c_{m_{i},\tilde{m}_{i}}\left(h_{1}+M-m_{i}\tilde{m}_{i}\right)}+\frac{C_{(M-m_{i}\tilde{m}_{i},m_{k},\tilde{m}_{k}),\left(n_{1}\tilde{n}_{1},n_{1},\tilde{n}_{1}\right),2}}{c_{n_1,\tilde{n}_1}\left(h_{2}\right)-c_{m_{i},\tilde{m}_{i}}\left(h_{1}+M-m_{i}\tilde{m}_{i}\right)}\right]}}}\nonumber 
\end{align}
and 
\begin{equation}
C_{\left(M,m_{i},\tilde{m}_{i}\right),\left(n_{1}\tilde{n}_{1},n_{1},\tilde{n}_{1}\right),2}=\frac{S_{n_{1},\tilde{n}_{1}}\left(h_{1}+M,h_{2}\right)C_{(M,m_{i},\tilde{m}_{i}),\left(0,0,0\right),1}}{c_{m_{i},\tilde{m}_{i}}\left(h_{1}+M-m_{i}\tilde{m}_{i}\right)-c_{n_{1},\tilde{n}_{1}}\left(h_{2}\right)}.
\end{equation}
And similarly for the case with $N-n_{j}\tilde{n}_{j}\ge2$ and $M=m_{1}\tilde{m}_{1}$. 
\item For the general case with $m-m_{i}\tilde{m}_{i}\ge2$ and $n-n_{j}\tilde{n}_{j}\ge2$,
we have 
\begin{align}
 & C_{\left(M,m_{i},\tilde{m}_{i}\right),\left(N,n_{j},\tilde{n}_{j}\right),1}\nonumber \\
= & \sum_{2\le m_{k}\tilde{m}_{k}\le M-m_{i}\tilde{m}_{i}}R_{m_{i},\tilde{m}_{i}}\left(h_{1}+M-m_{i}\tilde{m}_{i},h_{2}+N\right)\\
 & \times \resizebox{0.9\textwidth}{!}{\text{\ensuremath{\left[\frac{C_{(M-m_{i}\tilde{m}_{i},m_{k},\tilde{m}_{k}),\left(N,n_{j},\tilde{n}_{j}\right),1}}{c_{m_{k},\tilde{m}_{k}}\left(h_{1}+M-m_{i}\tilde{m}_{i}-m_{k}\tilde{m}_{k}\right)-c_{m_{i},\tilde{m}_{i}}\left(h_{1}+M-m_{i}\tilde{m}_{i}\right)}+\frac{C_{(M-m_{i}\tilde{m}_{i},m_{k},\tilde{m}_{k}),\left(N,n_{j},\tilde{n}_{j}\right),2}}{c_{n_{j},\tilde{n}_{j}}\left(h_{2}+N-n_{j}\tilde{n}_{j}\right)-c_{m_{i},\tilde{m}_{i}}\left(h_{1}+M-m_{i}\tilde{m}_{i}\right)}\right]}}}\nonumber 
\end{align}
and 
\begin{align}
 & C_{\left(M,m_{i},\tilde{m}_{i}\right),\left(N,n_{j},\tilde{n}_{j}\right),2}\nonumber \\
= & \sum_{2\le n_{k}\tilde{n}_{k}\le N-n_{j}\tilde{n}_{j}}S_{n_{j},\tilde{n}_{j}}\left(h_{1}+M,h_{2}+N-n_{j}\tilde{n}_{j}\right)\\
 & \times \resizebox{0.9\textwidth}{!}{\text{\ensuremath{\left[\frac{C_{(M,m_{i},\tilde{m}_{i}),\left(N-n_{j}\tilde{n}_{j},n_{k},\tilde{n}_{k}\right),2}}{c_{n_{k},\tilde{n}_{k}}\left(h_{2}+M-n_{i}\tilde{n}_{i}-n_{k}\tilde{n}_{k}\right)-c_{n_{j},\tilde{n}_{j}}\left(h_{2}+N-n_{j}\tilde{n}_{j}\right)}+\frac{C_{(M,m_{i},\tilde{m}_{i}),\left(N-n_{j}\tilde{n}_{j},n_{k},\tilde{n}_{k}\right),1}}{c_{m_{i},\tilde{m}_{i}}\left(h_{1}+M-m_{i}\tilde{m}_{i}\right)-c_{n_{j},\tilde{n}_{j}}\left(h_{2}+N-n_{j}\tilde{n}_{j}\right)}\right]}}}\nonumber 
\end{align}
\end{enumerate}
Using the above equations, we can compute $C_{\left(M,m_{i},\tilde{m}_{i}\right),\left(N,n_{j},\tilde{n}_{j}\right),1}$
and $C_{\left(M,m_{i},\tilde{m}_{i}\right),\left(N,n_{j},\tilde{n}_{j}\right),2}$
from small $\left(M,N\right)$ to larger $\left(M,N\right)$ up to
the order we want for $C_{M,N}$. The Mathematica code for this algorithm
is attached with this paper.

\section{Multi-Trace Contributions and Bulk Fields}
\label{app:GravMultiTrace}
In this appendix, we will discuss some of the differences between correlators of the proto-field $\phi$ and a full bulk field $\varphi$ that can be seen within perturbation theory in a low-energy EFT description.  In particular, consider a bulk theory with only $\varphi$ and gravity as low-energy fields:
\be
S = \int d^3 x \sqrt{g} \left( M_p R + \frac{1}{2} (\nabla \varphi)^2 - \frac{1}{2} m^2 \varphi^2-2\Lambda \right).
\ee
The bulk field $\varphi$ will contain contributions from multi-trace $\CO^n$ operators in the CFT due to gravitational interactions, even at tree-level.  This occurs because $\varphi$ is dressed by the bulk gravitational field $h_{\mu\nu}$, which in turn picks up contributions from multi-trace operators at the boundary.  It is easier to see this effect on $h_{\mu\nu}$, which is what we will calculate in this section. 

Fortunately,  the main content of the necessary computations were done in \cite{Howtozintegrals}. In the presence of two boundary scalar operators $\CO(x_1)$ and $\CO(x_3)$, the bulk field $h_{\mu\nu}$ is given by
\be
h_{\mu\nu} &=& |x_{13}|^{-2\Delta} \frac{1}{(w^2)^2} J_{\mu\lambda}(w) J_{\nu\rho}(w) I_{\lambda \rho}(w'- x_{13}') , \nn\\
I_{\mu\nu}(w'-x_{13}') &=& f(t) \left( \frac{1}{1-d} g'_{\mu\nu} + \frac{\delta_{0 \mu} \delta_{0\nu}}{w_0'^2} \right) + \dots,
\ee
where $f(t)$ is the solution to a differential equation to be presented below. The $\dots$ here are pure diffeomorphism terms, which depend on the choice of gauge; when $h_{\mu\nu}$ is part of an internal graviton line in a bulk correlator, e.g. $\< \varphi \CO \CO \CO\>$, these gauge-dependent  contributions vanish, and we will neglect them.  The notation of the above equation is that $w = (w_0, \vec{w})$ is the bulk position of $h_{\mu\nu}$, where $w_0$ is the radial Poincar\'e patch direction.  Translation invariance has been used to set the operators at $\CO(0)$ and $\CO(x_{13})$. The primes denote inversion,
\be
w' = \frac{w}{w^2} = \frac{w}{w_0^2 + \vec{w}^2}, \qquad
x_{13}' = \frac{x_{13}}{x_{13}^2},
\ee
and $g_{\mu\nu}' \equiv \frac{\delta_{\mu\nu}}{w_0'^2}$. 
The argument $t$ is
\be
t = \frac{w_0'^2}{w_0'^2 + (\vec{w}'-\vec{x}_{13}')^2} .
\ee
The polarization vectors are
\be
J_{\mu\nu}(w) = \delta_{\mu\nu} - 2 \frac{w_\mu w_\nu}{w^2} .
\ee
The boundary operator content of $h_{\mu\nu}$ can be read off by taking the limit $x_{13}\rightarrow 0$.  Using the OPE, we have
\be
\<h_{\mu\nu} \CO(x_1) \CO(x_3)\> \stackrel{x_1, x_3 \rightarrow 0}{\sim} \<h_{\mu\nu} \> x_{13}^{-2\Delta} + \< h_{\mu\nu} \CO^2\> + \dots
\label{eq:GravBulkOPE}
\ee
In the limit $x_1, x_3 \rightarrow 0$, we have $x_{13}' \rightarrow \infty$, so
\be
t \rightarrow \frac{w_0'^2}{x_{13}'^2} = \frac{x_{13}^2 w_0^2}{(w^2)^2} \sim 0
\ee.
The differential equation determining the  function $f(t)$ is
\be
4t (1-t) f'(t) -2 (d-2) f(t) = 2 \Delta t^\Delta,
\ee
and its solution in $d=2$ is
\be
f(t) = c_1+\frac{t^{\Delta } (\Delta +\Delta  t \, _2F_1(1,\Delta +1;\Delta +2;t)+1)}{2 (\Delta
   +1)}
   \ee
   where $c_1$ is fixed by an appropriate boundary condition.  We are interested in the limit $t\rightarrow 0$, where
   \be
   f(t) \sim c_1 + \frac{t^\Delta}{2} + \CO(t^{\Delta+1})
   \ee
Therefore,
\be
I_{\mu\nu} \approx \left( c_1 + \frac{1}{2}\left( \frac{x_{13}^{2\Delta} w_0^{2\Delta}}{(w^2)^{2\Delta}}\right) \right) \left( -\frac{\delta_{\mu\nu}}{w_0'^2} + \frac{\delta_{0\mu} \delta_{0\nu}}{w_0'^2} \right) 
\ee
and
\begin{small}
\begin{equation}
\< h_{\mu\nu} \CO(x_1) \CO(x_2) \> \approx |x_{13}|^{-2\Delta} \frac{1}{(w^2)^2} J_{\mu\lambda}(w) J_{\nu \rho}(w)\left( c_1 + \frac{1}{2}\left( \frac{x_{13}^{2\Delta} w_0^{2\Delta}}{(w^2)^{2\Delta}}\right) \right) \left( -\frac{\delta_{\lambda \rho}}{w_0'^2} + \frac{\delta_{0\lambda} \delta_{0\rho}}{w_0'^2} \right)
\end{equation}
\end{small}Comparing to (\ref{eq:GravBulkOPE}), we read off from the leading term at small $x_{13}$ that $c_1$ is the vacuum expectation value of the stress tensor, and so should be taken to vanish. From the subleading term, we obtain the double-trace $\CO^2$ content of the bulk gravitational field:
\be
\< h_{\mu\nu} | \CO^2(0)\> \approx \left( \frac{w_0}{(w^2)}\right)^{2\Delta} \left( \frac{- \delta_{\mu\nu} + J_{\mu 0} J_{\nu 0}}{w_0^{2}}  \right) .
\ee
Near the boundary, $w_0 \rightarrow 0$, this contributions vanishes,\footnote{For $\Delta<2$, the computation should be modified to use an alternate boundary condition for $\varphi$.} as it must since the bulk field $w_0^{-2} h_{\mu\nu}$ becomes the boundary stress tensor in this limit. However, it is clearly nonzero at $w_0>0$, and this effect implies that each bulk graviton that dresses a bulk field $\varphi$ brings (at least) two boundary $\CO$s along with it. For instance, in the tree-level diagram with one-graviton exchange for $\< \varphi \CO \CO \CO\>$, this effect produces a contribution from $\CO^3$ to the bulk field $\varphi$.  

\section{Non-Holomorphic Bulk Monodromy Method}
\label{app:nonchiralMonoBulk}

In this appendix, we describe how to apply the monodromy method to the full bulk block for $\< \phi_L \CO_L \CO_H \CO_H\>$. The analysis is complicated by the fact that both $T$ and $\bar{T}$ get contributions from the heavy background, and so both the holomorphic and the anti-holomorphic Schrodinger equations for $\psi, \bar{\psi}$ are difficult to solve and must be solved simultaneously.  We will again work only to first order in $h_L/c$, and show how to solve the monodromy equation order-by-order in a small $y$ expansion.  It would be much preferable to have a method to solve directly at any $y$. However, we will see that it is already somewhat nontrivial that the monodromy method contains enough information to solve for the bulk block, so the fact that it can be solved order-by-order in $y$ is a useful proof of principle.

The potential $T$ for the Schrodinger equation is again derived using the singular terms of the $T \times \CO$ and the $T \times \phi$ OPEs, for the latter see (\ref{eq:TphiOPE}).  Since we are expanding in $h_L/c$, we divide the potential $T(z)$ of the Schrodinger equation for $\psi$ into a ``heavy'' piece and a ``light'' piece:
\be
T(z) = T_H(z) + T_L(z),
\ee
where the heavy piece is just $T_H(z) = \frac{h_H}{z^2}$, the stress tensor in the heavy state background.  In the $T \times \phi$ OPE, at leading order in $h_L/c$, only $T_H$ contributes on the RHS of (\ref{eq:TphiOPE}), so ambiguities related to the singularities of $T_L(z)$ at the location of the light operators do not arise at this order.  We use conformal invariance again by demanding that $T(z)$ and $\bar{T}(\bar{z})$ decay like $z^{-4}, \bar{z}^{-4}$ at large $z,\bar{z}$.  In this section, it will be more convenient to work with the configuration
\be
z_1 = \infty, \quad z_2 = 0, \quad z_4 = 1 .
\ee
After making these simplifications and performing some straightforward but tedious manipulations, the light piece $T_L(z)$ is 
\be
T_L(z) &=& \frac{h_L}{z (1-z)^2} + \frac{1}{2} c_{y_3} y_3 \left( \frac{1}{z(1-z)} + \frac{1}{(z-z_3)^2} \right) + \frac{c_{z_3}(1-z_3) z_3}{(1-z) z (z-z_3)}  - \frac{c_{\bar{z}_3} y_3^2}{(z-z_3)^3}  \nn\\
 && - \frac{y_3^4 (c_{z_3} + c_{\bar{z}_3}  y_3^2 \frac{6 T_H(z_3)}{c}) \frac{6 \bar{T}_H(\bar{z}_3)}{c}}{(z-z_3)^2 (1-y_3^4 \frac{36}{c^2} T_H(z_3) \bar{T}_H(z_3))} + \dots , 
\ee
where $\dots$ are higher order in $h_L/c$, coming from the evaluation of $T(z_3), \bar{T}(\bar{z}_3)$ inside the $T \times \phi$ OPE.
As before, $c_X \equiv \partial_X g$.  At zeroth order in $h_L/c$, only $T_H$ contributes to the Schrodinger euqation, and the solutions for $\psi$ are
\be
\psi^{(1)}(z) = z^{\frac{1-\alpha_H}{2}} , \qquad 
\psi^{(2)}(z) = z^{\frac{1+\alpha_H}{2}}.
\ee
At next order, we apply the method of separation of variables, which ultimately gives the monodromy matrix $M$ as the residues of a matrix $m_{ij}$ 
\be
m_{ij} = \frac{T_L(z)}{\psi'^{(2)}(z) \psi^{(1)}(z) - \psi^{(2)}(z) \psi'^{(1)}(z)} \psi^{(i)}(z) \tilde{\psi}^{(j)}(z),
\ee
in terms of which $M$ is just
\be
M_{ij} &=& 2 \pi i \left( {\rm res}_{z \rightarrow 1} m_{ij} + {\rm res}_{z\rightarrow z_3} m_{ij} \right).
\ee
The diagonal components vanish, and the off-diagonal components are
\begin{align}
M_{12} &=\resizebox{0.91\textwidth}{!}{\text{\ensuremath{
 -\frac{i \pi  z_3^{-\alpha _H-1} \left(y_3^2 \left(\alpha _H-1\right) \alpha _H
   c_{\bar{z}_3}+y_3 z_3 c_{y_3} \left(\alpha _H+z_3^{\alpha _H}-1\right)+2 z_3^2
   c_{z_3} \left(z_3^{\alpha _H}-1\right)+2 h_L \alpha _H z_3^{\alpha
   _H+1}\right)}{\alpha _H}}}}\nn\\
   M_{21} &=  M_{12} ( \alpha_H \rightarrow - \alpha_H)
\end{align}
The eigenvalues vanishing requires $M_{12} M_{21}=0$, i.e. either $M_{12}$ or $M_{21}$ must vanish.

To solve for the ``action'' $g$ order-by-order in $y_3$, we take
\be
\frac{c}{6} g = 2 h_L \log(y_3) + h_L \sum_{n=0}^{\infty} y_3^{2n} g_{2n}(z_3, \bar{z}_3)
\ee
If we demand that $M_{12}$ and $\bar{M}_{12}$ vanish, we find two differential equations for $g_0$: 
\be
g_0{}^{(0,1)}\left(z_3,\bar{z}_3\right)&=&\frac{-\alpha _H \bar{z}_3^{\alpha
   _H}-\bar{z}_3^{\alpha _H}-\alpha _H+1}{\bar{z}_3 \left(\bar{z}_3^{\alpha _H}-1\right)} \nn\\
    g_0{}^{(1,0)}\left(z_3,\bar{z}_3\right)&=&\frac{-\alpha _H-\alpha _H z_3^{\alpha
   _H}-z_3^{\alpha _H}+1}{z_3 \left(z_3^{\alpha _H}-1\right)}
   \ee
   These are solved by 
   \be
   g_0(z_3, \bar{z}_3) &=& 2 \log \left(\frac{\alpha _H}{\left(z_3^{\alpha _H}-1\right) \left(\bar{z}_3^{\alpha
   _H}-1\right)}\right)+\left(\alpha _H-1\right) \log \left(z_3 \bar{z}_3\right)
   \ee
which just reproduces the boundary block in this large $c$, small $h_L/c$ limit.

  Next, we solve for $g_2$.  At this order, the equations we find for $g_2$ reduce to
  \begin{smaller}
   \be
   g_2{}^{(1,0)}\left(z_3,\bar{z}_3\right)&=&\frac{\left(\alpha _H-1\right) \alpha _H
   \left(\left(\alpha _H+1\right) \bar{z}_3^{\alpha _H}+\alpha _H-1\right)-2 z_3
   \bar{z}_3 \left(\alpha _H+z_3^{\alpha _H}-1\right) g_2\left(z_3,\bar{z}_3\right)
   \left(\bar{z}_3^{\alpha _H}-1\right)}{2 z_3^2 \bar{z}_3 \left(z_3^{\alpha _H}-1\right)
   \left(\bar{z}_3^{\alpha _H}-1\right)} \nn\\
g_2{}^{(0,1)}\left(z_3,\bar{z}_3\right)&=&\frac{\left(\alpha _H-1\right) \alpha _H
   \left(\alpha _H+\left(\alpha _H+1\right) z_3^{\alpha _H}-1\right)-2 z_3 \bar{z}_3
   \left(z_3^{\alpha _H}-1\right) g_2\left(z_3,\bar{z}_3\right) \left(\bar{z}_3^{\alpha
   _H}+\alpha _H-1\right)}{2 z_3 \bar{z}_3^2 \left(z_3^{\alpha _H}-1\right)
   \left(\bar{z}_3^{\alpha _H}-1\right)}    \nn\\
   \ee
   \end{smaller}
Each of these equations can be viewed as an ordinary differential equation, with $z_3$ or $\bar{z}_3$ treated as a constant.  Solving these ODEs, one therefore obtains two equations for $g_2$, one with an integration function of $z_3$, one with an integration function of $\bar{z}_3$:
   \be
   g_2\left(z_3,\bar{z}_3\right)&=&-\frac{2 \bar{z}_3 c_1\left(\bar{z}_3\right) z_3^{\alpha
   _H} \left(\bar{z}_3^{\alpha _H}-1\right)+\left(\alpha _H-1\right) \left(\left(\alpha
   _H+1\right) \bar{z}_3^{\alpha _H}+\alpha _H-1\right)}{2 z_3 \bar{z}_3
   \left(z_3^{\alpha _H}-1\right) \left(\bar{z}_3^{\alpha _H}-1\right)} \nn\\
   g_2\left(z_3,\bar{z}_3\right)&=&-\frac{2 z_3 c_1\left(z_3\right) \left(z_3^{\alpha
   _H}-1\right) \bar{z}_3^{\alpha _H}+\left(\alpha _H-1\right) \left(\alpha
   _H+\left(\alpha _H+1\right) z_3^{\alpha _H}-1\right)}{2 z_3 \bar{z}_3
   \left(z_3^{\alpha _H}-1\right) \left(\bar{z}_3^{\alpha _H}-1\right)} \nn\\
   \ee
   We can solve for $c_1(\bar{z}_3)$ in terms of $c_1(z_3)$:
   \be
   c_1\left(\bar{z}_3\right)=\frac{z_3^{-\alpha _H} \left(2 c_1\left(z_3\right) z_3^{\alpha
   _H+1} \bar{z}_3^{\alpha _H}-2 z_3 c_1\left(z_3\right) \bar{z}_3^{\alpha _H}-\alpha
   _H^2 \bar{z}_3^{\alpha _H}+\bar{z}_3^{\alpha _H}+\alpha _H^2 z_3^{\alpha
   _H}-z_3^{\alpha _H}\right)}{2 \bar{z}_3 \left(\bar{z}_3^{\alpha _H}-1\right)} \nn\\
   \ee
   Since the RHS cannot depend on $z_3$, we can take $z_3$ to be any value we want.  Naively, we can just take $z_3=1$, but this is too fast since this causes $c_1(z_3)$ to be multiplied by $z_3^{\alpha_H+1} -1 \rightarrow 0$, and in the correct answer $c_1$ has a singularity at $z_3 =1$.  In the correct answer, the singularity cancels the zero, but to extract the correct answer will define a new function as
   \be
   c_1(t) \equiv \frac{b_1(t) }{t(t^{\alpha_H}-1)}
   \ee
Now, $b_1(t)$ is regular at $t=1$.  The above equation for $c_1(\bar{z}_3)$ in terms of $c_1(z_3)$ becomes
\be
b_1\left(\bar{z}_3\right)=\frac{1}{2} z_3^{-\alpha _H} \left(2 b_1\left(z_3\right)
   \bar{z}_3^{\alpha _H}+\left(\alpha _H^2-1\right) \left(z_3^{\alpha
   _H}-\bar{z}_3^{\alpha _H}\right)\right)
   \ee
Setting $z_3=1$, we obtain an equation for $b_1(\bar{z}_3)$ in terms of $b_1(1)$.  The solution is
\be
b_1(t) = \frac{\beta_1 t^{\alpha_H}  + \alpha_H^2-1}{2}
\ee
where $\beta_1$ is an undetermined integration constant. 

We fix $\beta_1$ by substituting back into $g_2(z_3, \bar{z_3})$ and demanding that the result be a holomorphic times antiholomorphic function.  We compute
\be
\partial_{z_3} \partial_{\bar{z}_3} \log g_2(z_3, \bar{z}_3) &=&\frac{\left(\alpha _H-1\right){}^2 \alpha _H^2 \left(\left(\alpha
   _H+1\right){}^2-\beta_1\right) z_3^{\alpha _H-1} \bar{z}_3^{\alpha
   _H-1}}{\left(z_3^{\alpha _H} \left(\beta_1 \bar{z}_3^{\alpha _H}+\alpha
   _H^2-1\right)-\left(\alpha _H-1\right) \left(\left(\alpha _H+1\right)
   \bar{z}_3^{\alpha _H}+\alpha _H-1\right)\right){}^2}\nn\\
\ee
and therefore
\be
\beta_1 = (1+\alpha_H)^2
\ee
This procedure can be continued recursively to any order in $y$, and we have explicitly checked that it works up to and including $g_4$.

\bibliographystyle{utphys} 
\bibliography{VirasoroBib}

\providecommand{\href}[2]{#2}\begingroup\raggedright\begin{thebibliography}{10}

\bibitem{HartmanLargeC}
T.~Hartman, ``{Entanglement Entropy at Large Central Charge},''
\href{http://arxiv.org/abs/1303.6955}{{\ttfamily arXiv:1303.6955 [hep-th]}}.

\bibitem{DongGravityRenyi}
T.~Barrella, X.~Dong, S.~A. Hartnoll, and V.~L. Martin, ``{Holographic
  entanglement beyond classical gravity},''
  \href{http://dx.doi.org/10.1007/JHEP09(2013)109}{{\em JHEP} {\bfseries 1309}
  (2013) 109},
\href{http://arxiv.org/abs/1306.4682}{{\ttfamily arXiv:1306.4682 [hep-th]}}.

\bibitem{Fitzpatrick:2014vua}
A.~L. Fitzpatrick, J.~Kaplan, and M.~T. Walters, ``{Universality of
  Long-Distance AdS Physics from the CFT Bootstrap},''
  \href{http://dx.doi.org/10.1007/JHEP08(2014)145}{{\em JHEP} {\bfseries 1408}
  (2014) 145},
\href{http://arxiv.org/abs/1403.6829}{{\ttfamily arXiv:1403.6829 [hep-th]}}.

\bibitem{Roberts:2014ifa}
D.~A. Roberts and D.~Stanford, ``{Two-dimensional conformal field theory and
  the butterfly effect},''
\href{http://arxiv.org/abs/1412.5123}{{\ttfamily arXiv:1412.5123 [hep-th]}}.

\bibitem{Fitzpatrick:2015zha}
A.~L. Fitzpatrick, J.~Kaplan, and M.~T. Walters, ``{Virasoro Conformal Blocks
  and Thermality from Classical Background Fields},''
\href{http://arxiv.org/abs/1501.05315}{{\ttfamily arXiv:1501.05315 [hep-th]}}.

\bibitem{Fitzpatrick:2015foa}
A.~L. Fitzpatrick, J.~Kaplan, M.~T. Walters, and J.~Wang, ``{Hawking from
  Catalan},''
\href{http://arxiv.org/abs/1510.00014}{{\ttfamily arXiv:1510.00014 [hep-th]}}.

\bibitem{Anous:2016kss}
T.~Anous, T.~Hartman, A.~Rovai, and J.~Sonner, ``{Black Hole Collapse in the
  1/c Expansion},''
\href{http://arxiv.org/abs/1603.04856}{{\ttfamily arXiv:1603.04856 [hep-th]}}.

\bibitem{Asplund:2015eha}
C.~T. Asplund, A.~Bernamonti, F.~Galli, and T.~Hartman, ``{Entanglement
  Scrambling in 2d Conformal Field Theory},''
  \href{http://dx.doi.org/10.1007/JHEP09(2015)110}{{\em JHEP} {\bfseries 09}
  (2015) 110},
\href{http://arxiv.org/abs/1506.03772}{{\ttfamily arXiv:1506.03772 [hep-th]}}.

\bibitem{Asplund:2014coa}
C.~T. Asplund, A.~Bernamonti, F.~Galli, and T.~Hartman, ``{Holographic
  Entanglement Entropy from 2d CFT: Heavy States and Local Quenches},''
  \href{http://dx.doi.org/10.1007/JHEP02(2015)171}{{\em JHEP} {\bfseries 02}
  (2015) 171},
\href{http://arxiv.org/abs/1410.1392}{{\ttfamily arXiv:1410.1392 [hep-th]}}.

\bibitem{TakayanagiExcitedStates}
P.~Caputa, J.~Simon, A.~Stikonas, and T.~Takayanagi, ``{Quantum Entanglement of
  Localized Excited States at Finite Temperature},''
\href{http://arxiv.org/abs/1410.2287}{{\ttfamily arXiv:1410.2287 [hep-th]}}.

\bibitem{Beccaria}
M.~Beccaria, A.~Fachechi, and G.~Macorini, ``{Virasoro vacuum block at
  next-to-leading order in the heavy-light limit},''
\href{http://arxiv.org/abs/1511.05452}{{\ttfamily arXiv:1511.05452 [hep-th]}}.

\bibitem{KrausBlocks}
E.~Hijano, P.~Kraus, and R.~Snively, ``{Worldline approach to semi-classical
  conformal blocks},''
\href{http://arxiv.org/abs/1501.02260}{{\ttfamily arXiv:1501.02260 [hep-th]}}.

\bibitem{Hijano:2015qja}
E.~Hijano, P.~Kraus, E.~Perlmutter, and R.~Snively, ``{Semiclassical Virasoro
  Blocks from AdS$_3$ Gravity},''
\href{http://arxiv.org/abs/1508.04987}{{\ttfamily arXiv:1508.04987 [hep-th]}}.

\bibitem{Alkalaev:2015wia}
K.~B. Alkalaev and V.~A. Belavin, ``{Classical conformal blocks via AdS/CFT
  correspondence},'' \href{http://dx.doi.org/10.1007/JHEP08(2015)049}{{\em
  JHEP} {\bfseries 08} (2015) 049},
\href{http://arxiv.org/abs/1504.05943}{{\ttfamily arXiv:1504.05943 [hep-th]}}.

\bibitem{Alkalaev:2015lca}
K.~B. Alkalaev and V.~A. Belavin, ``{Monodromic vs geodesic computation of
  Virasoro classical conformal blocks},''
\href{http://arxiv.org/abs/1510.06685}{{\ttfamily arXiv:1510.06685 [hep-th]}}.

\bibitem{Chen:2016dfb}
B.~Chen, J.-q. Wu, and J.-j. Zhang, ``{Holographic Description of 2D Conformal
  Block in Semi-classical Limit},''
\href{http://arxiv.org/abs/1609.00801}{{\ttfamily arXiv:1609.00801 [hep-th]}}.

\bibitem{Lashkari:2017hwq}
N.~Lashkari, A.~Dymarsky, and H.~Liu, ``{Universality of Quantum Information in
  Chaotic CFTs},'' \href{http://dx.doi.org/10.1007/JHEP03(2018)070}{{\em JHEP}
  {\bfseries 03} (2018) 070},
\href{http://arxiv.org/abs/1710.10458}{{\ttfamily arXiv:1710.10458 [hep-th]}}.

\bibitem{Maxfield:2017rkn}
H.~Maxfield, ``{A view of the bulk from the worldline},''
\href{http://arxiv.org/abs/1712.00885}{{\ttfamily arXiv:1712.00885 [hep-th]}}.

\bibitem{Kusuki:2018nms}
Y.~Kusuki, ``{Large $c$ Virasoro Blocks from Monodromy Method beyond Known
  Limits},'' \href{http://dx.doi.org/10.1007/JHEP08(2018)161}{{\em JHEP}
  {\bfseries 08} (2018) 161},
\href{http://arxiv.org/abs/1806.04352}{{\ttfamily arXiv:1806.04352 [hep-th]}}.

\bibitem{Hikida:2018dxe}
Y.~Hikida and T.~Uetoko, ``{Conformal blocks from Wilson lines with loop
  corrections},'' \href{http://dx.doi.org/10.1103/PhysRevD.97.086014}{{\em
  Phys. Rev.} {\bfseries D97} no.~8, (2018) 086014},
\href{http://arxiv.org/abs/1801.08549}{{\ttfamily arXiv:1801.08549 [hep-th]}}.

\bibitem{Kraus:2017kyl}
P.~Kraus, A.~Sivaramakrishnan, and R.~Snively, ``{Black holes from CFT:
  Universality of correlators at large c},''
  \href{http://dx.doi.org/10.1007/JHEP08(2017)084}{{\em JHEP} {\bfseries 08}
  (2017) 084},
\href{http://arxiv.org/abs/1706.00771}{{\ttfamily arXiv:1706.00771 [hep-th]}}.

\bibitem{Cotler:2018zff}
J.~Cotler and K.~Jensen, ``{A theory of reparameterizations for AdS$_3$
  gravity},''
\href{http://arxiv.org/abs/1808.03263}{{\ttfamily arXiv:1808.03263 [hep-th]}}.

\bibitem{Fitzpatrick:2016ive}
A.~L. Fitzpatrick, J.~Kaplan, D.~Li, and J.~Wang, ``{On Information Loss in
  AdS$_3$/CFT$_2$},''
\href{http://arxiv.org/abs/1603.08925}{{\ttfamily arXiv:1603.08925 [hep-th]}}.

\bibitem{Chen:2017yze}
H.~Chen, C.~Hussong, J.~Kaplan, and D.~Li, ``{A Numerical Approach to Virasoro
  Blocks and the Information Paradox},''
  \href{http://dx.doi.org/10.1007/JHEP09(2017)102}{{\em JHEP} {\bfseries 09}
  (2017) 102},
\href{http://arxiv.org/abs/1703.09727}{{\ttfamily arXiv:1703.09727 [hep-th]}}.

\bibitem{Kusuki:2018wcv}
Y.~Kusuki, ``{New Properties of Large-$c$ Conformal Blocks from Recursion
  Relation},'' \href{http://dx.doi.org/10.1007/JHEP07(2018)010}{{\em JHEP}
  {\bfseries 07} (2018) 010},
\href{http://arxiv.org/abs/1804.06171}{{\ttfamily arXiv:1804.06171 [hep-th]}}.

\bibitem{Fitzpatrick:2016mjq}
A.~Liam~Fitzpatrick and J.~Kaplan, ``{On the Late-Time Behavior of Virasoro
  Blocks and a Classification of Semiclassical Saddles},''
\href{http://arxiv.org/abs/1609.07153}{{\ttfamily arXiv:1609.07153 [hep-th]}}.

\bibitem{Kusuki:2018wpa}
Y.~Kusuki, ``{Light Cone Bootstrap in General 2D CFTs \&amp; Entanglement from
  Light Cone Singularity},''
\href{http://arxiv.org/abs/1810.01335}{{\ttfamily arXiv:1810.01335 [hep-th]}}.

\bibitem{Kraus:2018zrn}
P.~Kraus, A.~Sivaramakrishnan, and R.~Snively, ``{Late time Wilson lines},''
\href{http://arxiv.org/abs/1810.01439}{{\ttfamily arXiv:1810.01439 [hep-th]}}.

\bibitem{Anand:2017dav}
N.~Anand, H.~Chen, A.~L. Fitzpatrick, J.~Kaplan, and D.~Li, ``{An Exact
  Operator That Knows Its Location},''
\href{http://arxiv.org/abs/1708.04246}{{\ttfamily arXiv:1708.04246 [hep-th]}}.

\bibitem{Chen:2017dnl}
H.~Chen, A.~L. Fitzpatrick, J.~Kaplan, and D.~Li, ``{The AdS$_{3}$ propagator
  and the fate of locality},''
  \href{http://dx.doi.org/10.1007/JHEP04(2018)075}{{\em JHEP} {\bfseries 04}
  (2018) 075},
\href{http://arxiv.org/abs/1712.02351}{{\ttfamily arXiv:1712.02351 [hep-th]}}.

\bibitem{Balasubramanian:2007qv}
V.~Balasubramanian, B.~Czech, V.~E. Hubeny, K.~Larjo, M.~Rangamani, and
  J.~Simon, ``{Typicality versus thermality: An Analytic distinction},''
  \href{http://dx.doi.org/10.1007/s10714-008-0606-8}{{\em Gen. Rel. Grav.}
  {\bfseries 40} (2008) 1863--1890},
\href{http://arxiv.org/abs/hep-th/0701122}{{\ttfamily arXiv:hep-th/0701122
  [hep-th]}}.

\bibitem{Liu}
H.~Liu, ``{Scattering in anti-de Sitter space and operator product
  expansion},'' \href{http://dx.doi.org/10.1103/PhysRevD.60.106005}{{\em Phys.
  Rev.} {\bfseries D60} (1999) 106005},
\href{http://arxiv.org/abs/hep-th/9811152}{{\ttfamily arXiv:hep-th/9811152}}.

\bibitem{Heemskerk:2009pn}
I.~Heemskerk, J.~Penedones, J.~Polchinski, and J.~Sully, ``{Holography from
  Conformal Field Theory},''
  \href{http://dx.doi.org/10.1088/1126-6708/2009/10/079}{{\em JHEP} {\bfseries
  0910} (2009) 079},
\href{http://arxiv.org/abs/0907.0151}{{\ttfamily arXiv:0907.0151 [hep-th]}}.

\bibitem{Unitarity}
A.~L. Fitzpatrick and J.~Kaplan, ``{Unitarity and the Holographic S-Matrix},''
  \href{http://dx.doi.org/10.1007/JHEP10(2012)032}{{\em JHEP} {\bfseries 1210}
  (2012) 032},
\href{http://arxiv.org/abs/1112.4845}{{\ttfamily arXiv:1112.4845 [hep-th]}}.

\bibitem{Pappadopulo:2012jk}
D.~Pappadopulo, S.~Rychkov, J.~Espin, and R.~Rattazzi, ``{OPE Convergence in
  Conformal Field Theory},''
\href{http://arxiv.org/abs/1208.6449}{{\ttfamily arXiv:1208.6449 [hep-th]}}.

\bibitem{Hartman:2014oaa}
T.~Hartman, C.~A. Keller, and B.~Stoica, ``{Universal Spectrum of 2d Conformal
  Field Theory in the Large c Limit},''
  \href{http://dx.doi.org/10.1007/JHEP09(2014)118}{{\em JHEP} {\bfseries 1409}
  (2014) 118},
\href{http://arxiv.org/abs/1405.5137}{{\ttfamily arXiv:1405.5137 [hep-th]}}.

\bibitem{Cardy:2017qhl}
J.~Cardy, A.~Maloney, and H.~Maxfield, ``{A new handle on three-point
  coefficients: OPE asymptotics from genus two modular invariance},''
  \href{http://dx.doi.org/10.1007/JHEP10(2017)136}{{\em JHEP} {\bfseries 10}
  (2017) 136},
\href{http://arxiv.org/abs/1705.05855}{{\ttfamily arXiv:1705.05855 [hep-th]}}.

\bibitem{JP}
I.~Heemskerk, J.~Penedones, J.~Polchinski, and J.~Sully, ``{Holography from
  Conformal Field Theory},''
  \href{http://dx.doi.org/10.1088/1126-6708/2009/10/079}{{\em JHEP} {\bfseries
  10} (2009) 079},
\href{http://arxiv.org/abs/0907.0151}{{\ttfamily arXiv:0907.0151 [hep-th]}}.

\bibitem{AdSfromCFT}
A.~L. Fitzpatrick and J.~Kaplan, ``{AdS Field Theory from Conformal Field
  Theory},'' \href{http://dx.doi.org/10.1007/JHEP02(2013)054}{{\em JHEP}
  {\bfseries 1302} (2013) 054},
\href{http://arxiv.org/abs/1208.0337}{{\ttfamily arXiv:1208.0337 [hep-th]}}.

\bibitem{Hamilton:2006az}
A.~Hamilton, D.~N. Kabat, G.~Lifschytz, and D.~A. Lowe, ``{Holographic
  representation of local bulk operators},''
  \href{http://dx.doi.org/10.1103/PhysRevD.74.066009}{{\em Phys. Rev.}
  {\bfseries D74} (2006) 066009},
\href{http://arxiv.org/abs/hep-th/0606141}{{\ttfamily arXiv:hep-th/0606141}}.

\bibitem{Kabat:2011rz}
D.~Kabat, G.~Lifschytz, and D.~A. Lowe, ``{Constructing local bulk observables
  in interacting AdS/CFT},''
  \href{http://dx.doi.org/10.1103/PhysRevD.83.106009}{{\em Phys. Rev.}
  {\bfseries D83} (2011) 106009},
\href{http://arxiv.org/abs/1102.2910}{{\ttfamily arXiv:1102.2910 [hep-th]}}.

\bibitem{Kabat:2016zzr}
D.~Kabat and G.~Lifschytz, ``{Locality, bulk equations of motion and the
  conformal bootstrap},'' \href{http://dx.doi.org/10.1007/JHEP10(2016)091}{{\em
  JHEP} {\bfseries 10} (2016) 091},
\href{http://arxiv.org/abs/1603.06800}{{\ttfamily arXiv:1603.06800 [hep-th]}}.

\bibitem{ZamolodchikovRecursion}
A.~Zamolodchikov, ``{Conformal Symmetry in Two-Dimensions: An Explicit
  Recurrence Formula for the Conformal Partial Wave Amplitude},''
\href{http://dx.doi.org/10.1007/BF01214585}{{\em Commun.Math.Phys.} {\bfseries
  96} (1984) 419--422}.

\bibitem{Zamolodchikovq}
A.~Zamolodchikov, ``{Conformal Symmetry in Two-dimensional Space: Recursion
  Representation of the Conformal Block},''
{\em Teoreticheskaya i Matematicheskaya Fizika} {\bfseries 73} (1987) 103--110.

\bibitem{Czech:2016xec}
B.~Czech, L.~Lamprou, S.~McCandlish, B.~Mosk, and J.~Sully, ``{A Stereoscopic
  Look into the Bulk},'' \href{http://dx.doi.org/10.1007/JHEP07(2016)129}{{\em
  JHEP} {\bfseries 07} (2016) 129},
\href{http://arxiv.org/abs/1604.03110}{{\ttfamily arXiv:1604.03110 [hep-th]}}.

\bibitem{Fitzpatrick:2016mtp}
A.~L. Fitzpatrick, J.~Kaplan, D.~Li, and J.~Wang, ``{Exact Virasoro Blocks from
  Wilson Lines and Background-Independent Operators},''
\href{http://arxiv.org/abs/1612.06385}{{\ttfamily arXiv:1612.06385 [hep-th]}}.

\bibitem{Besken:2018zro}
M.~Besken, E.~D'Hoker, A.~Hegde, and P.~Kraus, ``{Renormalization of
  gravitational Wilson lines},''
\href{http://arxiv.org/abs/1810.00766}{{\ttfamily arXiv:1810.00766 [hep-th]}}.

\bibitem{KeskiVakkuri:1998nw}
E.~Keski-Vakkuri, ``{Bulk and boundary dynamics in BTZ black holes},''
  \href{http://dx.doi.org/10.1103/PhysRevD.59.104001}{{\em Phys. Rev.}
  {\bfseries D59} (1999) 104001},
\href{http://arxiv.org/abs/hep-th/9808037}{{\ttfamily arXiv:hep-th/9808037
  [hep-th]}}.

\bibitem{Banks:1998dd}
T.~Banks, M.~R. Douglas, G.~T. Horowitz, and E.~J. Martinec, ``{AdS dynamics
  from conformal field theory},''
\href{http://arxiv.org/abs/hep-th/9808016}{{\ttfamily arXiv:hep-th/9808016
  [hep-th]}}.

\bibitem{Bena:1999jv}
I.~Bena, ``{On the construction of local fields in the bulk of AdS(5) and other
  spaces},'' \href{http://dx.doi.org/10.1103/PhysRevD.62.066007}{{\em Phys.
  Rev.} {\bfseries D62} (2000) 066007},
\href{http://arxiv.org/abs/hep-th/9905186}{{\ttfamily arXiv:hep-th/9905186
  [hep-th]}}.

\bibitem{Verlinde:2015qfa}
H.~Verlinde, ``{Poking Holes in AdS/CFT: Bulk Fields from Boundary States},''
\href{http://arxiv.org/abs/1505.05069}{{\ttfamily arXiv:1505.05069 [hep-th]}}.

\bibitem{Lewkowycz:2016ukf}
A.~Lewkowycz, G.~J. Turiaci, and H.~Verlinde, ``{A CFT Perspective on
  Gravitational Dressing and Bulk Locality},''
\href{http://arxiv.org/abs/1608.08977}{{\ttfamily arXiv:1608.08977 [hep-th]}}.

\bibitem{Roberts:2012aq}
M.~M. Roberts, ``{Time evolution of entanglement entropy from a pulse},''
  \href{http://dx.doi.org/10.1007/JHEP12(2012)027}{{\em JHEP} {\bfseries 12}
  (2012) 027},
\href{http://arxiv.org/abs/1204.1982}{{\ttfamily arXiv:1204.1982 [hep-th]}}.

\bibitem{HarlowLiouville}
D.~Harlow, J.~Maltz, and E.~Witten, ``{Analytic Continuation of Liouville
  Theory},'' \href{http://dx.doi.org/10.1007/JHEP12(2011)071}{{\em JHEP}
  {\bfseries 1112} (2011) 071},
\href{http://arxiv.org/abs/1108.4417}{{\ttfamily arXiv:1108.4417 [hep-th]}}.

\bibitem{Maldacena:2015iua}
J.~Maldacena, D.~Simmons-Duffin, and A.~Zhiboedov, ``{Looking for a bulk
  point},''
\href{http://arxiv.org/abs/1509.03612}{{\ttfamily arXiv:1509.03612 [hep-th]}}.

\bibitem{Perlmutter:2015iya}
E.~Perlmutter, ``{Virasoro conformal blocks in closed form},''
  \href{http://dx.doi.org/10.1007/JHEP08(2015)088}{{\em JHEP} {\bfseries 08}
  (2015) 088},
\href{http://arxiv.org/abs/1502.07742}{{\ttfamily arXiv:1502.07742 [hep-th]}}.

\bibitem{Cho:2017oxl}
M.~Cho, S.~Collier, and X.~Yin, ``{Recursive Representations of Arbitrary
  Virasoro Conformal Blocks},''
\href{http://arxiv.org/abs/1703.09805}{{\ttfamily arXiv:1703.09805 [hep-th]}}.

\bibitem{Alkalaev:2015fbw}
K.~B. Alkalaev and V.~A. Belavin, ``{From global to heavy-light: 5-point
  conformal blocks},'' \href{http://dx.doi.org/10.1007/JHEP03(2016)184}{{\em
  JHEP} {\bfseries 03} (2016) 184},
\href{http://arxiv.org/abs/1512.07627}{{\ttfamily arXiv:1512.07627 [hep-th]}}.

\bibitem{Fitzpatrick:2015dlt}
A.~L. Fitzpatrick and J.~Kaplan, ``{Conformal Blocks Beyond the Semi-Classical
  Limit},''
\href{http://arxiv.org/abs/1512.03052}{{\ttfamily arXiv:1512.03052 [hep-th]}}.

\bibitem{Fitzpatrick:2012yx}
A.~L. Fitzpatrick, J.~Kaplan, D.~Poland, and D.~Simmons-Duffin, ``{The Analytic
  Bootstrap and AdS Superhorizon Locality},''
  \href{http://dx.doi.org/10.1007/JHEP12(2013)004}{{\em JHEP} {\bfseries 1312}
  (2013) 004},
\href{http://arxiv.org/abs/1212.3616}{{\ttfamily arXiv:1212.3616 [hep-th]}}.

\bibitem{KomargodskiZhiboedov}
Z.~Komargodski and A.~Zhiboedov, ``{Convexity and Liberation at Large Spin},''
  \href{http://dx.doi.org/10.1007/JHEP11(2013)140}{{\em JHEP} {\bfseries 1311}
  (2013) 140},
\href{http://arxiv.org/abs/1212.4103}{{\ttfamily arXiv:1212.4103 [hep-th]}}.

\bibitem{Mathur:2009hf}
S.~D. Mathur, ``{The Information paradox: A Pedagogical introduction},''
  \href{http://dx.doi.org/10.1088/0264-9381/26/22/224001}{{\em Class. Quant.
  Grav.} {\bfseries 26} (2009) 224001},
\href{http://arxiv.org/abs/0909.1038}{{\ttfamily arXiv:0909.1038 [hep-th]}}.

\bibitem{Almheiri:2012rt}
A.~Almheiri, D.~Marolf, J.~Polchinski, and J.~Sully, ``{Black Holes:
  Complementarity or Firewalls?},''
\href{http://arxiv.org/abs/1207.3123}{{\ttfamily arXiv:1207.3123 [hep-th]}}.

\bibitem{Howtozintegrals}
E.~D'Hoker, D.~Z. Freedman, and L.~Rastelli, ``{AdS/CFT 4-point functions: How
  to succeed at z-integrals without really trying},''
  \href{http://dx.doi.org/10.1016/S0550-3213(99)00526-X}{{\em Nucl. Phys.}
  {\bfseries B562} (1999) 395--411},
\href{http://arxiv.org/abs/hep-th/9905049}{{\ttfamily arXiv:hep-th/9905049}}.

\end{thebibliography}\endgroup

\end{document}